\documentclass[authoryear,3p,times,preprint,review,12pt,fleqn]{elsarticle}

\usepackage{amsmath,natbib,mathrsfs,stmaryrd,titletoc,subfigure}
\usepackage[ruled,vlined]{algorithm2e}
\usepackage[scaled=0.92]{helvet}	

\usepackage[pdftex,
CJKbookmarks=false, bookmarksnumbered=true, bookmarksopen=true,
colorlinks=true, 
citecolor=blue, linkcolor=red, anchorcolor=green,
urlcolor=blue]{hyperref}

\usepackage[capitalise]{cleveref}
\usepackage{tabularx}
\usepackage{booktabs}
\usepackage[usenames]{color}
\usepackage[textsize=tiny]{todonotes}

\usepackage[section]{placeins}

\newcommand{\bfsigma}{\boldsymbol{\sigma}}
\newcommand{\bfepsilon}{\boldsymbol{\epsilon}}

\newcommand{\td}{\text{d}}

\newcommand{\trace}{\text{tr}}

\newcommand{\itw}{\mathrm{\mathit{w}}}

\numberwithin{equation}{section}

\newcounter{remark}[section]
\numberwithin{remark}{section}
\newcommand{\remark}{\noindent \textbf{Remark} \refstepcounter{remark} \textbf{\theremark} \;}

\SetAlCapSkip{1em} 
\crefname{figure}{Figure}{Figure}
\crefname{equation}{Eq.}{Eqs.}
\definecolor{darkgray}{rgb}{0.95,0.95,0.95}

\begin{document}

\begin{frontmatter}



\title{Unified analysis of phase-field models for cohesive fracture}



\author{Jian-Ying Wu\corref{cor1}}
\cortext[cor1]{Tel.: (+86) 20-87112787} \ead{jywu@scut.edu.cn}
\address{State Key Laboratory of Subtropical Building and Urban Science, South China University of Technology, 510641 Guangzhou, China.}

%
%

\begin{abstract}

Due to the great success achieved in the modeling of brittle fracture, the phase-field approach to cohesive fracture becomes popular. Despite the recent noteworthy contributions, a unified theoretical framework is lacking, imposing difficulty in selecting proper models and for further improvement. To this end, we address in this review unified analysis of phase-field models for cohesive fracture. Aiming to regularize the \cite{Barenblatt1959} cohesive zone model, all the discussed models are distinguished by three characteristic functions, i.e., the geometric function dictating the crack profile, the degradation function for the constitutive relation and the dissipation function defining the crack driving force. The latter two functions coincide in the associated formulation, while in the non-associated one they are designed to be different. Distinct from the counterpart for brittle fracture, in the phase-field model for cohesive fracture the regularization length parameter has to be properly incorporated into the dissipation and/or degradation functions such that the failure strength and traction--separation softening curve are both  well-defined. Moreover, the resulting crack bandwidth needs to be non-decreasing during failure in order that imposition of the crack irreversibility condition does not affect the anticipated traction--separation law (TSL). With a truncated degradation function that is proportional to the length parameter, the \cite{CFI2016} model and the latter improved versions can deal with crack nucleation only in the vanishing limit and capture cohesive fracture only with a particular TSL.  Owing to a length scale dependent degradation function of rational fraction, these deficiencies are largely overcome in the phase-field cohesive zone model (\texttt{PF-CZM}). Among many variants in the literature, only with the optimal geometric function, can the associated \texttt{PF-CZM} apply to general non-concave softening laws and the non-associated \texttt{PF-CZM} to (almost) any arbitrary one. Some mis-interpretations are clarified and representative numerical examples are presented. 

\end{abstract}

\begin{keyword}



Fracture; phase-field model; cohesive zone model; traction--separation law; softening.

\end{keyword}

\end{frontmatter}

%
\section{Introduction}

Ever since the pioneering work of \cite{FM1998}, the variational approach to fracture has become very popular in the community of fracture mechanics. Being the numerically more amenable counterpart, the variational phase-field model (PFM) \citep{BFM2000} is one of the most promising tools in capturing fracture induced failure of solids, e.g., crack nucleation, propagation, branching and merging, etc., in a standalone framework; see \cite{BFM2008,AGL2015,WNNSBS2018}. 

Motivated by the \cite{AT1990} elliptic regularization of the \cite{MS1989} functional, in the variational PFM the sharp crack is geometrically regularized into a localized crack band with the bandwidth measured by a small but finite length scale parameter. The crack surface area can be evaluated through a volume integral in terms of a spatially continuous field variable, i.e., the crack phase-field $d (\boldsymbol{x}) \in [0, 1]$, and its spatial gradient $\nabla d$. With the strain and surface energy functions of the cracking solid properly defined, a set of coupled partial differential equations (PDEs) governing the displacement field and the crack phase-field are derived from either the variational principle \citep{BFM2008} or the irreversible thermodynamics \citep{MWH2010a}. The resulting PFM is usually characterized by two characteristic functions of the crack phase-field, i.e., the geometric function determining the crack profile and the degradation function defining the strain energy potential \citep{Wu2017}. 

Initially the PFM was applied to linear elastic or brittle fracture \citep{Griffith1921} in which the surface energy density (i.e., the energy dissipation per surface area) is treated as a material property --- fracture energy. After the great successes achieved in brittle fracture, e.g., the popular models \texttt{AT2} \citep{BFM2000,MWH2010a} and \texttt{AT1} \citep{PAMM2011}, research efforts were exerted to regularize the cohesive zone model (CZM) \citep{Barenblatt1959}. As the surface energy density function is no longer a constant, one strategy is to quantify explicitly the crack opening \citep{BFM2008}, which is a challenging task in the context of smeared or regularized cracks. Along this direction, \citet{VdeB2013} introduced an auxiliary field variable, in addition to the displacement field and the crack phase-field, to approximate the crack opening; see also \citet{CdeB2022}.  \cite{NYZBC2016} avoided introducing such an auxiliary field by computing the crack opening at two opposing points at the interface, but the selection of these points is arbitrary and problem-dependent. In such PFMs though the crack opening can be directly evaluated, the crack propagation path has to be known \textit{a priori}, losing the merit of phase-field models in dealing with arbitrary crack propagation.

By means of only the displacement field and the crack phase-field, \cite{CFI2016,CFI2024} proposed a PFM for cohesive fracture. Despite adoption of the same quadratic geometric function as in the \texttt{AT2} model for brittle fracture \citep{BFM2000,MWH2010a}, an elegant modification is to use a degradation function that is proportional to the phase-field length scale parameter. It was proved that in the 1D scenario the resulting PFM $\varGamma$-converges to the CZM with a nonlinear softening curve, and the 2D fracture responses were numerically investigated in \cite{FI2017,LCM2023}. Alternatively, Lorentz and coworkers \citep{LG2011,LCK2012,Lorentz2017} proposed a PFM (in the name of gradient damage model or GDM) for cohesive fracture. Though the same linear geometric function is adopted as in the \texttt{AT1} model for brittle fracture \citep{PAMM2011,MBK2015}, the degradation function of rational fraction depends explicitly on the length scale parameter.  Application of this model to the 1D problem (usually uniaxial stretching or uniaxial straining) yields a semi-analytical traction--separation (crack opening) law with a particular nonlinear softening curve. 

Though the crack opening is not evaluate explicitly, the above two PFMs both converge to the \cite{Barenblatt1959} CZM upon a vanishing length scale, at least in the 1D scenario. As the geometric function is identical to that adopted for brittle fracture, this anticipated feature can only be attributed to the degradation function which depends not only on the crack phase-field but also on the incorporated length scale parameter. Importantly, these models are able to deal with arbitrary crack propagation as their counterparts for brittle fracture do. However, they also exhibit the following deficiencies:
\begin{itemize}
\item Firstly, the length-scale dependent degradation function is assumed \textit{a priori} and inflexible to be adapted for general softening curves. As a consequence, only rather limited traction–separation laws can be captured and the initial slope heavily affecting the fracture behavior cannot be well controlled. For instance, the \cite{CFI2016} model and the \cite{Lorentz2017} model both give particular nonlinear softening curves with finite ultimate crack opening, while the linear and exponential softening commonly adopted for cohesive fracture cannot be reproduced. 

\item Secondly and more importantly, the crack irreversibility condition $\dot{d} (\boldsymbol{x}) \ge 0$ was overlooked when seeking the analytical solution or proving the $\varGamma$-convergence in the 1D scenario. This neglect is not a severe issue for brittle fracture, but is unacceptable for cohesive fracture in which the post-peak softening regime has negligible effects on the global response. Irreversibility of the crack evolution implies that the localized crack bandwidth has to be a non-decreasing function during failure --- otherwise, some material points within the crack band unload and consequently the analytical traction--separation law no longer holds as the anticipated one. That is, the crack band has to be non-shrinking, which is, nevertheless, not always guaranteed.
\end{itemize}

The aforesaid issues were largely solved in the unified phase-field theory for fracture \citep{Wu2017}. With two parameterized characteristic functions, i.e., a second-order polynomial geometric function and a rational fraction degradation function, the \texttt{AT1/2} models for brittle fracture and the \cite{Lorentz2017} model for cohesive one are recovered as its particular instances. Moreover, the phase-field cohesive zone model (\texttt{PF-CZM}) \citep{Wu2018,WN2018} emerges naturally from this framework. The linear and commonly adopted convex softening curves can be reproduced or approximated with sufficient precision. 

Due to seamless incorporation of the stress-based criterion for crack nucleation, the energy-based criterion for crack propagation and the stability-based criterion for crack path selection, the \texttt{PF-CZM} rapidly becomes popular in the mechanics community \citep{CFI2024}. Ever since its birth, several variants have been developed in the literature. In \cite{WZF2020}, the quadratic geometric function was adopted as in the \texttt{AT2} model for brittle fracture, together with the parameterized degradation function of rational fraction, yielding a \texttt{PF-CZM} with infinite crack bandwidth. Similarly, in \cite{MBBPB2020} the averaging linear and quadratic geometric function was considered, resulting in a \texttt{PF-CZM} with positive-definite stiffness for the phase-field evolution equation.  \cite{GLML2023} introduced more parameters into the parameterized degradation function to approximate the concave \cite{PPR2009} softening. In \cite{FFL2021} the degradation function of rational fraction was assumed to connect directly with the geometric function, yielding a \texttt{PF-CZM} in which both characteristic functions are analytically determined for a given softening curve. As the condition for a non-shrinking crack band is not always guaranteed, these successors are incapable of reproducing the anticipated traction--separation softening law upon imposition of the crack irreversibility.

Nowadays almost all the PFMs for brittle fracture or cohesive one are associated, at least in the 1D scenario. That is, the stress--strain constitutive relation and the crack driving force are associated with a unique energy functional and characterized by a single degradation function. This coupling makes the original \texttt{PF-CZM} incapable of modeling concave softening curves while guaranteeing the condition for a non-shrinking crack band simultaneously. To remove the above limitation, \cite{Wu2024} extended the framework of the unified phase-field theory for fracture \citep{Wu2017} to non-associated cases and proposed a generalized phase-field cohesive zone model (\texttt{$\mu$PF-CZM}). 
Though the \texttt{$\mu$PF-CZM} is non-associated in general cases, it is still thermodynamically consistent. Moreover, it is also consistent with the ``local variational principle for fracture'' \citep{Larsen2024,LDLP2024} in which the governing equations for the displacement field and the crack phase-field are derived from distinct energy functionals. 

In this work, all these phase-field models for cohesive fracture are analyzed in a unified framework. The objectivity is fourfold: \textit{(i)} to clarify the mis-interpretation on the energy dissipation of the PFM for cohesive fracture \citep{CdeB2021} and to verify that the \texttt{PF-CZM} reproduces exactly the dissipated energy of the \cite{Barenblatt1959} CZM, \textit{(ii)} to establish the inter-relation between the \cite{CFI2016} model and the \texttt{PF-CZM}, and to show how the limitations exhibited by the former are overcome by the latter, \textit{(iii)} to present all the variants of the \texttt{PF-CZM} in a unified framework and to illustrate that some of them do not fulfill the condition for a non-shrinking crack band, and \textit{(iv)} to compare the capabilities of the associated \texttt{PF-CZM} and the non-associated $\mu$\text{PF-CZM} in the modeling of cohesive fracture through some representative benchmark examples. Note that it is generally assumed in all the existing PFMs for fracture in elastic solids that crack nucleation coincides with the peak stress (critical strength) of the material. Recently, \cite{ZM2024} found that crack nucleation could be retarded for some particular stress states, e.g., under the hydrostatic tension, to the softening regime. The similar phenomenon was previously pointed out in \cite{WC2016} regarding strain localization in inelastic solids. That is, some inelastic deformations need to be accumulated in order for the occurrence of bifurcation or loss of stability. As only fracture in elastic solids is interested, inelastic deformations prior to crack nucleation is not accounted for in this work and will be addressed elsewhere.

The remainder of this work is structured as follows. \Cref{sec:framework-pfczm} addresses the extended framework of the unified phase-field theory for fracture, with the non-associated crack driving force incorporated. In  \cref{sec:analytical-solution} the 1D analytical solution, e.g., the traction--separation softening law and crack bandwidth, etc., are presented. The conditions for length scale insensitive traction--separation laws and for a non-shrinking crack band are introduced. \Cref{sec:cfil-pfczm} focuses on the \cite{CFI2016} model for cohesive fracture, with the 1D fracture responses and its limitations analyzed. A particular \texttt{PF-CZM} is presented to show how the these limitations can be removed. \Cref{sec:pfczm} is devoted to the \texttt{PF-CZM} as general as possible. Both the associated formulation and the non-associated one, with the degradation function either assumed \textit{a priori} or solved analytically, are discussed. \Cref{sec:numerical-examples} presents some representative numerical examples. The capabilities of the associated \texttt{PF-CZM} and the non-associated \texttt{$\mu$PF-CZM} in the modeling of cohesive fracture are demonstrated. The conclusions are drawn in \cref{sec:conclusions}. For the sake of completeness, two appendices are attached in which the commonly adopted softening curves and a generic stress-based failure criterion are presented.

\section{A unified phase-field theory for fracture}
\label{sec:framework-pfczm}

Let us limit ourselves to the \cite{Barenblatt1959} cohesive fracture in solids under the quasi-static and infinitesimal strain regime. For the solid $\varOmega$ containing a sharp crack $\mathcal{S}$, the energy functional (with the potential energy of external forces omitted in this work) is expressed as
\begin{align}\label{eq:energy-functional-cohesive}
	\mathscr{E} (\boldsymbol{u}, \mathcal{S})
		= \int_{\varOmega} \psi_{0} (\bfepsilon^{\text{e}} (\boldsymbol{u})) 
			\; \td V
		+ \int_{\mathcal{S}} \mathcal{G} (\itw)	\; \td A
\end{align}
where $\boldsymbol{u} (\boldsymbol{x})$ denotes the displacement field. As usual, the elastic strain energy density $\psi_{0} (\bfepsilon^{\text{e}})$ is expressed in terms of the bulk (elastic) strain $\bfepsilon^{\text{e}}$
\begin{align}\label{eq:elastic-elastic-energy}
	\psi_{0} (\bfepsilon^{\text{e}})
		= \dfrac{1}{2} \bfepsilon^{\text{e}} : \mathbb{E}_{0} 
		: \bfepsilon^{\text{e}}
		= \dfrac{1}{2} \bfsigma : \mathbb{E}_{0}^{-1} : \bfsigma
	\qquad \text{with} \qquad
	\bfepsilon^{\text{e}} :
		= \mathbb{E}_{0}^{-1} : \bfsigma
\end{align}
for the stress tensor $\bfsigma$ and the elasticity tensor $\mathbb{E}_{0}$ of the material, respectively. For cohesive fracture the surface energy density $\mathcal{G} (\itw)$ is a concave function in terms of the crack opening $\itw$ and grows from zero to the \cite{Griffith1921} fracture energy $G_{\text{f}}$; see \ref{sec:softening-curves} for those commonly adopted and the corresponding softening curves. 


\subsection{Governing equations}

In phase-field models for fracture, the sharp crack $\mathcal{S}$ is regularized into a localized crack band $\mathcal{B} \subseteq \varOmega$ of finite measure. The cracking state of the solid is described by the crack phase-field $d(\boldsymbol{x}) \in [0, 1]$, which is a spatially continuous scalar field and fulfills the irreversibility condition $\dot{d} (\boldsymbol{x}) \ge 0$. The intact state and the completely broken one are represented by $d (\boldsymbol{x}) = 0$ and $d (\boldsymbol{x}) = 1$, respectively, with the intermediate value $d (\boldsymbol{x}) \in (0, 1)$ representing the partially broken material. 

The governing equations for the displacement field and the crack phase-field are expressed as
\begin{subequations}\label{eq:governing-equations-pfm}
\begin{align}
\label{eq:equilibrium-governing-equations-pfm}
&\begin{cases}
	\nabla \cdot \bfsigma
		+ \boldsymbol{b}^{\ast}
		= \boldsymbol{\mathit{0}} & \qquad \text{in} \; \varOmega \\
	\bfsigma \cdot \boldsymbol{n}^{\ast}
		= \boldsymbol{t}^{\ast} & \qquad \text{on} \; \partial \varOmega_{t}
\end{cases} \\
\label{eq:cracking-governing-equations-pfm}
&\begin{cases}
	\nabla \cdot \boldsymbol{q} + h \le 0 & \qquad \;\;\;
		\text{in} \; \mathcal{B} \\
	\boldsymbol{q} \cdot \boldsymbol{n}_{_\mathcal{B}} \ge 0 & \qquad \;\;\;
		\text{on} \; \partial \mathcal{B}
\end{cases}
\end{align}
\end{subequations}
where $\boldsymbol{n}^{\ast}$ and $\boldsymbol{n}_{_\mathcal{B}}$ denote the outward unit normal vectors of the boundary $\partial \varOmega$ and of the boundary $\partial \mathcal{B}$, respectively; $\boldsymbol{b}^{\ast}$ and $\boldsymbol{t}^{\ast}$ represent the body force distributed in the domain $\varOmega$ and the surface traction imposed on the boundary $\partial \varOmega_{t} \subset \partial \varOmega$, respectively. 


As usual, the Cauchy stress $\bfsigma$ in the cracking solid is given by
\begin{align}\label{eq:stress-pfm-nosplit}
	\bfsigma
		= \dfrac{\partial \psi}{\partial \bfepsilon}
		= \omega (d) \mathbb{E}_{0} : \bfepsilon
		= \omega (d) \bar{\bfsigma}
	\qquad \text{with} \qquad
	\bar{\bfsigma} :
		= \mathbb{E}_{0} : \bfepsilon	
\end{align}
where $\bar{\bfsigma}$ is the (undamaged) effective stress; the strain energy density function $\psi (\bfepsilon, d)$ is defined as
\begin{align}\label{eq:strain-energy}
	\psi (\bfepsilon, d)
		= \omega(d) \psi_{0} (\bfepsilon), \qquad
	\psi_{0} (\bfepsilon)
		= \dfrac{1}{2} \bfepsilon : \mathbb{E}_{0} : \bfepsilon
\end{align}
for the initial strain energy $\psi_{0} (\bfepsilon)$ and the degradation function $\omega (d)$. 

Regarding the crack phase-field, the flux $\boldsymbol{q}$ and the source term $h$ are expressed as
\begin{align}\label{eq:crack-flux-source}
	\boldsymbol{q}
		= \dfrac{2 b}{c_{_\alpha}} G_{\text{f}} \nabla d, \qquad
	h
		= Y (\bfepsilon, d) - \alpha' (d) \dfrac{1}{c_{_\alpha} b}
			G_{\text{f}}
\end{align}
where the geometric function $\alpha (d)$ dictates the ultimate profile of the crack phase-field, with $\alpha' (d) := \partial \alpha / \partial d$ being the derivative and $c_{\alpha} = 4 \displaystyle \int_{0}^{1} \sqrt{\alpha (\vartheta)} \; \td \vartheta$ the scaling constant, respectively; the length scale $b$ is a regularization parameter that measures the crack bandwidth; the crack driving force $Y$ will be addressed in \cref{sec:crack-driving-force}. 

\remark For the stress \eqref{eq:stress-pfm-nosplit}, the strain $\bfepsilon$ allows the following additive split into the elastic and inelastic components $(\bfepsilon^{\text{e}}, \bfepsilon^{\text{c}})$
\begin{align}
	\bfepsilon
		= \bfepsilon^{\text{e}} + \bfepsilon^{\text{c}}, \qquad
	\bfepsilon^{\text{c}}
		= \bfepsilon - \bfepsilon^{\text{e}}
		= \bigg[ \dfrac{1}{\omega (d)} - 1 \bigg] 
			\mathbb{E}_{0}^{-1} : \bfsigma
		= \phi (d) \mathbb{E}_{0}^{-1} : \bfsigma
\end{align}
for the \textit{cracking function} $\phi (d)$
\begin{align}\label{eq:cracking-characteristic-function}
	\phi (d)
		= \frac{1}{\omega (d)} - 1
	\qquad \Longrightarrow \qquad
	\omega (d)
		= \dfrac{1}{1 + \phi (d)}
\end{align}
which is the kernel function in evaluation of the crack opening \citep{Wu2024}. $\Box$ 

\subsection{Crack driving force}
\label{sec:crack-driving-force}

Regarding the crack driving force $Y$ in \cref{eq:crack-flux-source}, the existing phase-field models can be classified into either the associated formulation or the non-associated one.

\subsubsection{Associated formulation}

In most phase-field models \citep{BFM2000,BFM2008,MWH2010a} the crack driving force $Y$ is defined from the strain energy potential \eqref{eq:strain-energy}
\begin{align}\label{eq:crack-driving-force}
	Y :
    =-\dfrac{\partial \psi}{\partial d}
    =-\omega' (d) \bar{Y}
	\qquad \text{with} \qquad
	\omega' (d)
		=-\omega^{2} (d) \phi' (d) \le 0
\end{align}
for the derivatives $\omega' (d) = \partial \omega / \partial d$ and $\phi' (d) := \partial \phi / \partial d$. The effective crack driving force $\bar{Y} := \partial \psi / \partial \omega = \psi_{0}$ is addressed in details in \cref{rmk:effective-crack-driving-force-hybrid}. As the stress tensor \eqref{eq:stress-pfm-nosplit} and the crack driving force \eqref{eq:crack-driving-force} are defined in terms of an identical strain energy potential, the resulting model is associated \footnote{This is similar to the associated plasticity model in which the plastic potential function is coincident with the yield function.}. 

In this case, \cref{eq:governing-equations-pfm} are exactly the Euler-Lagrange equation of the following variational problem \citep{BFM2000,BFM2008}
\begin{align}\label{eq:minimization-pfm}
	(\boldsymbol{u}, d)
	  = \text{Arg} \Big\{ \min_{ \hat{\boldsymbol{u}}, \; \hat{d} }
	  	\mathscr{E} (\hat{\boldsymbol{u}}, \hat{d}) \Big\}
\end{align}
with the total energy functional $\mathscr{E} (\boldsymbol{u}, d)$ expressed as
\begin{align}\label{eq:total-energy-functional-pfm}
  \mathscr{E} (\boldsymbol{u}, d) 
    = \int_{\varOmega} \psi (\bfepsilon, d) \; \td V
    + \int_{\mathcal{B}} G_{\text{f}} \dfrac{1}{c_{\alpha}} 
    	\bigg[ \dfrac{1}{b} \alpha (d) + b \big| \nabla d \big|^{2} 
    	\bigg] \; \td V 
\end{align}
where the surface energy \`a la Griffith is approximated by the phase-field regularization of the sharp crack. 

\remark The energy functional \eqref{eq:total-energy-functional-pfm} is the \cite{AT1990} elliptic regularization of the \cite{Griffith1921} brittle fracture. However, it also applies to the \cite{Barenblatt1959} CZM, with the surface energy given by
\begin{align}\label{eq:energy-dissipation}
	\int_{\mathcal{S}} \mathcal{G} (\itw) \; \td A
	  = \int_{\mathcal{B}} G_{\text{f}} \dfrac{1}{c_{\alpha}} 
			\bigg[ \dfrac{1}{b} \alpha (d) + b \big| \nabla d \big|^{2} 
			\bigg]	\; \td V
	  + \int_{\varOmega} \phi (d) \dfrac{1}{2} \bfsigma : \mathbb{E}_{0}^{-1}
	  : \bfsigma \; \td V
\end{align}
Note that in \cite{CdeB2021} only the first term was accounted for, while the second one was missing. This neglect results in incorrect calculation of the energy dissipation. $\Box$

\subsubsection{Non-associated formulation}

The associated formulation exhibits some limitations. In particular, it is inapplicable to cohesive fracture with concave softening laws since the crack band shrinks during failure. In order to remove these limitations, the author \citep{Wu2024} proposed the generalized $\mu$\texttt{PF-CZM}, with the crack driving force $Y$ modified as
\begin{align}\label{eq:crack-driving-force-generalized}
	Y
    =-\dfrac{\partial \bar{\psi}}{\partial d}
		=-\varpi' (d) \bar{Y}
	\qquad \text{with} \qquad
  \bar{\psi} (\bfepsilon, d)
		= \varpi(d) \psi_{0} (\bfepsilon) 
\end{align}
for the \textit{dissipation function} $\varpi (d)$ with the following derivative
\begin{align}
	\varpi' (d)
		=-\omega^{2} (d) \mu' (d) \le 0
\end{align}
where $\mu (d)$ is an auxiliary function with the derivative $\mu' (d) := \partial \mu / \partial d$. In this case, the crack driving force is associated with an energy potential other than the one \eqref{eq:strain-energy} for the stress $\bfsigma$. As can be seen, the associated formulation \eqref{eq:crack-driving-force} is recovered provided the identity $\varpi' (d) = \omega' (d)$, or, equivalently, $\mu (d) = \phi (d)$, holds.


The above non-associated formulation corresponds to the ``local variational principle for fracture'' \citep{Wu2018b,Larsen2024,LDLP2024}
\begin{align}
\begin{cases}
	\boldsymbol{u} \displaystyle 
		= \text{Arg} \Big\{ \min_{\hat{\boldsymbol{u}}}
			\mathscr{E} (\hat{\boldsymbol{u}}, \hat{d}) \Big\} \vspace{2mm} \\
	d \displaystyle 
		= \text{Arg} \Big\{ \min_{\hat{d}}
			\bar{\mathscr{E}} (\hat{\boldsymbol{u}}, \hat{d}) \Big\}
\end{cases}
\end{align}
where the modified strain energy density functional $\bar{\mathscr{E}} (\boldsymbol{u}, d)$ is defined by
\begin{align}
  \bar{\mathscr{E}} (\boldsymbol{u}, d) 
    = \int_{\varOmega} \bar{\psi} (\bfepsilon (\boldsymbol{u}), d) \; \td V
    + \int_{\mathcal{B}} G_{\text{f}} \dfrac{1}{c_{\alpha}} 
    	\bigg[ \dfrac{1}{b} \alpha (d) + b \big| \nabla d \big|^{2} 
    	\bigg] \; \td V 
\end{align}

\remark \label{rmk:effective-crack-driving-force-hybrid} For the isotropic constitutive relation \eqref{eq:stress-pfm-nosplit}, either the associated formulation or the non-associated one predicts unrealistic symmetric tensile and compressive behavior. The simplest strategy addressing this issue might be using the hybrid formulation \citep{AGL2015}. That is, in \cref{eq:crack-driving-force} or \eqref{eq:crack-driving-force-generalized} the effective crack driving force $\bar{Y}$ is modified with the failure mode accounted for properly. For instance, the following expression usually applies \citep{Wu2017,WHN2020}
\begin{align}
	\bar{Y} :
		= \dfrac{\bar{\sigma}_{\text{eq}}^{2}}{2 E_{0}} 
\end{align}
with
\begin{align}\label{eq:equivalent-effective-stress}
	\bar{\sigma}_{\text{eq}} (\bar{\bfsigma})
		= \begin{cases}
				\langle \bar{\sigma}_{1} \rangle & 
					\quad \text{Rankine criterion} \vspace{2mm} \\
				\dfrac{\rho_{s} - 1}{2 \rho_{s}} \bar{I}_{1}
						+ \dfrac{1}{2 \rho_{s}} \sqrt{\big(\rho_{s} - 1 \big)^{2} \bar{I}_{1}^{2}
						+ 12 \rho_{s} \bar{J}_{2}} & \quad \text{modified von Mises criterion}
			\end{cases}						
\end{align}
where $E_{0}$ is Young's modulus of the material; $\rho_{s} := f_{\text{c}} / f_{\text{t}}$ is the strength ratio between the uniaxial compressive and tensile strengths. The equivalent effective stress $\bar{\sigma}_{\text{eq}}$ is defined in terms either of the major principal value $\bar{\sigma}_{1}$  of the effective stress tensor $\bar{\bfsigma}$ for mode-I fracture, or of the first invariant $\bar{I}_{1} = \trace \bar{\bfsigma}$ and the second invariant $\bar{J}_{2} := \frac{1}{2} \bar{\bfsigma} : \bar{\bfsigma} - \frac{1}{6} \bar{I}_{1}^{2}$ for tension-dominated mixed-mode failure. Note that other types of failure criteria, e.g., the Drucker-Prager one advocated in \cite{KBFLP2020,KKLP2024}, can also be incorporated as in \ref{sec:stress-based-failure-criterion}. $\Box$ 

\subsection{Characteristic functions}

One necessary step is to specify the involved characteristic functions, i.e., the \textit{geometric function} $\alpha (d)$, the \textit{degradation function} $\omega (d)$ or $\phi (d)$, and the \textit{dissipation function} $\varpi (d)$ or $\mu (d)$. 

\subsubsection{Geometric function}


The approximation \eqref{eq:total-energy-functional-pfm} relies on $\varGamma$-convergence \citep{Braides1998} of the phase-field regularization \eqref{eq:total-energy-functional-pfm}. Moreover, the anticipated \textit{bullet}-shaped crack profile demands a monotonically increasing geometric function \citep{Wu2024}. The above considerations transform into the following conditions
\begin{align}\label{eq:geometric-function-conditions}
	\alpha (0) 
		= 0, \quad
	\alpha' (d) > 0
	\qquad \Longrightarrow \qquad
	\alpha (1) > \alpha (0) = 0	
\end{align}
A stronger condition $\alpha (1) > 0$, e.g., $\alpha (1) = 1$, is generally assumed. For the sake of simplicity and without loss of generality, the author suggested using the following parameterized second-order polynomial \citep{Wu2017}
\begin{align}\label{eq:crack-geometric-function}
	\alpha (d)
		= \xi d + \big( 1 - \xi \big) d^{2}, \qquad
	\alpha' (d)
		= \xi + 2 \big( 1 - \xi \big) d
\end{align}
for the parameter $\xi \in [0, 2]$ in order to achieve a monotonically increasing geometric function $\alpha (d) \in [0, 1]$. In particular, the quadratic function $\alpha (d) = d^{2}$ (i.e., $\xi = 0$) results in a crack band of infinite support, while the others with $\xi \in (0, 2]$ yield a localized crack band of finite width.

\subsubsection{Degradation and dissipation functions}


The degradation function $\omega (d)$ describes the effect of the crack phase-field on the stress--strain relation. In accordance with previous studies \citep{MWH2010a,Wu2017}, it satisfies the following conditions
\begin{align}\label{eq:conditions-cracking-function}
	\omega (0) 
		= 1, \quad
	\omega (1) 
		= 0; \quad
	\omega' (d) < 0
	\qquad \Longleftrightarrow \qquad
	\phi (0)
		= 0, \quad
	\phi (1)
		=+\infty, \quad
	\phi' (d) > 0
\end{align}
The auxiliary function $\mu (d)$ needs to fulfill the similar conditions \citep{Wu2024}
\begin{align}\label{eq:conditions-degradation-functions}
	\mu (0)
		= 0, \quad
	\mu (1)
		=+\infty, \quad
	\mu' (d) 
		= O (\omega^{2} (d)) > 0 \qquad \Longrightarrow \qquad
	\varpi' (1)
		= 0
\end{align}
The last condition, which implies a vanishing crack driving force \eqref{eq:crack-driving-force-generalized} for completely broken solids (i.e., $d = 1$), is introduced to avoid unlimited widening of the crack bandwidth.

As will be shown, in order for a phase-field model to be applicable to cohesive fracture one needs to incorporate properly the length scale $b$ into the degradation functions $\omega (d)$ and the dissipation function $\varpi (d)$. 

\section{1-D analytical solution}\label{sec:analytical-solution}

In this section we apply the above phase-field model to analyze a softening bar subjected to uniaxial stretching. Consider a bar $x \in [-L, L]$ with a cross sectional area $A$. The bar is sufficiently long such that the boundary effects do not influence crack evolution. It is loaded at both ends by two equally increasing displacements in opposite directions. We assume that a crack is initiated at the centroid $x_{0} = 0$ and that the crack band is localized within the domain $\mathcal{B} := \Big\{ x \big| x \in [-D, D] \Big\}$. The half crack bandwidth $D \ll L$ may vary during the failure process. Distributed body forces are neglected in the analysis. 




As long as the crack maintains in the loading state, i.e., $\dot{d} (x) > 0$ \textit{anywhere} within the crack band $\forall x \in \mathcal{B}$, the phase-field evolution law \eqref{eq:cracking-governing-equations-pfm} becomes an identity \citep{Wu2017,WNNSBS2018,Wu2024}
\begin{align}\label{eq:governing-equations-1D}
	\dfrac{\sigma^{2}}{2 E_{0}} \mu' (d) 
		- \frac{G_{\text{f}}}{c_{\alpha} b} \Big[ \alpha' (d) 
		- 2 b^{2} \Delta d \Big]
		= 0
	\qquad \Longrightarrow \qquad
	\dfrac{\sigma^{2}}{2 E_{0}} \mu (d) 
		- \frac{G_{\text{f}}}{c_{\alpha} b} \Big[ \alpha (d) 
		- \big( b \nabla d \big)^{2} \Big]
	 	= 0
\end{align}
where the stress field $\sigma (x)$ is uniformly distributed along the bar. 

\subsection{Traction--separation law (TSL)}

The load level of the softening bar can be measured by the maximum value $d_{\ast} := d (x = x_{0})$ of the crack phase-field $d (x)$ at the centroid $x = x_{0}$ of the crack band $\mathcal{B}$. After some straightforward mathematical manipulation \citep{Wu2017,WNNSBS2018} to the governing equation \eqref{eq:governing-equations-1D}, the traction (stress in 1D) $\sigma$ and the \textit{apparent} separation $\itw$ across the crack band $\mathcal{B}$ are derived as \citep{Wu2024}
\begin{subequations}\label{eq:stress-opening-1D}
\begin{align}\label{eq:localised-stress-1D}
 	\sigma (d_{\ast})
 	 &= \sqrt{\frac{2 E_{0} G_{\text{f}}}{c_{\alpha}} \eta (d_{\ast})}
 	  = \sigma_{\text{c}} \sqrt{\frac{1}{\eta_{0}} \eta (d_{\ast})}
 	\qquad \quad\; \qquad \text{with} \qquad    
	\eta (d)
		= \dfrac{1}{b} \dfrac{\alpha (d)}{\mu (d)} \\
 	\label{eq:displacement-jump-apparent}
  \itw (d_{\ast}) 
	 &= \dfrac{2 \itw_{\text{cL}}}{c_{\alpha}} \sqrt{\eta_{0} 
		 	\eta (d_{\ast})} \int_{0}^{d_{\ast}} b \phi (\vartheta)
		 	\mathscr{H} (\vartheta; d_{\ast}) \; \td \vartheta
	\qquad \text{with} \qquad
  \mathscr{H} (d; d_{\ast})
		= \sqrt{\dfrac{\eta (d) / \alpha (d)}{\eta (d) - \eta (d_{\ast})}}	
\end{align}
\end{subequations}
where the critical stress $\sigma_{\text{c}}$ upon crack initiation is given by
\begin{align}\label{eq:critical-stress-general}
	\sigma_{\text{c}}
		= \lim_{d_{\ast} \to 0} \sigma (d_{\ast})
 		= \sqrt{\frac{2 E_{0} G_{\text{f}}}{c_{\alpha}} \eta_{0}}
 	\qquad \text{with} \qquad
 	\eta_{0}
		= \lim_{d_{\ast} \to 0} \eta (d_{\ast})
\end{align}
and $\itw_{\text{cL}} := 2 G_{\text{f}} / \sigma_{\text{c}}$ denotes the ultimate crack opening of the linear softening law \eqref{eq:linear-softening-cohesive}. 

In order for \cref{eq:stress-opening-1D} to constitute the TSL of the \cite{Barenblatt1959} CZM, the traction \eqref{eq:localised-stress-1D}, the separation \eqref{eq:displacement-jump-apparent} and the failure strength \eqref{eq:critical-stress-general} all have to be independent of (or $\varGamma$--convergent with respect to) the length scale parameter $b$. These considerations transform into the following conditions
\citep{Wu2024}
\begin{align}\label{eq:length-scale-insensitivity-condition}
\begin{cases}
	\sigma_{\text{c}}
		= f_{\text{t}} \in (0, +\infty) & 
	\qquad \Longleftrightarrow \qquad
	\eta_{0} 
		= \dfrac{c_{\alpha}}{2 l_{\text{ch}}} 
			\in (0, +\infty) \vspace{1mm} \\
	\mu (d) \propto \dfrac{1}{b}, \quad 
	\phi (d) \propto \dfrac{1}{b} & 
	\qquad \Longleftrightarrow \qquad
	\exists \; \bar{\mu} (d) 
		= b \mu (d), \quad
	\exists \; \bar{\phi} (d) 
		= b \phi (d)
\end{cases}
\end{align}
for the Irwin internal length $l_{\text{ch}} := E_{0} G_{\text{f}} / f_{\text{t}}^{2}$ of the material. As will be shown, the conditions \eqref{eq:length-scale-insensitivity-condition} can be satisfied either in the strong sense (length scale \textit{insensitive}) or in the weak one with the vanishing limit $b \to 0$ (length scale \textit{convergent}).

\remark For the traction--separation law \eqref{eq:stress-opening-1D}, the surface energy density function $\mathcal{G} (\itw)$ is given by
\begin{align}
	\mathcal{G} (\itw (d_{\ast}))
		= \int_{0}^{\itw (d_{\ast})} \sigma (\itw) \; \td \itw
		= \int_{0}^{d_{\ast}} \sigma (\vartheta) 
			\dfrac{\partial \itw}{\partial \vartheta} \; \td \vartheta
\end{align}
Regarding the associated formulation with $\phi (d) = \mu (d)$,  the surface energy \eqref{eq:energy-dissipation} becomes
\begin{align}\label{eq:fracture-energy-CZM}
	\mathcal{G} (\itw) A
	  =	\dfrac{2 G_{\text{f}}}{c_{\alpha}} \frac{1}{b} 
	 	  \int_{\mathcal{B}} \alpha (d) \; \td V
	\qquad \Longrightarrow \qquad
	\mathcal{G} (\itw (d_{\ast}))
	  = \dfrac{4 G_{\text{f}}}{c_{\alpha}} 
	  	\int_{0}^{d^{\ast}} \sqrt{\dfrac{\alpha (\vartheta) 
	  	\eta (\vartheta)}{\eta (\vartheta) - \eta (d_{\ast})} } 
	  	\; \td \vartheta
\end{align}
upon the identity \eqref{eq:governing-equations-1D}$_{2}$. As will be shown, not only the TSL but also the energy dissipation can be reproduced. $\Box$

\subsection{Condition for a non-shrinking crack band}
\label{sec:condition-non-shrinking-crack-band}

The identity \eqref{eq:governing-equations-1D}, based upon which the parameterized TSL \eqref{eq:stress-opening-1D} is analytically derived, holds only if the crack band does not shrink during the failure process. As the maximum value $d_{\ast}$ is always increasing, it implies that
\begin{align}\label{eq:non-shrinking-crack-band}
	\dot{D} (d_{\ast})
		= \dfrac{\partial D}{\partial d_{\ast}} \dot{d}_{\ast} \ge 0
	\qquad \Longrightarrow \qquad
	\dfrac{\partial D}{\partial d_{\ast}} \ge 0
\end{align}
where the crack bandwidth $D (d_{\ast})$ is expressed as
\begin{align}\label{eq:half-bandwidth-1D}
  D (d_{\ast}) 
    = b \int_{0}^{d_{\ast}} \mathscr{H} (\vartheta; d_{\ast}) 
    	\; \td \vartheta
\end{align}
with the function $\mathscr{H} (d; d_{\ast})$ given in  \cref{eq:displacement-jump-apparent}.

For general characteristic functions, usually neither the closed-form nor the monotonicity of the crack bandwidth \eqref{eq:half-bandwidth-1D} is unavailable. Fortunately, it allows determining the initial and ultimate crack bandwidths as \citep{Wu2017,WNNSBS2018,Wu2024}
\begin{align}\label{eq:crack-bandwidths-general}
	D_{0} :
	  = D (d_{\ast} = 0)
		= \pi b \sqrt{\dfrac{2}{\eta_{0} b \mu'' (0) - \alpha'' (0)}}, 
	\qquad
	D_{\text{u}} :
	  = D (d_{\ast} = 1)
		= b \int_{0}^{1} \dfrac{1}{\sqrt{\alpha (\vartheta)}} \; \td \vartheta
\end{align}
for the second derivatives $\alpha'' (d) = \partial^{2} \alpha / \partial d^{2}$ and $\mu'' (d) := \partial^{2} \mu / \partial d^{2}$, respectively. Therefore, the condition \eqref{eq:non-shrinking-crack-band} demands that the initial crack bandwidth $D_{0}$ has to be not larger than the ultimate one $D_{\text{u}}$, i.e.,
\begin{align}\label{eq:irreversibility-condition}
	\dfrac{\partial D}{\partial d_{\ast}} \ge 0 
	\qquad \Longrightarrow \qquad
  D_{0} \le D_{\text{u}}
\end{align}
which can be employed to determine the optimal geometric function $\alpha (d)$.

\section{Length scale convergent \textit{versus} insensitive models}
\label{sec:cfil-pfczm}

In this section, two particular PFMs for cohesive fracture are discussed. Though the extension with other geometric functions is straightforward, the quadratic one
\begin{align}\label{eq:geometric-crack-function-quadratic}
	\alpha (d) 
		= d^{2} \qquad \Longrightarrow \qquad
	c_{\alpha}
		= 2, \qquad
	D_{\text{u}}
		=+\infty 
\end{align}
is adopted in both models. Moreover, the associated formulation is considered, i.e.,
\begin{align}
	\mu (d)
		= \phi (d), \qquad
	\varpi (d)
		= \omega (d)
\end{align}
We will show how the phase-field length scale parameter $b$ is incorporated into the dissipation/degradation function $\varpi (d) = \omega (d)$ such that cohesive fracture can be dealt with. 

\subsection{The \cite{CFI2016,CFI2024} model}

Mimicking the \texttt{AT2} model for brittle fracture \citep{BFM2000,BFM2008}, \citet{CFI2016,CFI2024} proposed a PFM for cohesive fracture. In this model, the degradation function $\omega (d)$ is assumed to be proportional to the length scale parameter $b$, i.e.,
\begin{align}\label{eq:degradation-function-cfil}
	\omega (d)
		= \min \Big\{ b \bar{\omega}^{2} (d), 1 \Big\}
	\qquad \text{with} \qquad
	\lim_{d \to 0} d \cdot \bar{\omega} (d) \in (0, + \infty)	
\end{align}
in terms of a monotonically decreasing function $\bar{\omega} (d)$. In \cite{CFI2016}, the following simple function was postulated
\begin{align}
	\bar{\omega} (d)
		= \ell \cdot \dfrac{1 - d}{d} 
\end{align}
for the parameter $\ell \in (0, +\infty)$. Upon the above setting, $\varGamma$-convergence of the regularized energy functional \eqref{eq:total-energy-functional-pfm} to \cref{eq:energy-functional-cohesive} for cohesive fracture has been proved. Specifically, for the strain energy \eqref{eq:strain-energy} with an artificial elastic modulus $E_{0} = 2$ and the fracture energy $G_{\text{f}} = 1$, the failure strength $f_{\text{t}} = \ell$ is reproduced. 

\cite{FI2017} extended the \cite{CFI2016,CFI2024} model to linearized elasticity problems with the real material properties $E_{0}$ and $G_{\text{f}}$. However, upon assuming $\ell = 1$ the resulting failure strength $\sigma_{\text{c}} = \sqrt{E_{0} G_{\text{f}} / 2}$ is not an independent model parameter, losing the flexibility in reproducing general material properties. In \cite{LCM2023}, the above model was further improved with the following choice
\begin{align}\label{eq:energetic-function-CFIL}
	\ell 
		= \dfrac{1}{\sqrt{l_{\text{ch}}}}
	\qquad \Longrightarrow \qquad
	\omega (d)
		= \min \bigg\{ \bar{b} \dfrac{\big( 1 - d \big)^{2}}{d^{2}}, 1 
			\bigg\}
\end{align}
for the normalized parameter $\bar{b} := b / l_{\text{ch}}$. It was proved that the resulting surface energy density function $\varGamma$-converges to a monotonically increasing function $\mathcal{G} (\itw)$ which fulfills the following properties
\begin{align}\label{eq:fracture-energy-properties-cfil}
	\mathcal{G} (0)
		= 0, \qquad
	\lim_{\itw \to +\infty} \mathcal{G} (\itw) 
		= G_{\text{f}}, \qquad
	\mathcal{G}' (0)
		= \lim_{\itw \to 0} \dfrac{\mathcal{G} (\itw)}{\itw}
		= f_{\text{t}}
\end{align}
That is, the given fracture energy $G_{\text{f}}$ and failure strength  $f_{\text{t}}$ of the material are properly incorporated.

\subsubsection{Global responses}

To be specific, let us discuss the \cite{CFI2016,CFI2024} model in more details. The degradation function \eqref{eq:energetic-function-CFIL} corresponds to the following cracking function
\begin{align}\label{eq:characteristic-functions-CFIL}
	\phi (d)
		= \dfrac{1}{\bar{b}} \bigg\langle \dfrac{d^{2}}{(1 - d)^{2}}
		- \bar{b} \bigg\rangle 
	\qquad \Longrightarrow \qquad
	\bar{\phi} (d)
		= b \phi (d)
		= l_{\text{ch}} \bigg\langle \dfrac{d^{2}}{(1 - d)^{2}}
		- \bar{b} \bigg\rangle
\end{align}
such that 
\begin{align}
	\eta (d)
		= \dfrac{1}{l_{\text{ch}} } 
			\dfrac{d^{2} \big(1 - d \big)^{2}}{\big\langle d^{2}
		- \bar{b}	\big(1 - d \big)^{2} \big\rangle}, \qquad
	\eta_{0}
		= \dfrac{1}{l_{\text{ch}}} \lim_{d \to 0} \dfrac{d^{2}}{
			\big\langle d^{2} - \bar{b} \big\rangle}
		= \begin{cases}
				+\infty & \qquad b \ge l_{\text{ch}} d^{2} \\
				 \dfrac{1}{l_{\text{ch}}} & \qquad b < l_{\text{ch}} d^{2} \\
			\end{cases}			
\end{align}
with the McAuley brackets $\langle x \rangle := \max (0, x)$. 

\begin{figure}[h!] \centering
  \subfigure[Softening curve]{
  \includegraphics[width=0.48\textwidth]{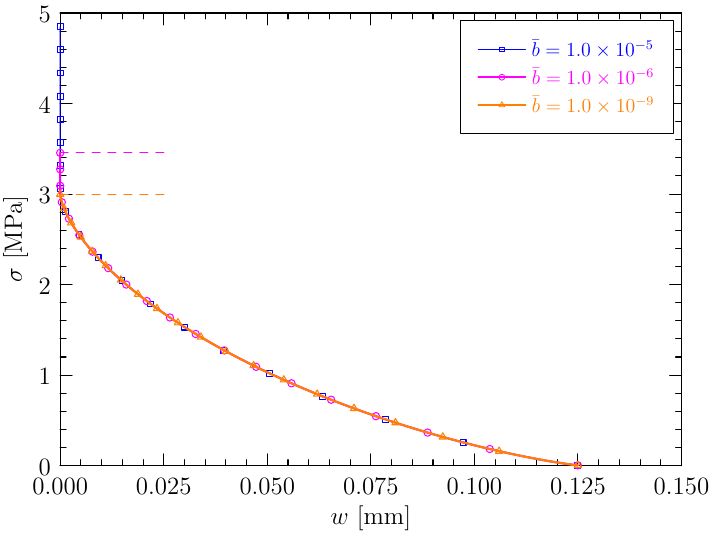}
  \label{fig:asymptotic-softening-curves-cfil}} \hfill
  \subfigure[Surface energy density function]{
  \includegraphics[width=0.48\textwidth]{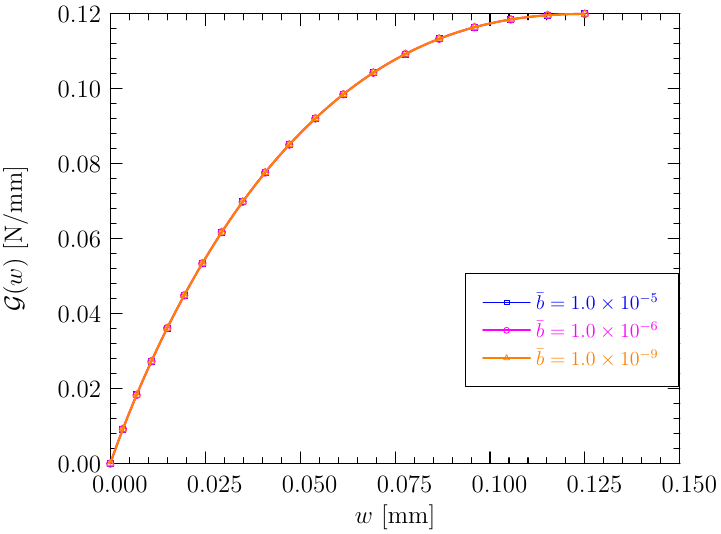}
  \label{fig:fracture-energy-cfil}}
  \caption{The softening curve and surface energy density function predicted by the \cite{CFI2016,CFI2024} model. The material properties $f_{\text{t}} = 3.0$ MPa and $G_{\text{f}} = 0.12$ N/mm are adopted.}
  \label{fig:cfil-model}  
\end{figure}

Upon the condition $\bar{b} < d^{2}$, \cref{eq:stress-opening-1D} gives the following TSL
\begin{subequations}\label{eq:stress-opening-1D-CFIL}
\begin{align}\label{eq:localised-stress-1D-CFIL}
 	\sigma (d_{\ast})
 	 &= f_{\text{t}} \sqrt{l_{\text{ch}} \eta (d_{\ast})}
 	\\
 	\label{eq:displacement-jump-apparent-CFIL}
  \itw (d_{\ast}) 
	 &= \itw_{\text{cL}} \sqrt{\eta (d_{\ast})} \int_{0}^{d_{\ast}} 
	 		\sqrt{ \dfrac{\bar{\phi} (\vartheta) / l_{\text{ch}}}{
	 		\eta (\vartheta) - \eta (d_{\ast})}} \; \td \vartheta
\end{align}
\end{subequations}
Moreover, the surface energy density function \eqref{eq:fracture-energy-CZM} particularizes into
\begin{align}
	\mathcal{G} (d_{\ast})
	  = 2 G_{\text{f}} \int_{0}^{d^{\ast}} \dfrac{\vartheta 
	  	\sqrt{\eta (\vartheta)}}{\sqrt{\eta (\vartheta) 
	  - \eta (d_{\ast})}} \; \td \vartheta
\end{align}
The above results are shown in \cref{fig:cfil-model}. 

As can be seen, the value $\eta_{0}$ is well-defined (finite) and the nonlinear softening curve $\sigma (\itw)$ with a concave monotonically increasing surface energy density $\mathcal{G} (\itw)$ applies, \textit{if and only if} $\bar{b} < d^{2}$ or, equivalently, $b < l_{\text{ch}} d^{2}$. Consequently, the failure strength $f_{\text{t}}$ upon crack initiation is over-estimated for a finite value of the length scale parameter $b$ and is recovered only in the vanishing limit $b \to 0$. In this sense, the \cite{CFI2016,CFI2024} model is length scale convergent.

\remark \label{rmk:cracking-function-cfil-limit} Note that the following identity holds for the function $\bar{\phi} (d)$ given in \cref{eq:characteristic-functions-CFIL}$_{2}$
\begin{align}\label{eq:cfil-functions-limit}
	\lim_{b \to 0} \bar{\phi} (d)
		= l_{\text{ch}} \dfrac{d^{2}}{\big(1 - d \big)^{2}}
	\qquad \Longrightarrow \qquad
	\lim_{b \to 0} \eta (d)
		= \dfrac{1}{l_{\text{ch}}} \big(1 - d \big)^{2}, \qquad		
	\lim_{b \to 0} \eta_{0}
		= \dfrac{1}{l_{\text{ch}}}
\end{align}
That is, in the vanishing limit $b \to 0$ the truncation in the degradation function \eqref{eq:energetic-function-CFIL} is inactive. $\Box$


\subsubsection{Limitations of the \cite{CFI2016,CFI2024} model}

Though $\varGamma$-convergence to the \cite{Barenblatt1959} CZM has been proved, the \cite{CFI2016,CFI2024} model exhibits some practical limitations.


\begin{figure}[ht] \centering
  \subfigure[Truncated degradation function $\omega (d)$]{
  \includegraphics[width=0.48\textwidth]{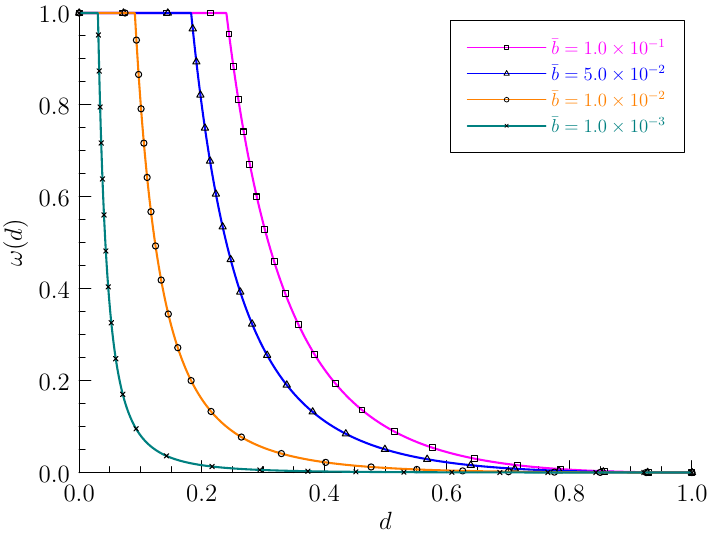}
  \label{fig:degradation-function-CFIL}} \hfill
  \subfigure[Piece-wisely smoothened degradation function $\omega (d)$]{
  \includegraphics[width=0.48\textwidth]{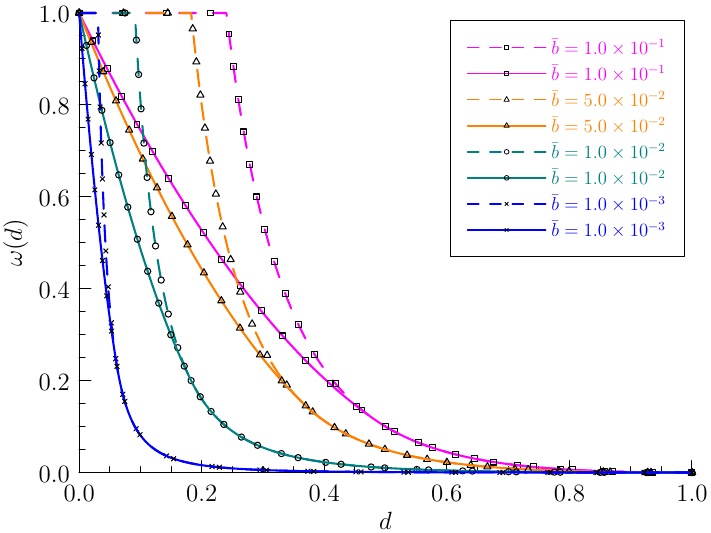}
  \label{fig:degradation-function-CFIL-smooth}}
  \caption{The degradation function $\omega (d)$ adopted in the \cite{CFI2016,CFI2024} model}
  \label{fig:degradation-functions-CFIL}  
\end{figure}

As shown in \cref{fig:degradation-function-CFIL}, the degradation function \eqref{eq:energetic-function-CFIL} is truncated at the limit $\omega (d) = 1$ and this truncation is activated except in the vanishing limit $b \to 0$. Due to the presence of this truncation, the resulting energy functional is neither differentiable nor convex with respect to the crack phase-field. Moreover, the crack driving force vanishes when the truncation is activated, i.e., $\omega' (d) = \varpi' (d) = 0$. 

In order to address the above issues, \cite{FI2017} suggested smoothening the truncated degradation function \eqref{eq:energetic-function-CFIL} by the following $\mathcal{C}^{1}$-approximation
\begin{align}\label{eq:degradation-function-cfil-smooth}
	\omega (d)
		= \begin{cases}
					\Big[ \hat{b}_{1} \big(d_{0} - d \big) + \hat{b}_{2} \Big]^{2} & \qquad
					0 \le d \le d_{0} \vspace{2mm} \\
				\bar{b} \dfrac{\big( 1 - d \big)^{2}}{d^{2}} & \qquad
					d_{0} \le d \le 1 
			\end{cases}
\end{align}
with the parameters 
\begin{align}
	d_{0}
		= \dfrac{2 \sqrt{\bar{b}}}{1 + \sqrt{\bar{b}}}, \qquad
	\hat{b}_{1}
		= \dfrac{\big(1 + \sqrt{\bar{b}} \; \big)^{2}}{4 \sqrt{\bar{b}}}, \qquad
	\hat{b}_{2}
		= \dfrac{1 - \sqrt{\bar{b}}}{2}
\end{align}
The above piece-wisely smoothened degradation function is depicted in \cref{fig:degradation-function-CFIL-smooth}. As can be seen, the difference between these two functions vanishes for a vanishing limit $b \to 0$. However, the smoothened degradation function \eqref{eq:degradation-function-cfil-smooth} does not fulfill the necessary condition \eqref{eq:degradation-function-cfil}
\begin{align}
	\lim_{d \to 0} d \cdot \bar{\omega} (d) 
		= \dfrac{1}{\sqrt{b}} \cdot \lim_{d \to 0} d \cdot
			\Big[ \hat{b}_{1} \big(d_{0} - d \big) + \hat{b}_{2} \Big]
		= 0 \notin (0, + \infty)	
\end{align}
Moreover, it follows that
\begin{align}
	\eta_{0}
		= \dfrac{1}{l_{\text{ch}}} \lim_{d \to 0} \dfrac{1}{\bar{b}}
			\dfrac{d^{2}}{1 - \big[ \hat{b}_{1} \big(d_{0} - d \big) 
		+ \hat{b}_{2} \big]^{2}}
		= \dfrac{1}{l_{\text{ch}}} \lim_{d \to 0}
			\dfrac{d}{\bar{b}	\hat{b}_{1}}
		= 0 \qquad \Longrightarrow \qquad
	\sigma_{\text{c}}
		= 0
\end{align}
That is, crack nucleation occurs at the very beginning once the load is applied no matter how small it is. As a consequence, though the $\mathcal{C}^{1}$-approximation \eqref{eq:degradation-function-cfil-smooth} restores the differentiability and convexity of the total energy potential with respect to the crack phase-field, the issue of crack nucleation still presents.

\subsection{Phase-field cohesive zone model (PF-CZM)}
\label{sec:pfczm-quadratic}

The aforesaid issues due to truncation in the degradation function \eqref{eq:energetic-function-CFIL} are overcome by the \texttt{PF-CZM} \citep{Wu2018,WN2018}. To this end, we present in this section a particular version of the \texttt{PF-CZM} corresponding to the \cite{CFI2016,CFI2024} model.

Noticing the identity \eqref{eq:cfil-functions-limit}, let us consider the following characteristic function $\bar{\phi} (d)$
\begin{align}\label{eq:pfczm-cfil-quadratic}
	\bar{\phi} (d)
		= l_{\text{ch}} \dfrac{d^{2}}{\big(1 - d \big)^{2}}
	\qquad \Longrightarrow \qquad
	\eta (d)
		= \dfrac{1}{l_{\text{ch}}} \big(1 - d \big)^{2}, \qquad
	\eta_{0}
		= \dfrac{1}{l_{\text{ch}}} 
\end{align}
such that the failure strength $\sigma_{\text{c}} = f_{\text{t}}$ is reproduced independently of the length scale parameter $b$.

\begin{figure}[ht] \centering
  \subfigure[Rational fraction degradation function $\omega (d)$]{
  \includegraphics[width=0.475\textwidth]{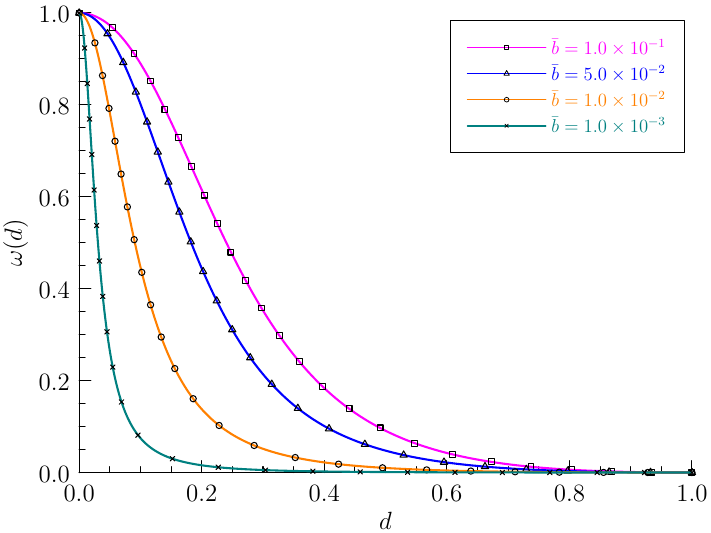}
  \label{fig:degradation-function-pfczm-quadratic}}  
	\subfigure[Comparison with the truncated degradation function]{
  \includegraphics[width=0.475\textwidth]{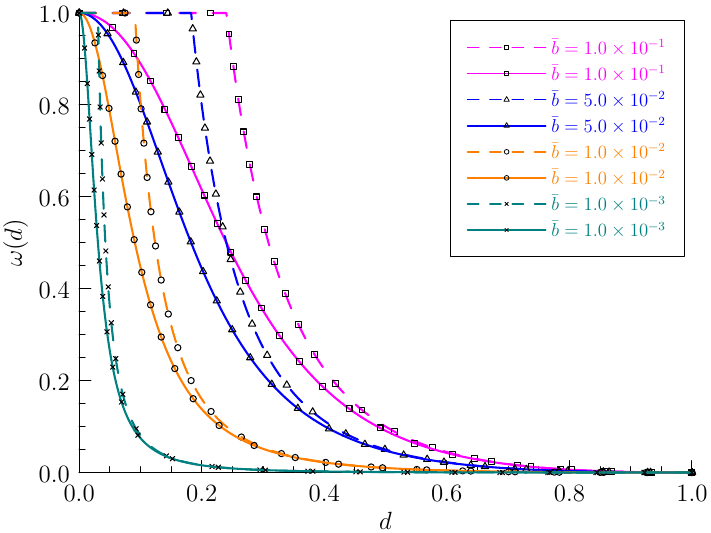}
  \label{fig:degradation-function-CFIL-pfczm-comparison}}
  \caption{The degradation function $\omega (d)$ of rational fraction adopted in the \texttt{PF-CZM}}
  \label{fig:degradation-functions-CFIL-smooth}  
\end{figure}

Accordingly, the cracking and dissipation functions are given by
\begin{align}\label{eq:pfczm-quadratic-degradation-function}
	\phi (d)
		= a_{0} \dfrac{d^{2}}{\big(1 - d \big)^{2}}, \qquad
	\omega (d)
		= \dfrac{\big(1 - d \big)^{2}}{\big(1 - d \big)^{2} 
		+ a_{0} d^{2}}
	\qquad \text{with} \qquad
	a_{0} 
		= \dfrac{1}{\bar{b}}
		= \dfrac{l_{\text{ch}}}{b}
\end{align}
which fulfills the following property
\begin{align}
	\lim_{b \to 0} \dfrac{\big(1 - d \big)^{2}}{\big(1 - d \big)^{2} 
		+ a_{0} d^{2}}
		= \lim_{b \to 0} \bar{b} \dfrac{\big(1 - d \big)^{2}}{
			\bar{b} \big(1 - d \big)^{2} + d^{2}}
		= \bar{b} \dfrac{\big(1 - d \big)^{2}}{d^{2}}
\end{align}
in the vanishing limit $b \to 0$. 

As shown in \cref{fig:degradation-functions-CFIL-smooth}, the degradation function \eqref{eq:pfczm-quadratic-degradation-function} of rational fraction is a continuous approximation of the truncated one \eqref{eq:energetic-function-CFIL} adopted in the \cite{CFI2016,CFI2024} model, and both degradation functions coincide in the vanishing limit $b \to 0$; see \cref{rmk:cracking-function-cfil-limit}.

\begin{figure}[h!] \centering
  \subfigure[Softening curve]{
  \includegraphics[width=0.48\textwidth]{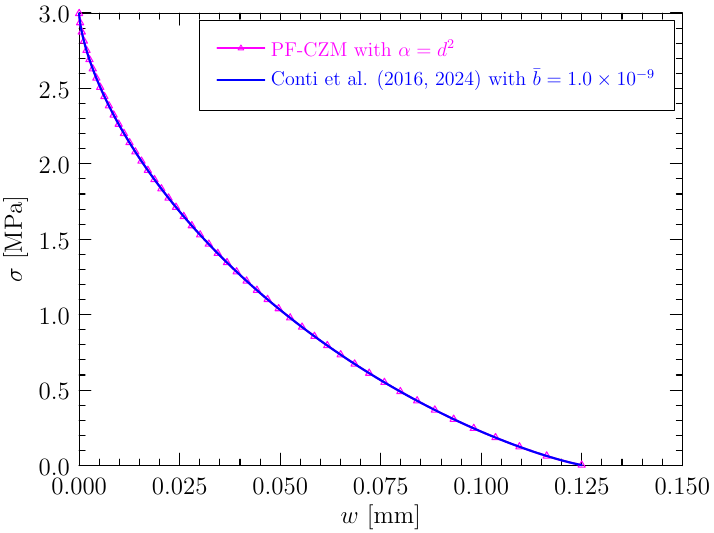}
  \label{fig:asymptotic-softening-curves-quadratic}} \hfill
  \subfigure[Surface energy density function]{
  \includegraphics[width=0.48\textwidth]{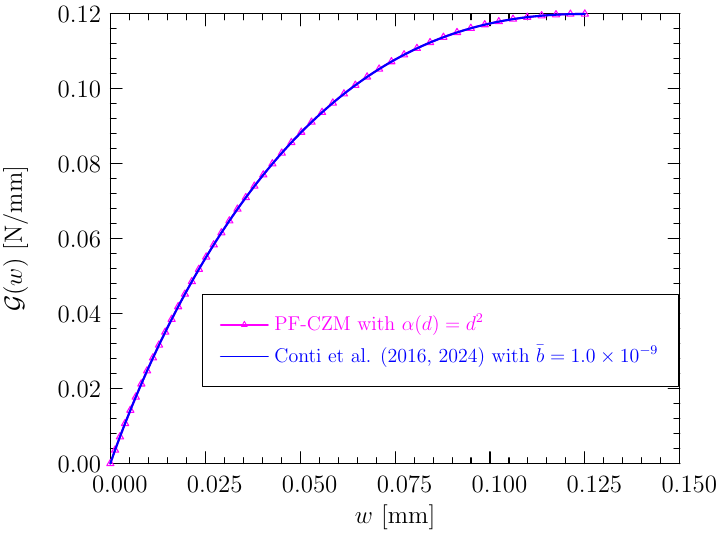}
  \label{fig:fracture-energy-pfczm-quadratic}}
  \caption{The softening curve and surface energy density function predicted by the \texttt{PF-CZM} with $\alpha (d) = d^{2}$. The material properties $f_{\text{t}} = 3.0$ MPa and $G_{\text{f}} = 0.12$ N/mm are adopted.}
  \label{fig:pfczm-quadratic} 
\end{figure}

The parameterized TSL \eqref{eq:stress-opening-1D} can be expressed as the following closed-form
\begin{subequations}\label{eq:tsl-pfczm-quadratic}
\begin{align}
	\sigma (d_{\ast})
	  &= f_{\text{t}} \big( 1 - d_{\ast} \big) \\
	\label{eq:crack-opening-pfczm-quadratic}
  \itw (d_{\ast})
   &= \itw_{\text{cL}} \Bigg[ \arccos \big(1 - d_{\ast} \big)
    - \big(1 - d_{\ast} \big) \ln \dfrac{1 
    + \sqrt{1 - (1 - d_{\ast})^{2}}}{1 - d_{\ast}} \Bigg]
\end{align}
\end{subequations}
Moreover, the surface energy density function \eqref{eq:fracture-energy-CZM} is given by
\begin{align}
	\mathcal{G} (d_{\ast})
	  = G_{\text{f}} \Bigg[ \sqrt{1 - (1 - d_{\ast})^{2}} 
	 	- \big( 1 - d_{\ast} \big)^{2} \ln \dfrac{1 
	 	+ \sqrt{1 - (1 - d_{\ast})^{2}}}{1 - d_{\ast}} \Bigg]
\end{align}
which fulfills the properties \eqref{eq:fracture-energy-properties-cfil} as expected. For the softening curve \eqref{eq:tsl-pfczm-quadratic}, the ultimate crack opening $\itw_{\text{c}}$ and crack bandwidth $D (d_{\ast})$ are given by
\begin{align}
	\itw_{\text{c}} :
		= \itw (d_{\ast} = 1)
		= \dfrac{\pi}{2} \itw_{\text{cL}}
		= \dfrac{\pi G_{\text{f}}}{f_{\text{t}}}, \qquad
  D (d_{\ast})
    =+\infty
\end{align}
As the crack band is of infinite support once fracture is initiated, the condition \eqref{eq:non-shrinking-crack-band} is automatically fulfilled.
The above global responses, characterized by a particular nonlinear softening curve with a finite ultimate crack opening, are insensitive to the phase-field length scale parameter $b$; see \cref{fig:pfczm-quadratic}.

\remark In the \cite{CFI2016,CFI2024} model and the \texttt{PF-CZM} counterpart, the degradation/dissipation function depends not only on the crack phase-field, but also on the length scale parameter. Though this dependence is introduced from different motivations, both models are closely inter-related and converges to the same CZM with a nonlinear softening curve in the vanishing limit $b \to 0$. The difference is that the conditions \eqref{eq:length-scale-insensitivity-condition} are fulfilled in the weak ($\varGamma$-convergent) sense for the \cite{CFI2016,CFI2024} model and in the strong (length scale insensitive) one for the \texttt{PF-CZM}, respectively.

\section{Length scale insensitive phase-field cohesive zone model}
\label{sec:pfczm}

In the \cite{CFI2016,CFI2024} model and the \texttt{PF-CZM} counterpart analyzed in \cref{sec:cfil-pfczm}, the associated formulation is considered together with the quadratic geometric function \eqref{eq:geometric-crack-function-quadratic}. In this section, the \texttt{PF-CZM} is discussed in the general form, with the non-associated formulation incorporated. 

\subsection{General formulation}

The simplest function $\mu (d)$ fulfilling the conditions \eqref{eq:conditions-degradation-functions} might be of the following rational fraction
\begin{align}
	\mu (d)
		= a_{0} \dfrac{Q (d)}{\big(1 - d \big)^{2 p}}
	\qquad \text{with} \qquad
	Q (0)
		= 0
\end{align}
for the parameter $a_{0} > 0$ and the exponent $p \ge 1$. Noticing the conditions $\alpha (0) = Q (0) = 0$, a natural choice is
\begin{align}\label{eq:assumed-function-Q}
	Q (d)
		= \alpha (d) P (d)
	\qquad \Longrightarrow \qquad
	\mu (d)
		= a_{0} \dfrac{\alpha (d)}{h (d)} \qquad
	\text{with} \qquad
	h (d)
		= \dfrac{\big(1 - d \big)^{2 p}}{P (d)}
\end{align}
with the function $P (d)$ satisfying the condition $P (0) = 1$. 

It then follows that
\begin{align}
	\eta (d)
		= \dfrac{\alpha (d)}{b \mu (d)}
		= \dfrac{1}{b a_{0}} h (d),
	\qquad
	\eta_{0}
		= \lim_{d \to 0} \eta (d)
		= \dfrac{1}{b a_{0}} \lim_{d \to 0} h (d)
		= \dfrac{1}{b a_{0}}
\end{align}
Accordingly, the length scale insensitivity conditions \eqref{eq:length-scale-insensitivity-condition} imply
\begin{align}\label{eq:pfczm-parameter-a0}
	\eta_{0}
		= \dfrac{1}{b a_{0}}
		= \dfrac{c_{\alpha}}{2 l_{\text{ch}}}
	\qquad \Longrightarrow \qquad
	a_{0}
		= \dfrac{2}{c_{\alpha}} \dfrac{l_{\text{ch}}}{b}
\end{align}
As can be seen, the parameter $a_{0}$ and the resulting function $\mu (d)$ are both inversely proportional to the length scale $b$. This fact intrinsically guarantees the length scale insensitivity of the \texttt{PF-CZM}.

With the above function $\mu (d)$, the parameterized TSL \eqref{eq:stress-opening-1D} becomes
\begin{subequations}\label{eq:pfczm-tsl-general}
\begin{align}\label{eq:pfczm-traction-general}
	\sigma (d_{\ast})
	 &= f_{\text{t}} \sqrt{h (d_{\ast})} \\
	\label{eq:pfczm-opening-general}
	\itw (d_{\ast})
	 &= \itw_{\text{cL}} \dfrac{2}{c_{\alpha} a_{0}} \sqrt{h (d_{\ast})}
	 		\int_{0}^{d_{\ast}} \dfrac{\phi (\vartheta) \sqrt{h (\vartheta) /
	 		\alpha (\vartheta)}}{\sqrt{h (\vartheta) - h (d_{\ast})}}
	 		\; \td \vartheta
\end{align}
\end{subequations}
Once the functions $\alpha (d)$ and $\phi (d)$ are known, the softening curve $\sigma (\itw)$ can be evaluated. 

Vice versa, for a given TSL $\sigma (\itw)$ is given, the cracking function $\phi (d)$ can be solved as \citep{PM2008,FFL2021,Wu2024} 
\begin{align}\label{eq:solution-Abel-equation-pfczm}
	\phi (d)
		= \dfrac{c_{\alpha} a_{0}}{\pi}
			\sqrt{\dfrac{\alpha (d)}{h (d)}} \; \bar{\varXi} (d)
	\qquad \text{with} \qquad
	\bar{\varXi} (d)
		= \dfrac{\partial}{\partial d} \Bigg(\int_{0}^{d} 
		- \frac{1}{2} \dfrac{\bar{\itw} (\vartheta) h' (\vartheta)
		/ \!\! \sqrt{h (\vartheta)}}{\sqrt{h (\vartheta) 
		- h (d_{\ast})}} \; \td \vartheta \Bigg)
\end{align}
such that the crack opening \eqref{eq:pfczm-opening-general} becomes
\begin{align}\label{eq:pfczm-opening-general-simplified}
	\itw (d_{\ast})
	 	= \itw_{\text{cL}} \bar{\itw} (d_{\ast}) 
	\qquad \text{with} \qquad
	\bar{\itw} (d_{\ast})
		= \dfrac{1}{\pi} \sqrt{h (d_{\ast})} \int_{0}^{d_{\ast}}	
			\dfrac{\bar{\varXi} (\vartheta)}{\sqrt{h (\vartheta) 
	 	- h (d_{\ast})}} \; \td \vartheta
\end{align}
for the normalized one $\bar{\itw} (d_{\ast}) := \itw (d_{\ast}) / \itw_{\text{cL}}$. As expected, the solved cracking function $\phi (d)$ also fulfills the conditions \eqref{eq:conditions-cracking-function} as the dissipation function $\mu (d)$ does. 

As stressed in \cref{sec:condition-non-shrinking-crack-band}, the above results hold only when the crack band does not shrink during the failure process, i.e., 
\begin{align}\label{eq:pfczm-crackbandwidth-general}
	\dfrac{\partial D}{\partial d_{\ast}} \ge 0
	\qquad \text{with} \qquad
  D (d_{\ast})
    = b \int_{0}^{d_{\ast}} \dfrac{\sqrt{h (\vartheta) / 
    	\alpha (\vartheta)}}{\sqrt{h (\vartheta) - h (d_{\ast})}}
    	\; \td \vartheta
\end{align}
If the closed-form of \cref{eq:pfczm-crackbandwidth-general} is not available, the condition \eqref{eq:irreversibility-condition} can be considered, i.e.,
\begin{align}\label{eq:irreversibility-condition-pfczm}
	D_{0}
		= \dfrac{\pi b}{\sqrt{\big[ 2 p + P' (0) \big] \alpha' (0)}} \le 
	D_{\text{u}}
		= b \int_{0}^{1} \dfrac{1}{\sqrt{\alpha (\vartheta)}} \; \td \vartheta
\end{align}
As can be seen, larger values $p \ge 1$, $P' (0)$ and $\alpha' (0)$ are favorable for satisfaction of the condition \eqref{eq:irreversibility-condition-pfczm}.

\subsection{Phase-field cohesive zone models with parameterized degradation function}
\label{sec:pfczm-assumed}

In this section, for the geometric function \eqref{eq:crack-geometric-function} the associated \texttt{PF-CZM} is addressed, i.e.,
\begin{subequations}\label{eq:degradation-function-pfczm}
\begin{align}
	\phi (d)
		= \mu (d)
		= a_{0} \dfrac{\alpha (d)}{\big(1 - d \big)^{2 p}} P (d)
\end{align}
For the function $P (d)$, \cite{Wu2017} suggested using the following parameterized polynomial
\begin{align}\label{eq:parameterized-polynomial-pfczm}
	P (d)
		= 1 + a_{1} d + a_{2} d^{2} + \cdots
\end{align}
\end{subequations}
with $a_{1}$, $a_{2}, \cdots$ being the parameters to be calibrated. 


In this case, the crack opening $\itw (d_{\ast})$ and surface energy density function $\mathcal{G} (d_{\ast})$ are evaluated as
\begin{subequations}\label{eq:pfczm-tsl-general-associated}
\begin{align}
	\label{eq:pfczm-crack-opening-general}
	\itw (d_{\ast})
	 &= \itw_{\text{cL}} \dfrac{2}{c_{\alpha}} \sqrt{h (d_{\ast})}
	 		\int_{0}^{d_{\ast}} \dfrac{\sqrt{\alpha (\vartheta) / 
	 		h (\vartheta)}}{\sqrt{h (\vartheta) - h (d_{\ast})}}
	 		\; \td \vartheta \\
	\label{eq:pfczm-energy-general}
	\mathcal{G} (d_{\ast})
	 &= \dfrac{4 G_{\text{f}}}{c_{\alpha}} 
	  	\int_{0}^{d_{\ast}} \bigg[ 1 - \dfrac{h (d_{\ast})}{h (\vartheta)} 
    	\bigg]^{-\frac{1}{2}}	\sqrt{\alpha (\vartheta)} \; \td \vartheta
\end{align} 
\end{subequations}
For the geometric function \eqref{eq:crack-geometric-function} and the cracking function \eqref{eq:degradation-function-pfczm}, it is rather difficult, if not impossible, to evaluate \cref{eq:pfczm-tsl-general-associated} analytically. Fortunately, the initial slope $k_{0}$ of the softening curve $\sigma (\itw)$ and the ultimate crack opening $\itw_{\text{c}}$ can be determined in closed-form
\begin{subequations}
\begin{align}
	\label{eq:initial-slope-pfczm-wu}
  k_{0} :
   &= \lim_{d_{\ast} \to 0} \dfrac{\partial \sigma}{\partial \itw}
		= \dfrac{c_{\alpha}}{2 \pi \sqrt{\xi}}
			\big(2 p + a_{1} \big)^{\frac{3}{2}} k_{\text{0L}} \\      
	\label{eq:ultimate-crack-opening-pfczm-wu}
	\itw_{\text{c}} :
	 &= \lim_{d_{\ast} \to 1} \itw
	  = \begin{cases}
	  		\dfrac{\pi}{c_{\alpha}} \itw_{\text{cL}} \sqrt{P (1)} & 
	  			\qquad p = 1 \\
	  		+\infty & \qquad p > 1 \\
	  	\end{cases}
\end{align}
\end{subequations}
for the initial slope $k_{\text{0L}} := -\frac{1}{2} f_{\text{t}}^{2} / G_{\text{f}}$ of the linear softening law \eqref{eq:linear-softening-cohesive}. 

Accordingly, for a given TSL $\sigma (\itw)$ the parameters $a_{1}$, $a_{2}$ ....., are determined as
\begin{subequations}\label{eq:pfczm-parameters-wu}
\begin{align}\label{eq:pfczm-parameter-a1}
 &a_{1}
	  = \bigg( \dfrac{2 \pi \sqrt{\xi}}{c_{\alpha}} \bar{k}_{0} 
			\bigg)^{\frac{2}{3}} - 2 p \\
 &a_{2} + \cdots 
	  = \begin{cases}
	  		0 & \qquad p > 1 \vspace{1mm} \\
	  		\bigg( \dfrac{c_{\alpha}}{\pi} \bar{\itw}_{\text{c}} 
	  			\bigg)^{2} - \big( 1 + a_{1} \big) & \qquad p = 1 \\
	  	\end{cases}
\end{align}
\end{subequations}
in terms of the ratios $\bar{k}_{0} := k_{0} / k_{\text{0L}}$ and $\bar{\itw}_{\text{c}} := \itw_{\text{c}} / \itw_{\text{cL}}$, respectively. 

The above results apply upon the condition \eqref{eq:irreversibility-condition-pfczm}
\begin{align}\label{eq:irreversibility-condition-pfczm-wu}
	D_{0}
	  = \dfrac{\pi b}{\sqrt{\xi \big(2 p + a_{1} \big)}} 
		= \pi b \bigg( \dfrac{2 \pi \xi^{2}}{c_{\alpha}} \bar{k}_{0}
			\bigg)^{-\frac{1}{3}} \le D_{\text{u}}
	\qquad \Longrightarrow \qquad
	\dfrac{2 \pi \xi^{2}}{c_{\alpha}} \bar{k}_{0}
			\ge \bigg( \dfrac{\pi b}{D_{\text{u}}} \bigg)^{3}
\end{align}
For the polynomial geometric function \eqref{eq:crack-geometric-function}, both the initial and ultimate crack bandwidths decrease monotonically with the parameter $\xi \in [0, 2]$, and the former decreases more rapidly \citep{Wu2017}. Dependent on the parameter $\xi \in [0, 2]$, the following \texttt{PF-CZM}s have been proposed in the literature:
\begin{align*}
\begin{cases}
	\text{PF$^{\; 0}$-CZM \citep{WZF2020}}: & \xi
		= 0, \; 
	\alpha (d)
		= d^{2} \\
	\text{PF$^{\frac{1}{2}}$-CZM \citep{Wu2017,MBBPB2020}}: & \xi
		= \frac{1}{2}, \; 
	\alpha (d)
		= \frac{1}{2} \big(d + d^{2} \big) \\
	\text{PF$^{\; 1}$-CZM \citep{Lorentz2017,Wu2017,GLHTD2019}}: & \xi
		= 1, \;\;
	\alpha (d)
		= d \\
	\text{PF$^{\; 2}$-CZM \citep{Wu2017,Wu2018,WN2018}}: & \xi
		= 2, \; 
	\alpha (d)
		= 2 d - d^{2} \\		
\end{cases}
\end{align*}
In the following subsections some of them are discussed in more details.


\subsubsection{PF$^{\; 0}$-CZM with the quadratic geometric function $\alpha (d) = d^{2}$}

For brittle fracture the quadratic geometric function $\alpha (d) = d^{2}$ yields the popular \texttt{AT2} model \citep{BFM2000,BFM2008}. It was also considered in the context of cohesive fracture \citep{WZF2020}
\begin{align}
	\alpha (d)
		= d^{2}
	\qquad \Longrightarrow \qquad
	c_{\alpha}
		= 2, \qquad
 	a_{0}
 		= \dfrac{l_{\text{ch}}}{b} 
\end{align}
In this case, it follows from \cref{eq:initial-slope-pfczm-wu} that 
\begin{align}
	\alpha' (0)
		= \xi
		= 0 \qquad \Longrightarrow \qquad
	k_{0}
		=-\infty
\end{align}
As the initial slope $k_{0}$ is negatively infinite,  \cref{eq:pfczm-parameter-a1} is indefinite. Therefore, in order to achieve a finite ultimate crack opening $\itw_{\text{c}}$, the exponent $p = 1$ is considered, leading to
\begin{align}
	\itw_{\text{c}} 
		= \dfrac{\pi}{2} \itw_{\text{cL}} \sqrt{P (1)}
	\qquad \Longrightarrow \qquad
	a_{1} + a_{2} + \cdots
		= \bigg( \dfrac{2}{\pi} \bar{\itw}_{\text{c}} \bigg)^{2} - 1
\end{align}
The \texttt{PF-CZM} presented in \cref{sec:pfczm-quadratic} is a particular instance with $a_{1} = a_{2} = \cdots = 0$, resulting in an ultimate crack opening $\itw_{\text{c}} = \frac{1}{2} \pi \itw_{\text{cL}}$, i.e., $\bar{\itw}_{\text{c}} = \frac{1}{2} \pi$.

\begin{figure}[h!] \centering
  \subfigure[Softening curve]{
  \includegraphics[width=0.48\textwidth]{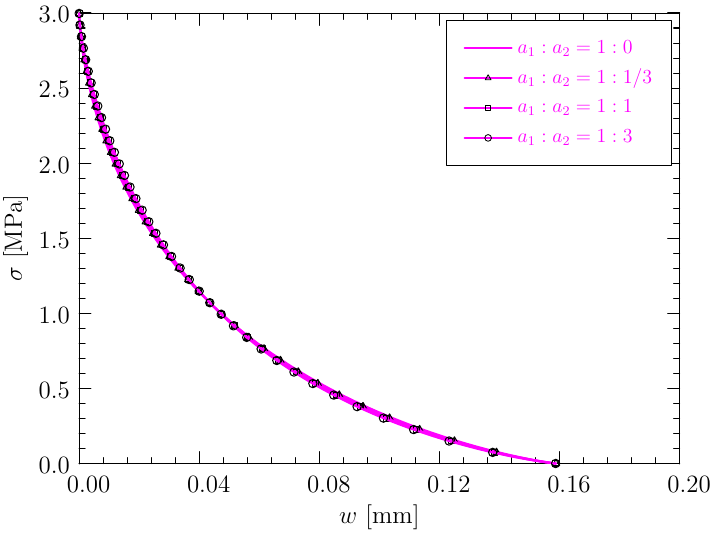}
  \label{fig:asymptotic-softening-curves-quadratic2}} \hfill
  \subfigure[Surface energy density function]{
  \includegraphics[width=0.48\textwidth]{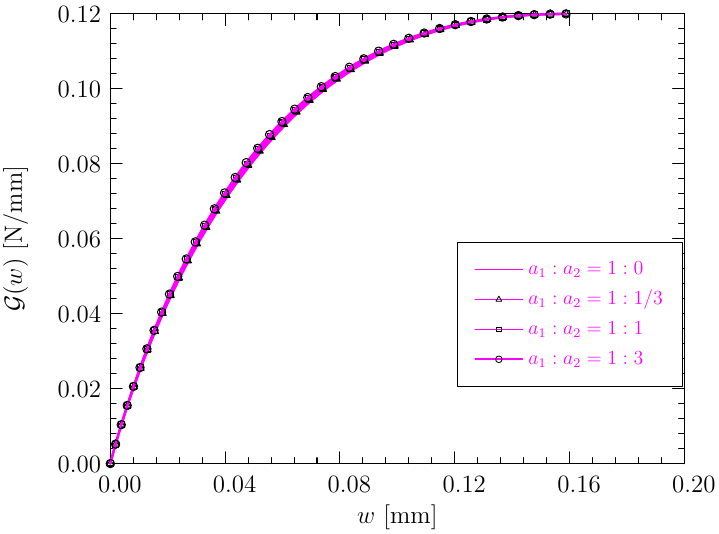}
  \label{fig:fracture-energy-pfczm-quadratic2}}
  \caption{The softening curve and surface energy density function given by the \texttt{PF$^{\; 0}$-CZM} with the quadratic geometric function $\alpha (d) = d^{2}$ and $\bar{\itw}_{\text{c}} = 2$. The material properties $f_{\text{t}} = 3.0$ MPa and $G_{\text{f}} = 0.12$ N/mm are adopted.}
  \label{fig:pfczm-quadratic2} 
\end{figure}

The softening curve and surface energy density function given by the above \texttt{PF$^{\; 0}$-CZM} are depicted in \cref{fig:pfczm-quadratic2}. As can be seen, the \texttt{PF$^{\; 0}$-CZM} is characterized by a nonlinear softening curve $\sigma (\itw)$ and a surface energy density function $\mathcal{G} (\itw)$ approaching the \cite{Griffith1921} fracture energy $G_{\text{f}}$ at the finite ultimate crack opening $\itw = \itw_{\text{c}}$. Remarkably, the specific values of the parameters $a_{1}$ and $a_{2}$ have negligible effects on the results.

\remark As in the \texttt{AT2} model for brittle fracture, the crack bandwidth $D (d_{\ast})$ is infinite, i.e.,
\begin{align}
	D (d_{\ast})
	  =+\infty
\end{align}
In this \texttt{PF$^{\; 0}$-CZM}, the crack phase-field needs to be solved in the whole domain. $\Box$




\subsubsection{PF$^{\; 1}$-CZM with the linear geometric function $\alpha (d) = d$}

The geometric function $\alpha (d) = d$ or, equivalently, $\xi = 1$, has been adopted in the \texttt{AT1} model for brittle fracture \citep{PAMM2011,MBK2015} and in the Lorentz model for cohesive one \citep{Lorentz2017,GLHTD2019}; see also \cite{Wu2017}. 

In this case, it follows that
\begin{align}
	\alpha (d)
		= d \qquad \Longrightarrow \qquad
	c_{\alpha}
		= \dfrac{8}{3}, \qquad
	a_{0}
		= \dfrac{3}{4} \dfrac{l_{\text{ch}}}{b}, \qquad
	D_{\text{u}}
		= 2 b
\end{align}
The condition \eqref{eq:irreversibility-condition-pfczm-wu} for a non-shrinking crack band becomes
\begin{align}
	D_{0}
		= \dfrac{\pi b}{\sqrt{2 p + a_{1}}} \le 
	D_{\text{u}}
		= 2 b
	\qquad \Longrightarrow \qquad
	2 p + a_{1} \ge \dfrac{1}{4} \pi^{2} \qquad \text{or} \qquad
	\bar{k}_{0} \ge \dfrac{\pi^{2}}{6}
\end{align}
That is, the \texttt{PF$^{\; 1}$-CZM} applies only to those softening curves with the initial slope ratio $\bar{k}_{0} \ge \frac{1}{6} \pi^{2}$. 

To be more specific, let us consider the following two softening curves.
\begin{itemize}
\item Linear softening ($\bar{k}_{0} = 1$): For a finite ultimate crack opening, the exponent $p = 1$ applies and \cref{eq:pfczm-parameters-wu} gives
\begin{align}
	a_{1}
		=-0.2293, \qquad
	a_{2}
		=-0.0502 \qquad \Longrightarrow \qquad
	D_{0}
		= 2.36 b \ge 
	D_{\text{u}}
		= 2 b
\end{align}
which violates the necessary condition \eqref{eq:irreversibility-condition-pfczm-wu}. In other words, the \texttt{PF$^{\; 1}$-CZM} should be used with cautions for linear softening.

\item Exponential softening ($\bar{k}_{0} = 2$): For this softening curve with an infinite ultimate crack opening, the exponent $p = 1.25 > 1$ applies and it then follows from \cref{eq:pfczm-parameters-wu} that
\begin{align}
	a_{1}
		= 0.3108, \qquad
	a_{2}
		= \cdots
		= 0 \qquad \Longrightarrow \qquad
	D_{0}
		= 1.874 b <
	D_{\text{u}}
		= 2 b
\end{align}
As expected, for cohesive fracture with exponential softening the crack band is non-shrinking during the failure process and the \texttt{PF$^{\; 1}$-CZM} applies.

\end{itemize}

\subsubsection{PF$^{\; 2}$-CZM with the optimal geometric function $\alpha (d) = 2 d - d^{2}$}

For cohesive fracture in brittle and quasi-brittle solids the most frequently adopted softening curves are either convex $\bar{k}_{0} > 1$ or linear $\bar{k}_{0} = 1$. Let us consider the least favorable case of linear softening with $\bar{k}_{0} = 1$, the condition \eqref{eq:irreversibility-condition-pfczm-wu} becomes
\begin{align}
	\dfrac{2 \pi \xi^{2}}{c_{\alpha}} 
		\ge \bigg( \dfrac{\pi b}{D_{\text{u}}} \bigg)^{3}	
		\qquad \Longrightarrow \qquad
	\xi
		= 2, \qquad
	\alpha (d)
		= 2 d - d^{2}, \qquad
	c_{\alpha}
		= \pi
\end{align}
That is, the geometric function $\alpha (d) = 2 d - d^{2}$ is optimal for the non-concave (linear and convex) softening curves with $\bar{k}_{0} \ge 1$ since the resulting \texttt{PF$^{\; 2}$-CZM} automatically guarantees a non-shrinking crack band \citep{Wu2017}. 


\begin{figure}[h!] \centering
  \subfigure[Linear softening]{
  \includegraphics[width=0.48\textwidth]{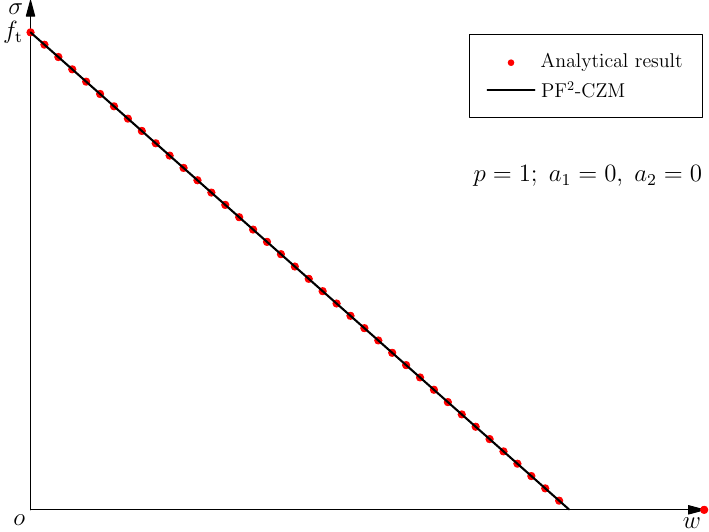}
  \label{fig:asymptotic-softening-curves-linear}} \hfill
  \subfigure[Exponential softening]{
  \includegraphics[width=0.48\textwidth]{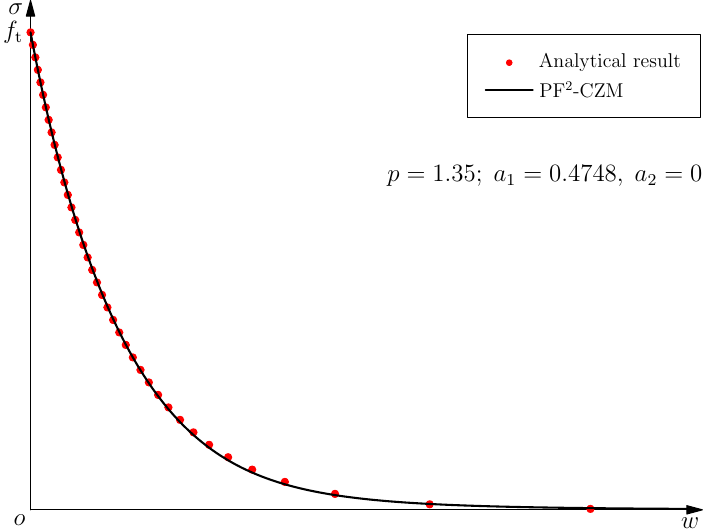}
  \label{fig:asymptotic-softening-curves-exponential}}
  \subfigure[\cite{CHR1986} softening]{
  \includegraphics[width=0.48\textwidth]{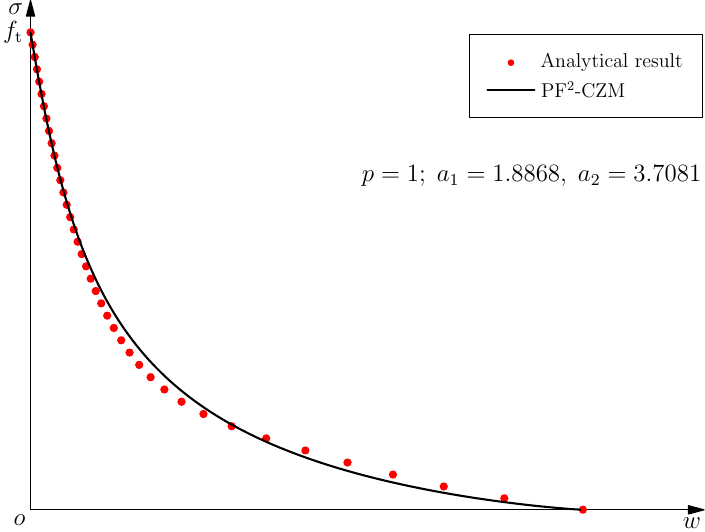}
  \label{fig:asymptotic-softening-curves-Hordijk}} \hspace{5.2mm}
  \subfigure[Evolution of the crack bandwidth]{
  \raisebox{-4.6mm}{\includegraphics[width=0.45\textwidth]{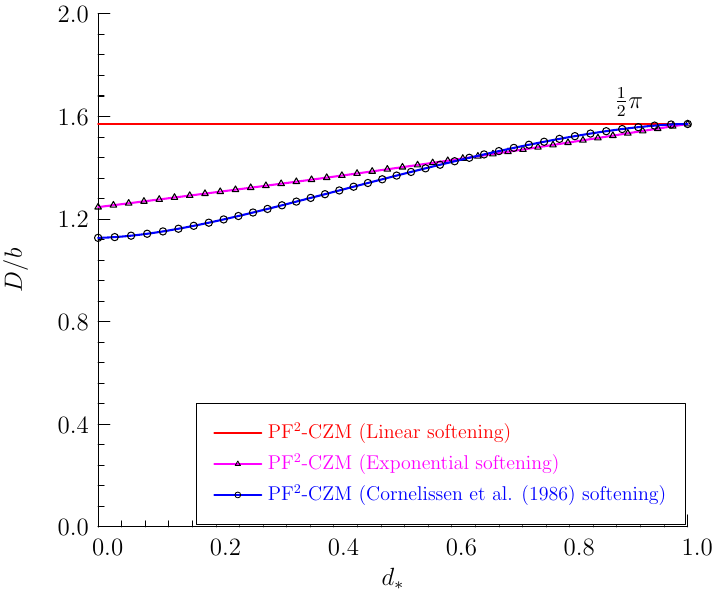}}
  \label{fig:pfczm2-bandwidths}}
  \caption{The softening curves given by the \texttt{PF$^{\; 2}$-CZM} with the optimal geometric function $\alpha (d) = 2 d - d^{2}$.}
  \label{fig:softening-curves-optimal-linear-exponential}  
\end{figure}

Accordingly, the parameters $a_{1}$, $a_{2}$, $\cdots$ are determined as
\begin{align}
  a_{1}
	  = 2 \big( \bar{k}_{0}^{\frac{2}{3}} - p \big), \qquad
  a_{2} + \cdots 
	  = \begin{cases}
	  		0 & \qquad p > 1 \\
	  		\bar{\itw}_{\text{c}}^{2} - \big( 1 + a_{1} \big) & 
	  		\qquad p = 1 \\
	  	\end{cases}
\end{align}
For those commonly adopted non-concave softening curves, the following parameters apply
\begin{align}
\begin{cases}
	p
		= 1, \phantom{.35} \qquad
	a_{1}
		= a_{2}
		= \cdots
		= 0 & \qquad \text{Linear softening} \\
	p
		= 1.35, \qquad
	a_{1}
		= 0.4748, \qquad
	a_{2}
		= \cdots
		= 0 & \qquad \text{Exponential softening} \\
	p
		= 1, \phantom{.35} \qquad
	a_{1}
		= 1.8868, \qquad
	a_{2}
		= 3.7081 & \qquad \text{\cite{CHR1986} softening} \\					
\end{cases}
\end{align}
\Cref{fig:softening-curves-optimal-linear-exponential} compares the softening curves predicted by \cref{eq:pfczm-traction-general,eq:pfczm-crack-opening-general} against the target ones. Among them, the linear softening curve is exactly reproduced, while the discrepancies for the exponential one are invisible. For the \citet{CHR1986} softening curve, the discrepancy is also acceptable and can be reduced by increasing the order of $P (d)$ as in \cite{WYL2023}. Other non-concave softening curves can also be considered. Importantly, the crack bandwidth $D (d_{\ast})$ monotonically increases to a finite value $D_{u} = \frac{1}{2} \pi b$ for any non-concave softening curve.

\subsection{Phase-field cohesive zone models with solved degradation function}
\label{sec:pfczm-solved}

In the \texttt{PF-CZM}s discussed in \cref{sec:pfczm-assumed}, the associated formulation is adopted and the degradation/dissipation function of rational fraction is postulated \textit{a priori} in terms of a parameterized polynomial function \eqref{eq:parameterized-polynomial-pfczm}. Though the \texttt{PF$^{\; 2}$-CZM} with the optimal geometric function applies to linear and general convex softening behavior, those concave softening curves, e.g., the \cite{PPR2009} one for adhesive fracture, cannot be considered. Aiming to address the above issues, the author \citep{Wu2024} recently proposed the generalized phase-field cohesive zone model ($\mu$\texttt{PF-CZM}) with different degradation and dissipation functions. Almost any arbitrary TSL softening law can be reproduced with the condition \eqref{eq:irreversibility-condition-pfczm} for a non-shrinking crack band intrinsically fulfilled. 


In the $\mu$\texttt{PF-CZM}, the simplest polynomial $P (d)$ is considered in \cref{eq:assumed-function-Q}
\begin{align}\label{eq:gpfczm-dissipation-function}
	P (d)
		= 1 \qquad \Longrightarrow \qquad
	Q (d)
		= \alpha (d), \qquad
	\mu (d)
		= a_{0} \dfrac{\alpha (d)}{h (d)}, \qquad
	h (d)
		= \big(1 - d \big)^{2 p}
\end{align}
In this case, the cracking function $\phi (d)$ given in \cref{eq:solution-Abel-equation-pfczm} simplifies into
\begin{align}\label{eq:cracking-function-Abel}
	\phi (d)
		= a_{0} p \dfrac{c_{\alpha}}{\pi} 
			\dfrac{\sqrt{\alpha (d)}}{\big(1 - d \big)^{p + 1}} \varXi (d)
	\qquad \text{with} \qquad \;
	\varXi (d)
		= \big(1 - d \big) \dfrac{\partial}{\partial d} 
	  	\Bigg[ \int_{0}^{d} \dfrac{\bar{\itw} (\vartheta) 
	  	\big(1 - \vartheta \big)^{p - 1}}{\sqrt{h (\vartheta) - h (d)}} 
	  	\; \td \vartheta	\Bigg]
\end{align}
where the function $\varXi (d)$ depends on the specific softening curve to be adopted. For instance, the following expressions were derived in \cite{Wu2024}
\begin{align}\label{eq:cracking-function-characteristics}
	\varXi (d)
		= \begin{cases}
				s (d) & \qquad \text{Linear softening} \\
				\dfrac{1}{2} \text{arctanh} \big( s(d) \big) & \qquad 
					\text{Exponential softening} \\
				\bar{c}_{1} s + \bar{c}_{3} s^{3}
				+ \bar{c}_{5} s^{5} + \big( \bar{c}_{2} s_{1}^{2}	
				+	\bar{c}_{4} s_{1}^{4} + \bar{c}_{6} s_{1}^{6} \big)
					\; \text{arctanh} \big( s(d) \big) & \qquad 
					\text{6th-order polynomial softening} \\
			\end{cases}
\end{align}
for the functions
\begin{align}
	s (d)
		= \sqrt{1 - h (d)}
		= \sqrt{1 - (1 - d)^{2p}}, \qquad
	s_{1} (d) 
		= \sqrt{h (d)}
		= (1 - d)^{p}
\end{align}
where the coefficients $\bar{c}_{i} \; (i = 1, 2, \cdots, 6)$ for the 6th-order polynomial fitting of the \cite{CHR1986} and \cite{PPR2009} softening curves are given in \ref{sec:softening-curves}.
%
%

\begin{figure}[h!] \centering
  \subfigure[Linear softening]{
  \includegraphics[width=0.48\textwidth]{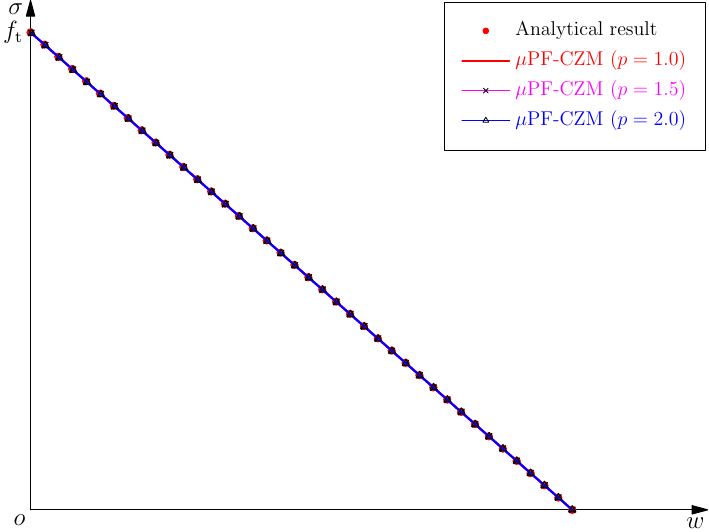}
  \label{fig:gpfczm-softening-curves-linear}} \hfill
  \subfigure[Exponential softening]{
  \includegraphics[width=0.48\textwidth]{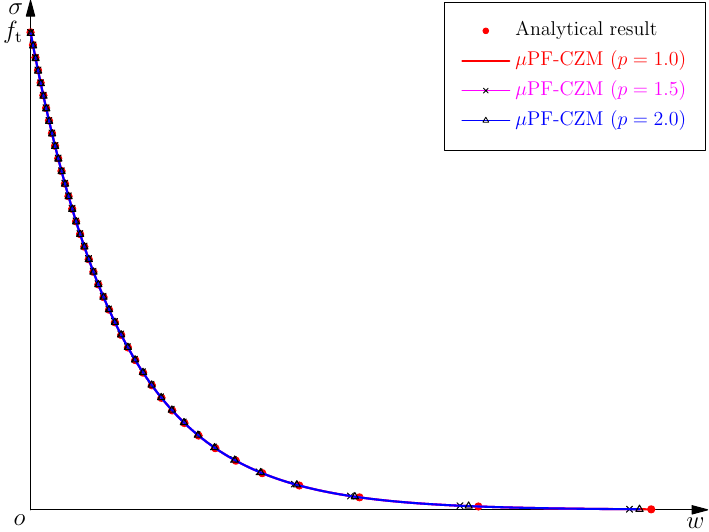}
  \label{fig:gpfczm-softening-curves-exponential}}
  \subfigure[\cite{CHR1986} softening]{
  \includegraphics[width=0.48\textwidth]{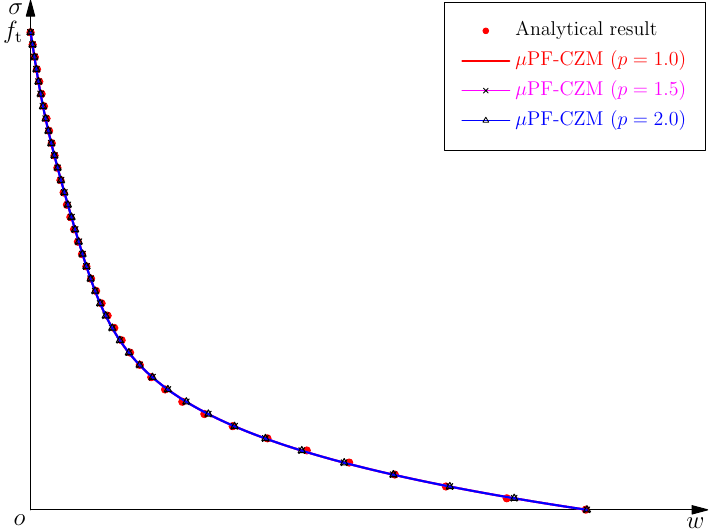}
  \label{fig:gpfczm-softening-curves-cornelissen}} \hfill
  \subfigure[\cite{PPR2009} softening]{
  \includegraphics[width=0.48\textwidth]{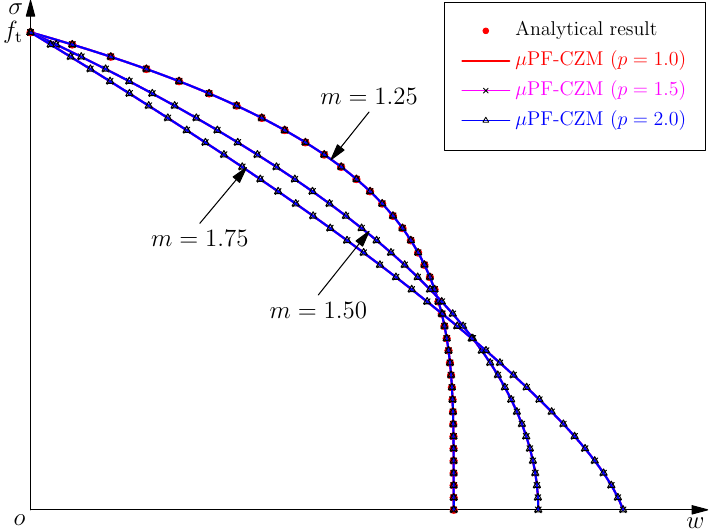}
  \label{fig:gpfczm-softening-curves-PPR}} 
  \caption{Softening curves predicted by the \texttt{$\mu$PF-CZM} \citep{Wu2024}.}
  \label{fig:gpfczm-softening-curves}  
\end{figure}

Accordingly, the traction--separation softening law parameterized in \cref{eq:pfczm-traction-general,eq:pfczm-opening-general-simplified} becomes
\begin{subequations}\label{eq:gpfczm-tsl}
\begin{align}
	\label{eq:traction-gpfczm}
	\sigma (d_{\ast})
	 &= f_{\text{t}} \big( 1 - d_{\ast} \big)^{p} \\
	\label{eq:crack-opening-gpfczm-general}
  \itw (d_{\ast}) 
	 &= \itw_{\text{cL}} \big(1 - d_{\ast} \big)^{p} \dfrac{1}{\pi} 
	 		\int_{0}^{d_{\ast}} \dfrac{2p \varXi (\vartheta) / (1 - \vartheta)}{\sqrt{h (\vartheta) - h (d_{\ast})}} \; \td \vartheta
\end{align}
\end{subequations}
The resulting softening curve is depicted in \cref{fig:gpfczm-softening-curves}. As expected, all the linear, convex and concave softening curves can be reproduced independently of the exponent $p \ge 1$ \citep{Wu2024}. 

Again, the above results holds upon the condition \eqref{eq:non-shrinking-crack-band}, with the crack bandwidth $D (d_{\ast})$ given by
\begin{align}\label{eq:pfczm-bandwidth-general}
	\dfrac{\partial D}{\partial d_{\ast}} \ge 0
	\qquad \text{with} \qquad
	D (d_{\ast})
	  = \int_{0}^{d_{\ast}} \dfrac{(1 - \vartheta)^{p}}{
	 		\sqrt{\alpha (\vartheta)}} \cdot \dfrac{1}{
	 		\sqrt{h (\vartheta) - h (d_{\ast})}} \; \td \vartheta
\end{align}
In particular, the condition \eqref{eq:irreversibility-condition-pfczm} for a non-shrinking crack band becomes
\begin{align}\label{eq:irreversibility-condition-gpfczm}
	D_{0}
		= \dfrac{\pi b}{\sqrt{2p \alpha' (0)}} \le 
	D_{\text{u}}
		= b \int_{0}^{1} \dfrac{1}{\sqrt{\alpha (\vartheta)}} \; \td \vartheta
\end{align}
from which the optimal geometric function $\alpha (d)$ can be determined.

Regarding whether the crack driving force is associated or non-associated, two strategies can be considered.



\subsubsection{Associated formulation}
\label{sec:pfczm-feng-li}

Let us first discuss the associated formulation, i.e., $\phi (d) = \mu (d)$. In this case, it follows from \cref{eq:gpfczm-dissipation-function,eq:cracking-function-Abel} that
\begin{align}
	a_{0} \dfrac{\alpha (d)}{\big(1 - d \big)^{2 p}}
		= a_{0} p \dfrac{c_{\alpha}}{\pi} 
			\dfrac{\sqrt{\alpha (d)}}{\big(1 - d \big)^{p + 1}} \; 
			\varXi (d) \qquad \Longrightarrow \qquad
	\sqrt{\alpha (d)}
		= p \dfrac{c_{\alpha}}{\pi} \big(1 - d \big)^{p - 1} \varXi (d)
\end{align}
or, equivalently,
\begin{align}\label{eq:geometric-function-pfczm-feng}
	\sqrt{\alpha (d)}
		= \big(1 - d \big)^{p - 1} \varXi (d), \qquad
	c_{\alpha}
		= 4 \int_{0}^{1} \sqrt{\alpha (\vartheta)} \; \td \vartheta
		= \dfrac{1}{p} \pi
\end{align}
Accordingly, the cracking function $\phi (d)$ is determined as
\begin{align}\label{eq:cracking-function-gpfczm-associated}
	\phi (d)
		= \mu (d)
		= a_{0} \dfrac{\varXi^{2} (d)}{\big(1 - d \big)^{2}}
	\qquad \text{with} \qquad
	a_{0}
		= \dfrac{2p}{\pi} \dfrac{l_{\text{ch}}}{b}
\end{align}
The TSL is still given by \cref{eq:gpfczm-tsl}, with the following surface energy density function $\mathcal{G} (d_{\ast})$
\begin{align}
	\mathcal{G} (d_{\ast})
	  = \dfrac{4 p}{\pi} G_{\text{f}} \int_{0}^{d^{\ast}} 
	  	\dfrac{\big(1 - \vartheta \big)^{2 p - 1} \varXi (\vartheta)}{
	  	\sqrt{h (\vartheta) - h (d_{\ast})}} \; \td \vartheta
\end{align}
Again, the above results hold provided the crack bandwidth $D (d_{\ast})$
\begin{align}\label{eq:crack-bandwidth-pfczm-feng}
  D (d_{\ast})
    = b \int_{0}^{d_{\ast}} \dfrac{1 - \vartheta}{\varXi (\vartheta)} 
    	\cdot \dfrac{1}{\sqrt{h (\vartheta) - h (d_{\ast})}} 
    	\; \td \vartheta
\end{align}
is monotonically non-decreasing. Note that the lowest-order case of linear traction (i.e., $p = 1$) recovers the particular version of the \texttt{PF-CZM} developed in \cite{FFL2021}. 

The associated $\mu$\texttt{PF-CZM} has the merit that all the geometric and degradation/dissipation functions can be analytically determined in closed-form. However, the geometric function \eqref{eq:geometric-function-pfczm-feng} is not optimal since the condition \eqref{eq:irreversibility-condition-gpfczm} cannot be \textit{a priori} fulfilled. In order for better clarification, the following two particular softening curves are discussed.
\begin{itemize}
\item Linear softening. For the parameterized traction \eqref{eq:traction-gpfczm}, linear softening gives
\begin{align}
	\bar{\itw} (d_{\ast})
		= 1 - \big(1 - d_{\ast} \big)^{p}, \qquad
	\mathcal{G} (\itw (d_{\ast}))
		= G_{\text{f}} \Big[ 1 - \big(1 - d_{\ast} \big)^{2p} \Big] 
		= G_{\text{f}} \bar{\itw} \big( 2 - \bar{\itw} \big)		
\end{align}
As shown in \cref{fig:fracture-energy-pfczm-linear-feng}, the surface energy density function $\mathcal{G} (\itw)$ approaches the fracture energy $G_{\text{f}}$ at the ultimate crack opening $\itw = \itw_{\text{cL}}$ (i.e., $\bar{\itw} = 1$) \footnote[4]{In \cite{CdeB2021} only the first term in the energy dissipation \eqref{eq:energy-dissipation} was accounted for, leading to the conclusion ``\textit{the phase field method can partially produce the response of cohesive zone model, including the traction-separation law, but not entirely reproduce the cohesive response, which for instance holds for the dissipated energy.}'' However, with the missed second term in \cref{eq:energy-dissipation} was accounted for, the \texttt{PF-CZM} and $\mu$\texttt{PF-CZM} can reproduce not only the TSL but also the energy dissipation, so long as the condition \eqref{eq:pfczm-bandwidth-general} is fulfilled.}.

\Cref{fig:cohesive_pfczm_linear_feng_bandwidth} presents evolution of the crack bandwidth \eqref{eq:crack-bandwidth-pfczm-feng} for various value of the traction order parameter $p \ge 1$. 
As expected, the condition \eqref{eq:irreversibility-condition-gpfczm} is fulfilled for linear softening.

\begin{figure}[h!] \centering
  \subfigure[Surface energy density function $\mathcal{G} (\itw)$]{
  \includegraphics[width=0.48\textwidth]{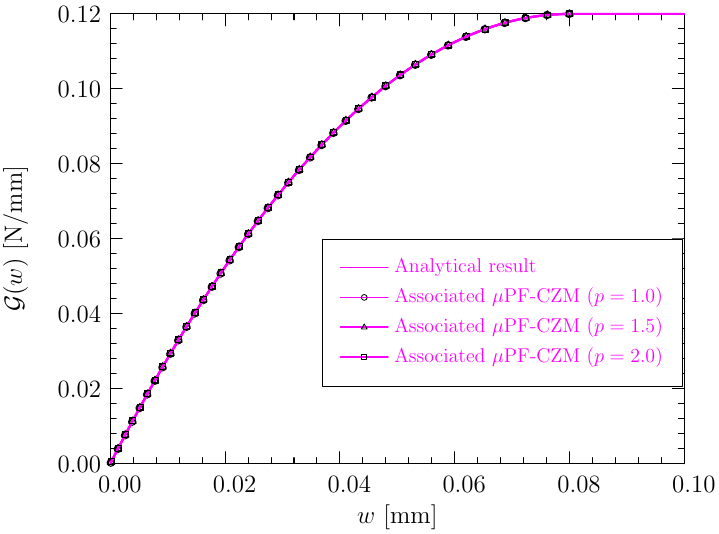}
  \label{fig:fracture-energy-pfczm-linear-feng}} \hfill
  \subfigure[Evolution of the crack band width $D (d_{\ast})$]{
  \includegraphics[width=0.48\textwidth]{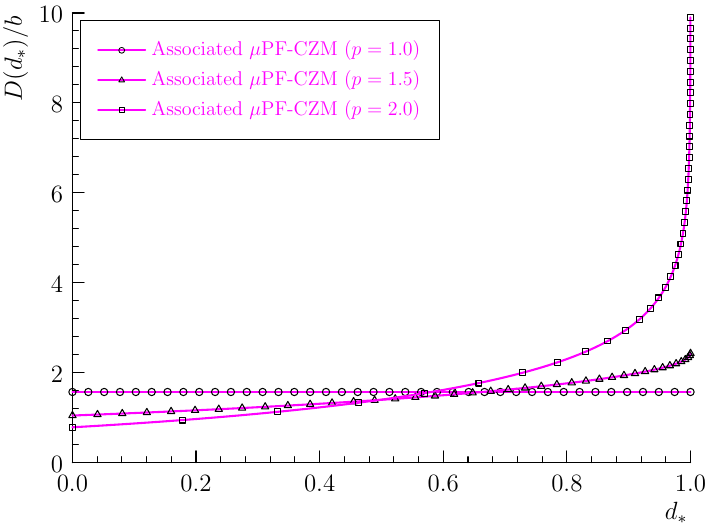}
  \label{fig:cohesive_pfczm_linear_feng_bandwidth}}
  \caption{The surface energy density function and crack band width predicted by the associated $\mu$\texttt{PF-CZM} with the solved degradation function for linear softening. The material properties $f_{\text{t}} = 3.0$ MPa and $G_{\text{f}} = 0.12$ N/mm are adopted.}
  \label{fig:cohesive_pfczm_linear_feng} 
\end{figure}

%
\item Exponential softening. For the parameterized traction \eqref{eq:traction-gpfczm}, exponential softening gives
\begin{align}
	\bar{\itw} (d_{\ast}) 
		=-\dfrac{1}{2} \ln \big(1 - d_{\ast} \big)^{p}, \qquad
	\mathcal{G} (\itw (d_{\ast}))
		= G_{\text{f}} \Big[ 1 - \big(1 - d_{\ast} \big)^{p} \Big]
		= G_{\text{f}} \Big[ 1 - \exp \big( -2 \bar{\itw} \big) \Big]
\end{align}
As shown in \cref{fig:fracture-energy-pfczm-exponential-feng}, the surface energy density function increases asymptotically to the fracture energy $G_{\text{f}}$. 

\begin{figure}[h!] \centering
  \subfigure[Surface energy density function $\mathcal{G} (\itw)$]{
  \includegraphics[width=0.475\textwidth]{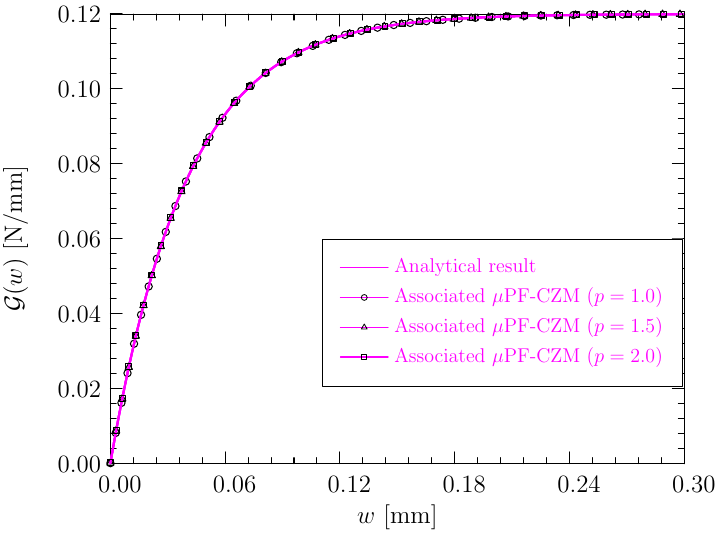}
  \label{fig:fracture-energy-pfczm-exponential-feng}} \hfill
  \subfigure[Half crack band width $D (d_{\ast})$]{
  \includegraphics[width=0.475\textwidth]{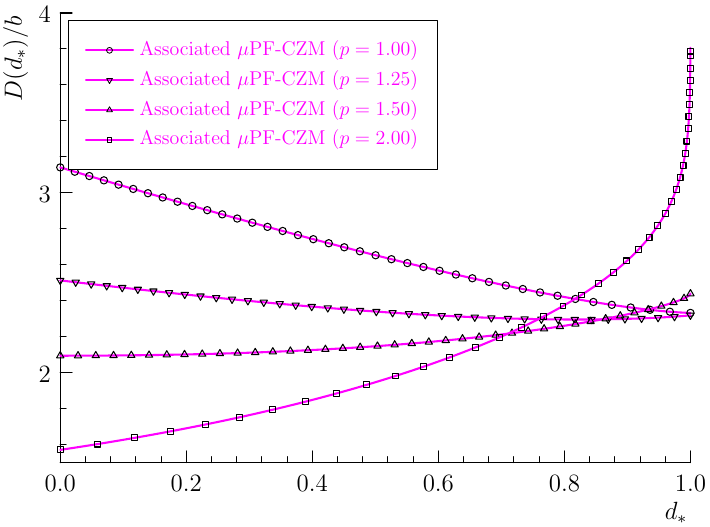}
  \label{fig:cohesive_pfczm_exponential_feng_bandwidth}}
  \caption{The surface energy density function and crack band width predicted by the associated $\mu$\texttt{PF-CZM} with the solved degradation function for exponential softening. The material properties $f_{\text{t}} = 3.0$ MPa and $G_{\text{f}} = 0.12$ N/mm are adopted.}
  \label{fig:cohesive_pfczm_exponential_feng} 
\end{figure}

However, as shown in \cref{fig:cohesive_pfczm_exponential_feng_bandwidth}, the increasing monotonicity of the crack bandwidth $D (d_{\ast})$ can only be guaranteed for the traction order parameter $p \ge 1.5$. 
That is, for the traction order parameter $p < 1.5$, the crack band shrinks as cracking proceeds and some regions experience unloading. In such cases the crack irreversibility condition will affect the softening curve and the expected exponential one cannot be reproduced. This conclusion also applies to the \cite{CHR1986} softening; see \cite{Wu2024}. 

\end{itemize}

In summary, the associated $\mu$\texttt{PF-CZM} with solved geometric function cannot always guarantee a non-shrinking crack band during the failure process and should be used with cautions.

\remark Except for the case $p = 1$, the geometric function \eqref{eq:geometric-function-pfczm-feng} vanishes both at $d = 0$ and $d = 1$ such that the increasing monotonicity \eqref{eq:geometric-function-conditions}$_{2}$ is not satisfied. As a consequence,  the associated $\mu$\texttt{PF-CZM} with a higher-order exponent $p > 1$ does not yield the expected crack profile of the bullet shape \citep{Wu2024}, though it is beneficial for guaranteeing a non-shrinking crack band. $\Box$

\subsubsection{Non-associated formulation}
\label{sec:gPFCZM}

The aforesaid issues exhibited by the associated $\mu$\texttt{PF-CZM} were successfully removed in the recently proposed non-associated formulation \citep{Wu2024}. That is, the degradation function is different from the dissipation one, i.e., $\mu (d) \ne \phi (d)$ and $\varpi' (d) \ne \omega' (d)$, though in some particular cases both functions may coincide. 

\begin{figure}[h!] \centering
  \includegraphics[width=0.5\textwidth]{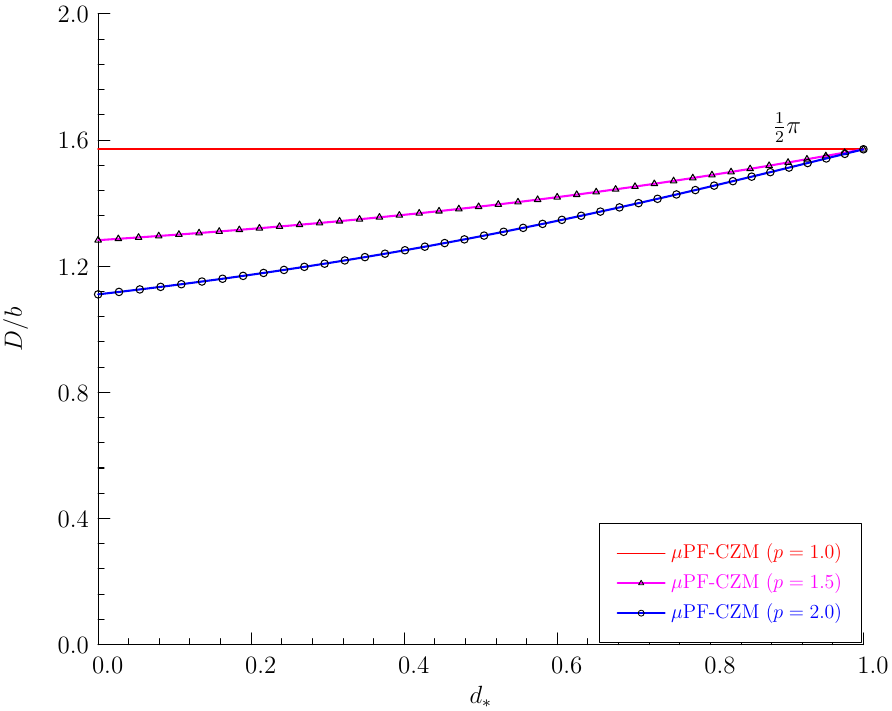}
  \caption{Evolution of the crack bandwidth given by the non-associated $\mu$\texttt{PF-CZM} with the optimal geometric function $\alpha (d) = 2d - d^{2}$.}
  \label{fig:gpfczm-bandwidth-orders}
\end{figure}

In this case, as the geometric function $\alpha (d)$ does not affect the parameterized traction--separation law \eqref{eq:gpfczm-tsl}, an optimal expression can be determined from the condition \eqref{eq:non-shrinking-crack-band} or \eqref{eq:irreversibility-condition}. As proved in \cite{Wu2024}, regarding the parameterized polynomial function \eqref{eq:crack-geometric-function} with the parameter $\xi \in [0, 2]$, only the geometric function
\begin{align}\label{eq:optimal-geometric-function}
	\xi
		= 2 \qquad \Longrightarrow \qquad
	\alpha (d)
		= 2 d - d^{2}, \qquad
	c_{\alpha}
		= \pi
\end{align}
is optimal for arbitrary softening curves and any traction order $p \ge 1$ in the sense that the resulting crack band is non-shrinking as expected; see \cref{fig:gpfczm-bandwidth-orders}. 

Correspondingly, the auxiliary dissipation function \eqref{eq:gpfczm-dissipation-function} and cracking function \eqref{eq:cracking-function-Abel} become
\begin{align}
	\mu (d)
		= a_{0} \dfrac{2 d - d^{2}}{\big(1 - d \big)^{2p}}, \qquad
	\phi (d)
		= a_{0} p \dfrac{\sqrt{2 d - d^{2}}}{\big(1 - d \big)^{p + 1}}
			\cdot \varXi (d), \qquad
	a_{0}
		= \dfrac{2}{\pi} \dfrac{l_{\text{ch}}}{b}
\end{align}
In particular, for the linear softening curve and the lowest traction order $p = 1$, it follows that
\begin{align}
	\varXi (d)
		= \sqrt{2 d - d^{2}} \qquad \Longrightarrow \qquad
	\phi (d)
		= \mu (d)
		= a_{0} \dfrac{2 d - d^{2}}{\big(1 - d \big)^{2}}
\end{align}
which is exactly the result of the \texttt{PF$^{2}$-CZM} \citep{Wu2017,Wu2018,WN2018} for linear softening. 


\remark Though the traction order parameter $p \ge 1$ does not affect the TSL $\sigma (\itw)$, it does affect the crack bandwidth $D (d_{\ast})$ as shown in \cref{fig:gpfczm-bandwidth-orders}. However, for various exponents $p \ge 1$ the ultimate crack bandwidths $D_{\text{u}}$ coincide since it depends only on the geometric function $\alpha (d)$. $\Box$

\section{Numerical examples}
\label{sec:numerical-examples}

In \cite{Wu2024}, the non-associated $\mu$\texttt{PF-CZM} was applied to the modeling of mode-I fracture in brittle and quasi-brittle solids. It was proved that the non-associated $\mu$\texttt{PF-CZM} is sensitive neither to the phase-field length scale parameters as the original associated \texttt{PF$^{2}$-CZM} nor to the traction order parameter $p \ge 1$. 

In this section, the non-associated $\mu$\texttt{PF-CZM} is further validated against several numerical examples of concrete under mode-I and mixed-mode failure. For the sake of comparison, the numerical results predicted by the \texttt{PF$^{2}$-CZM} are also presented. The \cite{CHR1986} softening law is adopted for concrete. Note the minor difference of the softening curves given in \cref{fig:asymptotic-softening-curves-Hordijk} by the \texttt{PF$^{2}$-CZM} and in \cref{fig:gpfczm-softening-curves-cornelissen} by the non-associated $\mu$\texttt{PF-CZM}, respectively. Moreover, an extra example, which involves mode-I fracture with the concave \cite{PPR2009} softening curve and thus defies the associated \texttt{PF-CZM}, is presented. 

The numerical simulations were performed by our in-house code \textsc{feFrac} based on the open-source library \textsc{jive} \citep{NNVMW2020}. Unstructured quadrilateral 4-node bilinear elements (Q4) generated by \textsc{Gmsh} \citep{gmsh2009} were used. Visualization was performed in \textsc{Paraview} \citep{Utkarsh2015}. 

\subsection{Koyna dam under overflow pressure}

\begin{figure}[hbtp]
  \centering
  \includegraphics[width=0.6\textwidth]{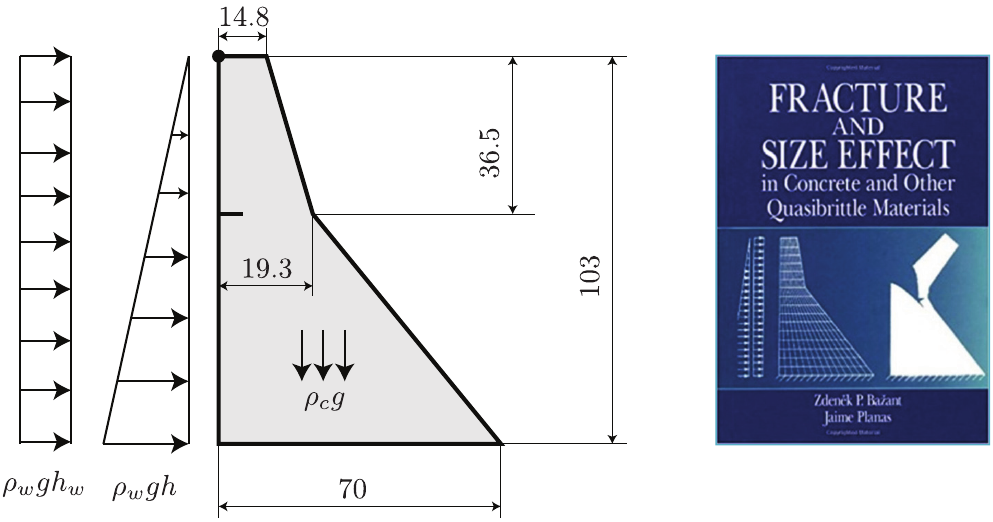}
  \caption{Koyna dam under overflow pressure. Left: Geometry (Unit of length: m), boundary and loading conditions; Right: Book cover of the monograph \citep{BP1997}}
  \label{fig:koyna-dam-geometry}
\end{figure}

The first example is mode-I fracture in the well-known Koyna dam under overflow pressure. As depicted in \cref{fig:koyna-dam-geometry}, it is a concrete gravity dam of height 103 m. The initial crack length is 1.93 m, $10\%$ of the dam width at the location with varying slopes on the downstream face. 

In the numerical simulation, the following material properties were adopted: Young's modulus $E_{0} = 2.5 \times 10^{4}$ MPa, Poisson's ratio $\nu_{0} = 0.20$, the failure strength $f_{\text{t}} = 1.0$ MPa, the fracture energy $G_{\text{f}} = 100$ J/m$^{2}$ and the mass density $\rho = 2450$ kg/m$^{3}$. The Rankine failure criterion \eqref{eq:equivalent-effective-stress}$_{1}$ was considered for mode-I fracture. The length scale parameter $b = 0.3$ m and the mesh size $h_{e} = 0.06$ m around the fracture process zone (FPZ) were considered. The loads were applied in three steps: in the first and second steps, the load control was used for the self weight and the full reservoir hydrostatic force, while in the third one for the overflow load with an increasing height, the indirect displacement control \citep{Wu2018b} based on the horizontal displacement of the dam crest was employed. The simulations were run up to a horizontal displacement of 70 mm with an incremental size 0.5 mm.

%

\begin{figure}[h!] \centering
  \subfigure[$\mu$PF-CZM ($p = 1.0$)]{
  \includegraphics[width=0.225\textwidth]{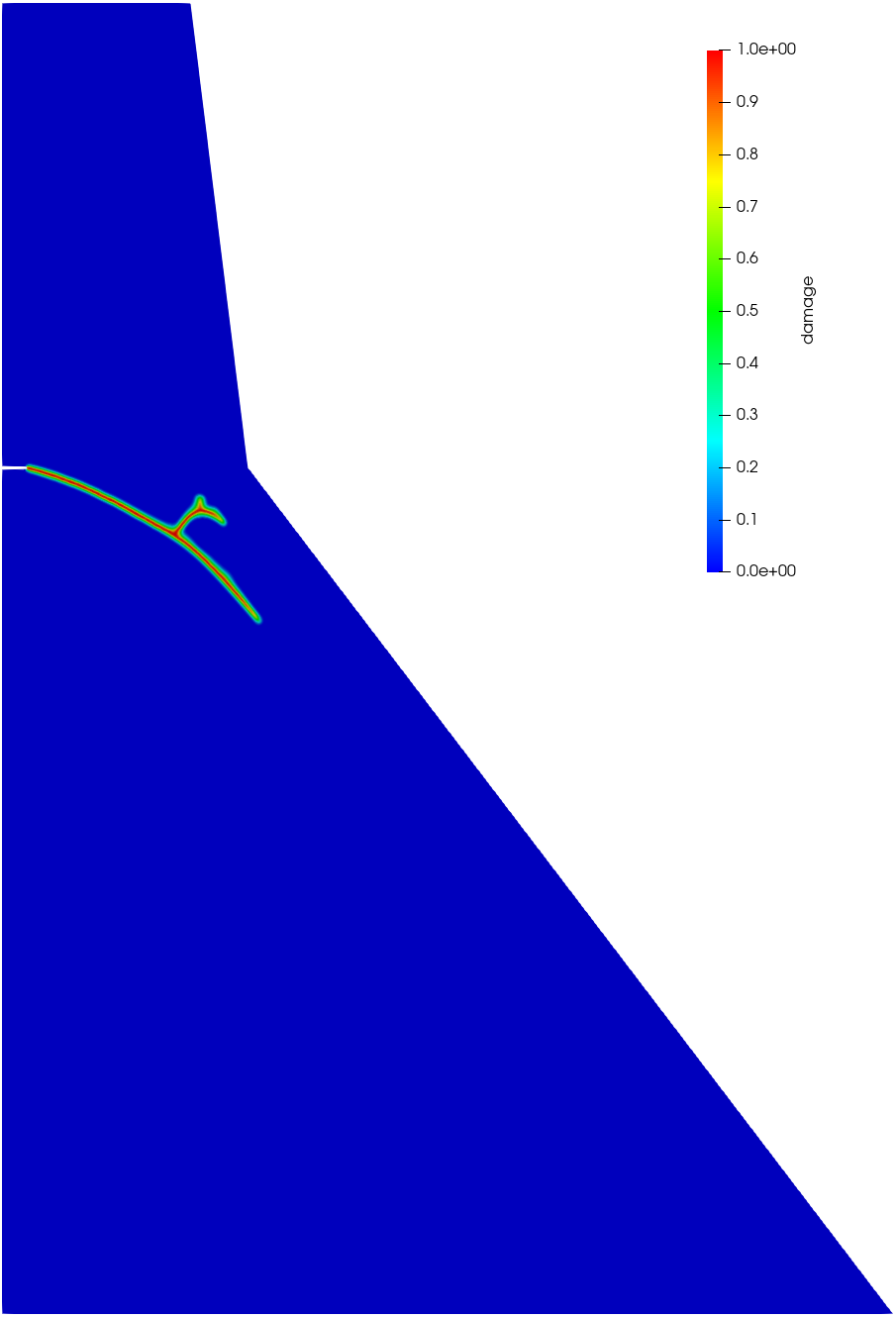}
  \label{fig:koyna-crack-gpfczm-p10}} \hfill
  \subfigure[$\mu$PF-CZM ($p = 1.5$)]{
  \includegraphics[width=0.225\textwidth]{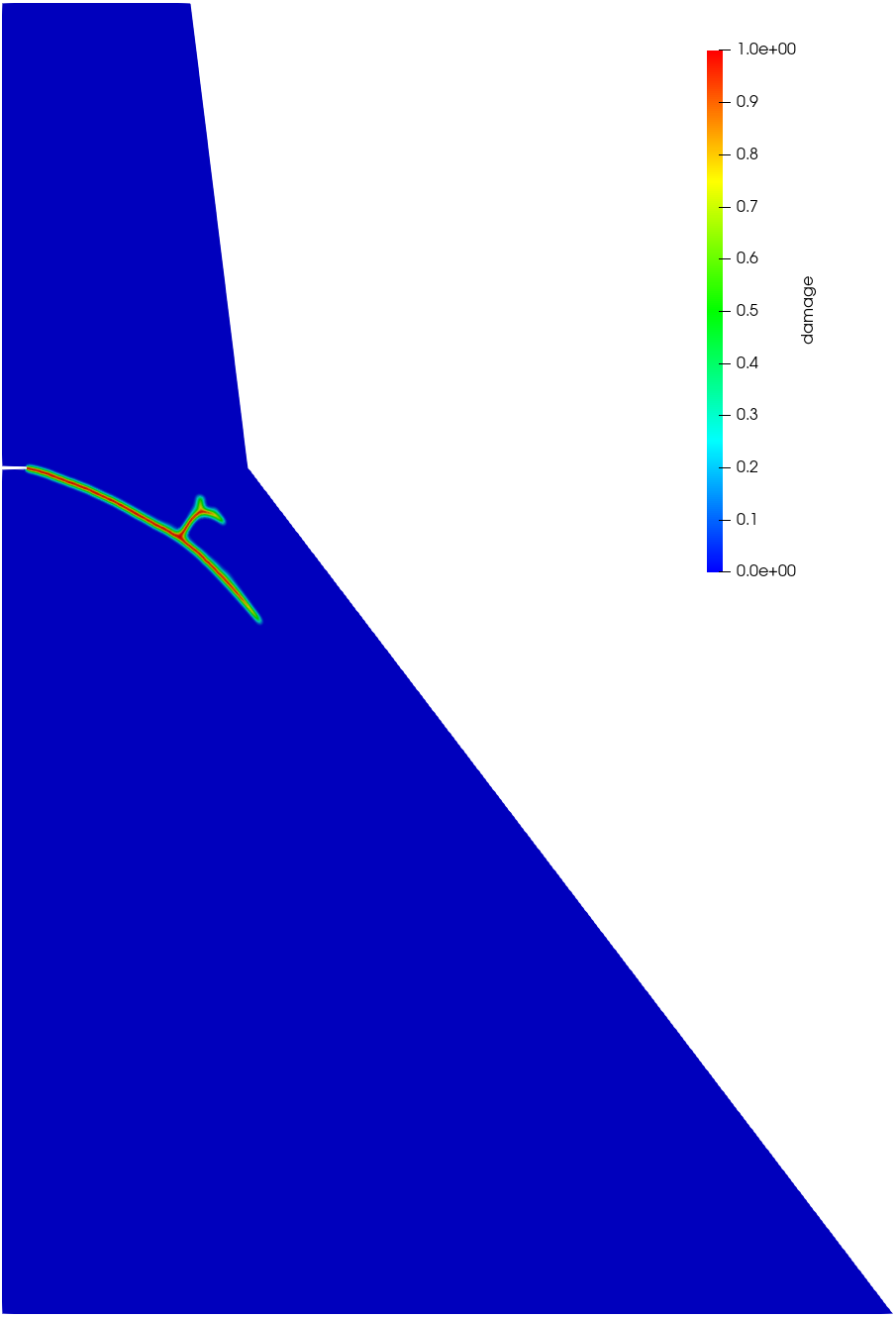}
  \label{fig:koyna-crack-gpfczm-p15}} \hfill
  \subfigure[$\mu$PF-CZM ($p = 2.0$)]{
  \includegraphics[width=0.225\textwidth]{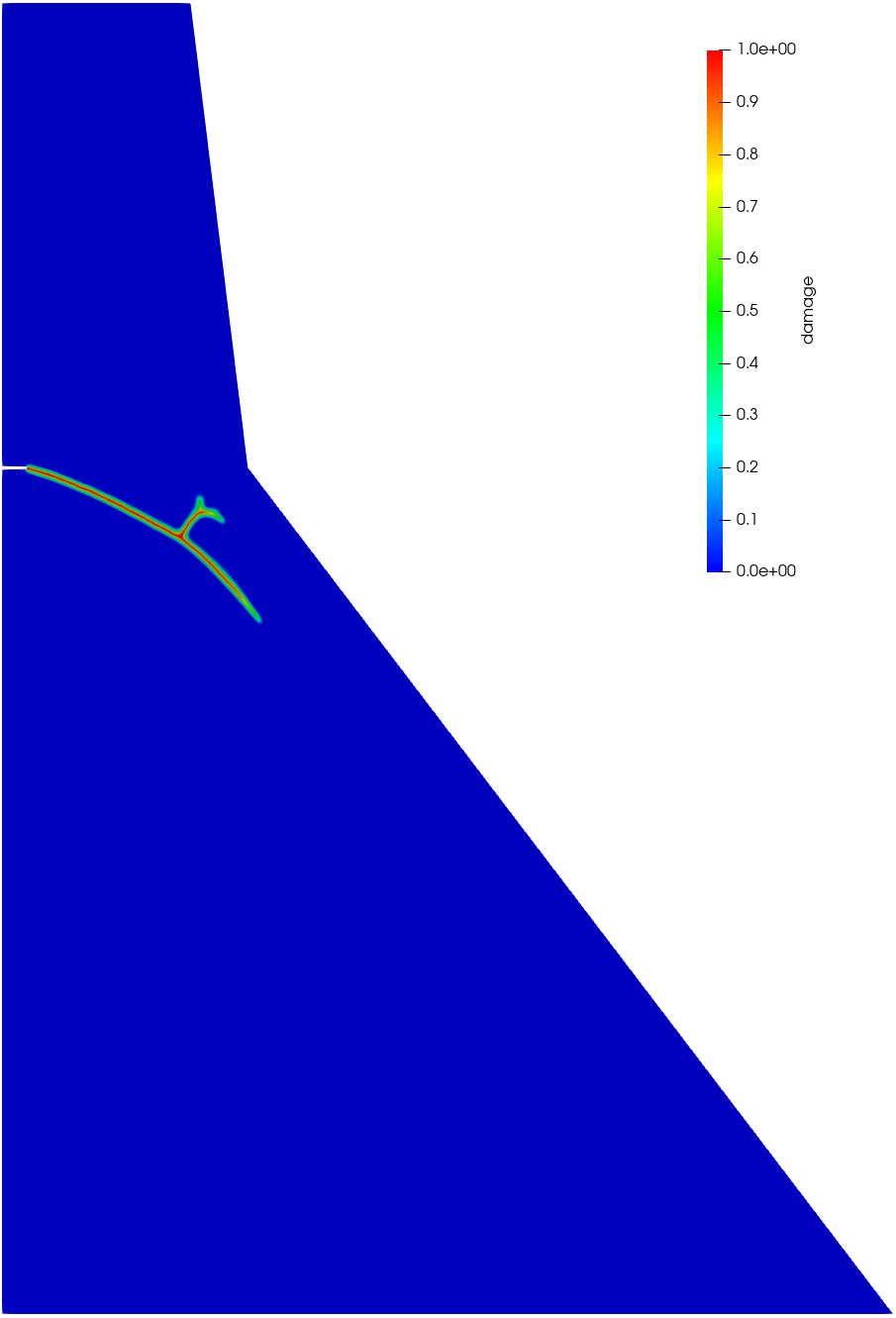}
  \label{fig:koyna-crack-gpfczm-p20}} \hfill
  \subfigure[PF$^{2}$-CZM]{
  \includegraphics[width=0.225\textwidth]{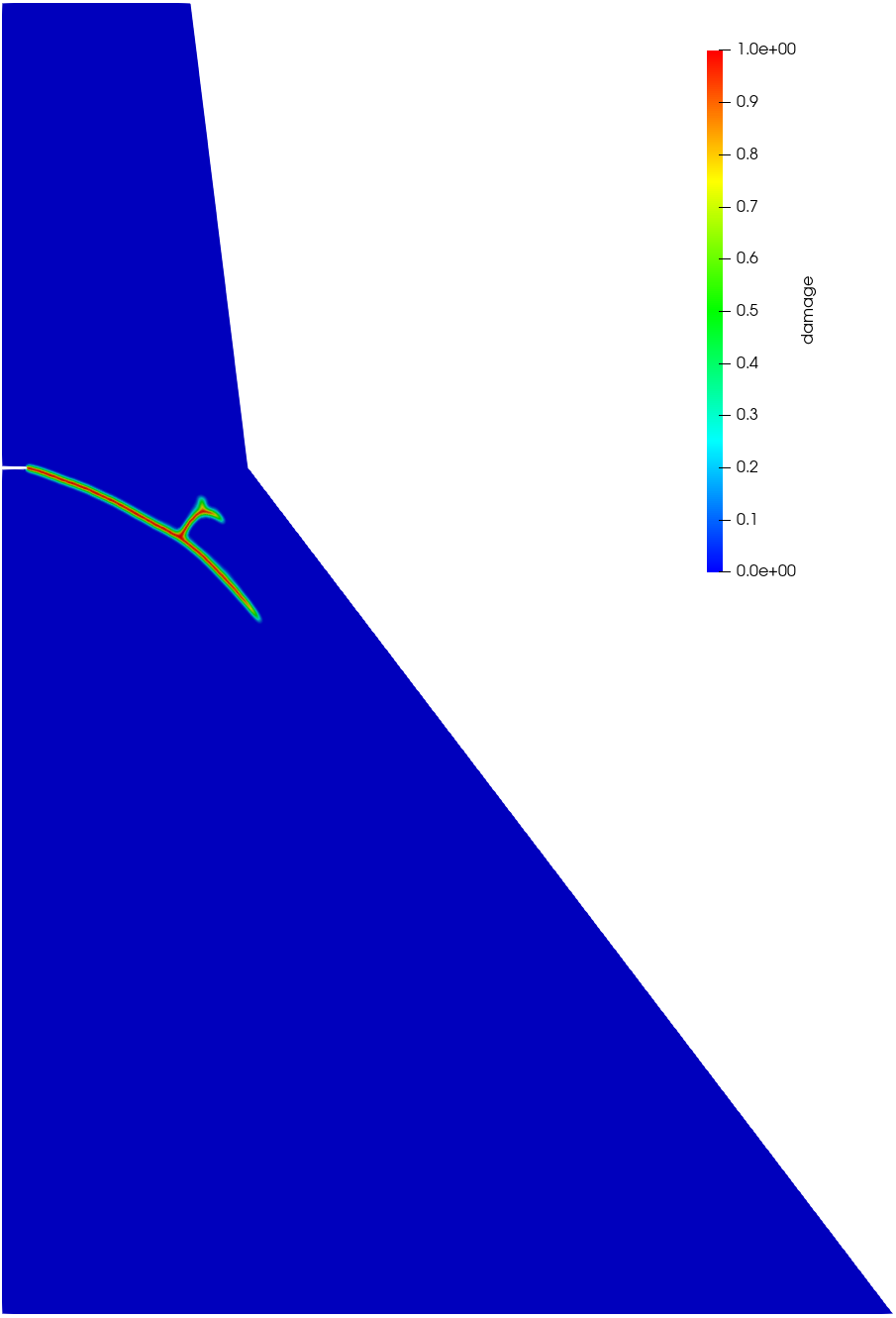}
  \label{fig:koyna-crack-pfczm}}
	\caption{Koyna dam under overflow pressure: Numerical crack patterns predicted by the $\mu$\texttt{PF-CZM} and \texttt{PF$^{2}$-CZM}.}
	\label{fig:koyna-cracks}
\end{figure}

As shown in \cref{fig:koyna-cracks}, the crack patterns predicted by the associated \texttt{PF$^{2}$-CZM} and the non-associated $\mu$\texttt{PF-CZM} with various values $p \ge 1$ almost coincide, and are fairly close to that printed on the book cover of the monograph \citep{BP1997}. In the early stage, a single crack initiates at the tip of the initial notch and extends downward to the downstream face. As the self-weight induced compressive stresses increases, the rate of crack propagation slows, leading to the occurrence of crack branching. A secondary crack then develops towards the intersecting point where the slope of the downstream face changes. The elevated compressive stresses in this region facilitate further crack branching.
  
\begin{figure}[h!]\centering  
  \includegraphics[width=0.55\textwidth]{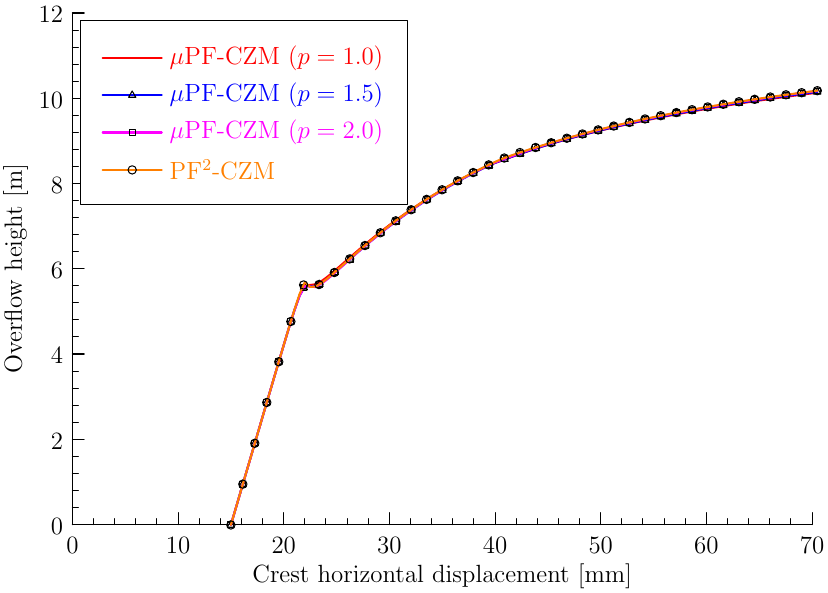}
  \caption{Koyna dam under overflow pressure: Numerical curves of overflow height \textit{versus} crest displacement.}
  \label{fig:koyna-overflow-displacement}
\end{figure}

The responses of the dam, illustrated by the curve of overflow height \textit{versus} crest horizontal displacement, are presented in \cref{fig:koyna-overflow-displacement}. As anticipated, the numerical results obtained from both the associated \texttt{PF$^{2}$-CZM} and the non-associated $\mu$\texttt{PF-CZM} with varying values of the traction order parameter $p \ge 1$ almost coincide. Notably, for overflow heights below approximately 5.5 m, fracture propagation is minimal, and the dam behaves in a nearly linear elastic manner. Following this, crack propagation occurs rapidly, leading to an initial decrease in the overflow level. A local maximum is observed in the overflow height versus displacement curve appears during the early stage, after which the relationship between crest displacement and overflow height becomes nonlinear.

\subsection{Single edge-notched beam under proportional loading}

\begin{figure}[h!] \centering
	\includegraphics[width=0.9\textwidth]{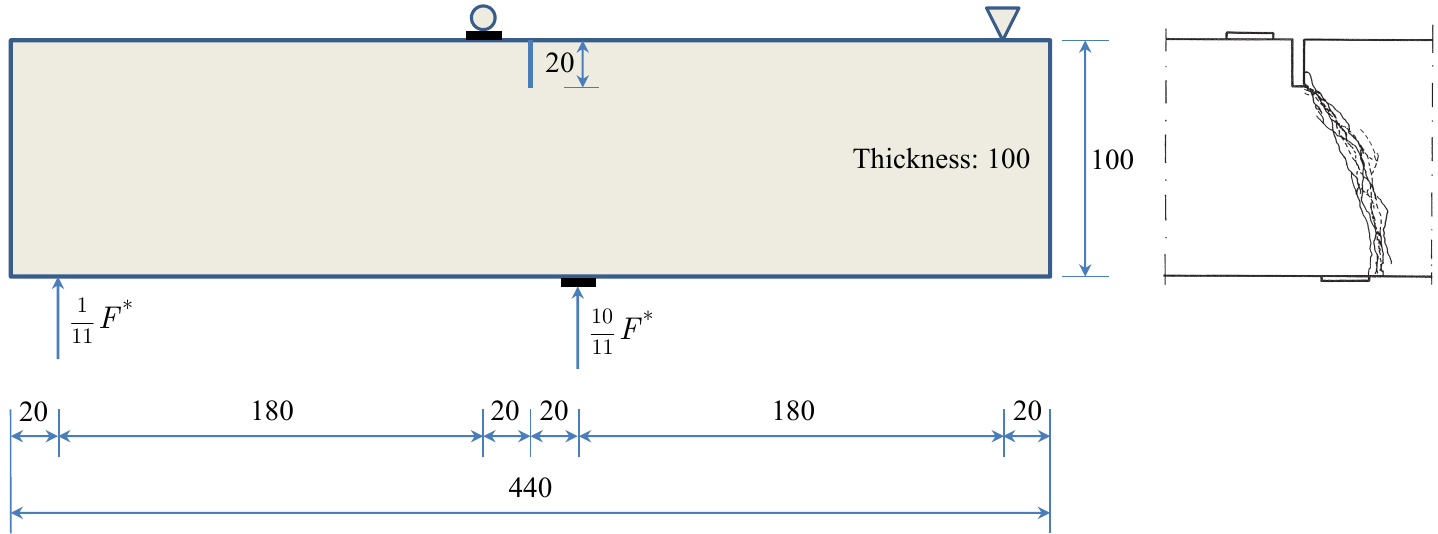}
	\caption{Single edge-notched beam \citep{Schlangen1993}. Left: Geometry (Unit of length: mm), loading and boundary conditions; Right: Experimentally observed crack paths.}
	\label{fig:SENB-test}
\end{figure}

The second example involves a single-edge notched beam (SENB) subjected to mixed-mode failure \citep{Schlangen1993}. As depicted in \cref{fig:SENB-test}, the beam has dimensions of 440 mm $\times$ 100 mm $\times$ 100 mm, with a notch of sizes 5 mm $\times$ 20 mm $\times$ 100 mm located at the top center. 
Two eccentrically proportional upward forces $10F^{\ast} / 11$ and $F^{\ast} / 11$ were exerted onto the two steel caps at the bottom of the beam. The relative vertical displacements on both sides of the notch, referred to as crack mouth sliding displacement (CMSD), was monitored to control the loading procedure. It was observed that a curved crack propagates from the notch downwards to the right hand side of the middle steel cap where the larger load was applied. 

In the numerical simulation, the following mechanical parameters for concrete were adopted: Young's modulus $E_{0} = 3.2 \times 10^{4}$ MPa, Poisson's ratio $\nu_{0} = 0.2$, the failure strength $f_{\text{t}} = 3.0$ MPa and the fracture energy $G_{\text{f}} = 0.1$ N/mm. The modified von Mises failure criterion \eqref{eq:equivalent-effective-stress}$_{2}$ with the strength ratio $\rho_{s} = 10.0$ was used in the modeling of mixed-mode failure. The phase-field length scale $b = 1.0$ mm and the mesh size $h_{e} = 0.2$ mm around the FPZ were considered. The simulation was performed using the CMSD based indirect displacement control \citep{Wu2018b}, with an incremental size 0.001 mm.

\begin{figure}[h!] \centering
  \subfigure[$\mu$PF-CZM ($p = 1.0$)]{
  \includegraphics[width=0.475\textwidth]{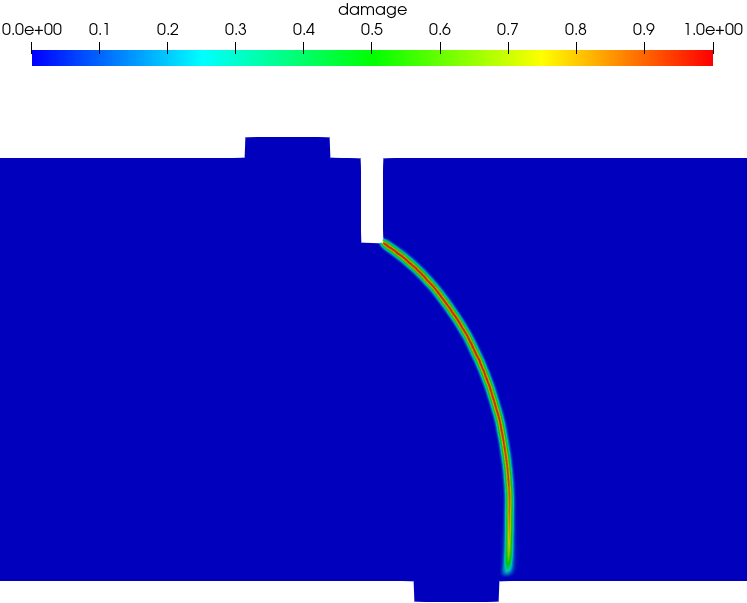}
  \label{fig:senb-crack-path-p10}} \hfill
  \subfigure[$\mu$PF-CZM ($p = 1.5$)]{
  \includegraphics[width=0.475\textwidth]{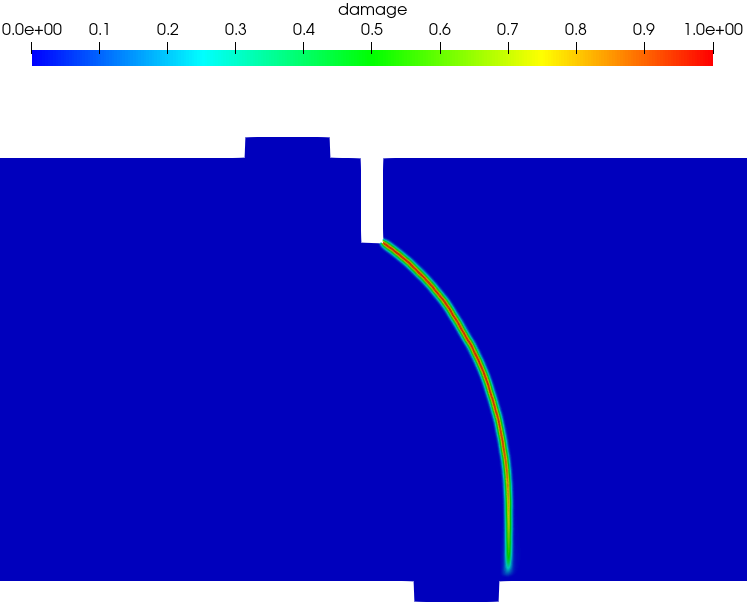}
  \label{fig:senb-crack-path-p15}} \\
  \subfigure[$\mu$PF-CZM ($p = 2.0$)]{
  \includegraphics[width=0.475\textwidth]{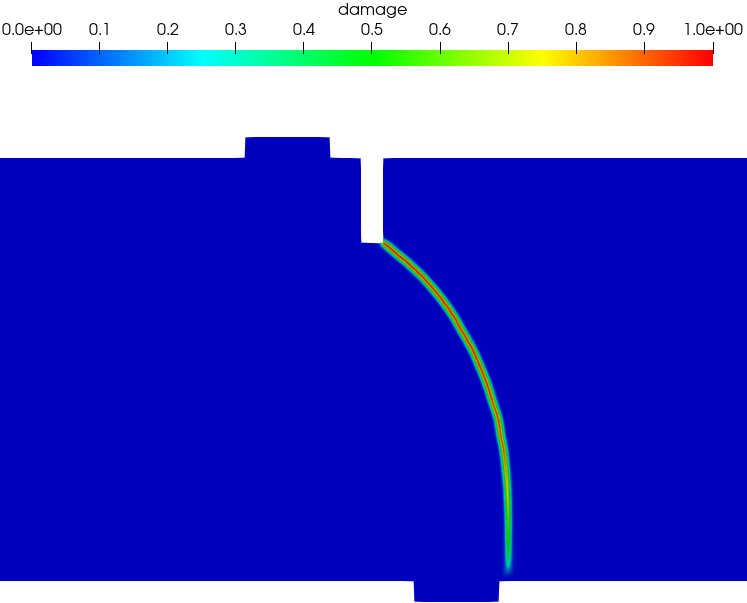}
  \label{fig:senb-crack-path-p20}} \hfill
  \subfigure[PF$^{2}$-CZM]{
  \includegraphics[width=0.475\textwidth]{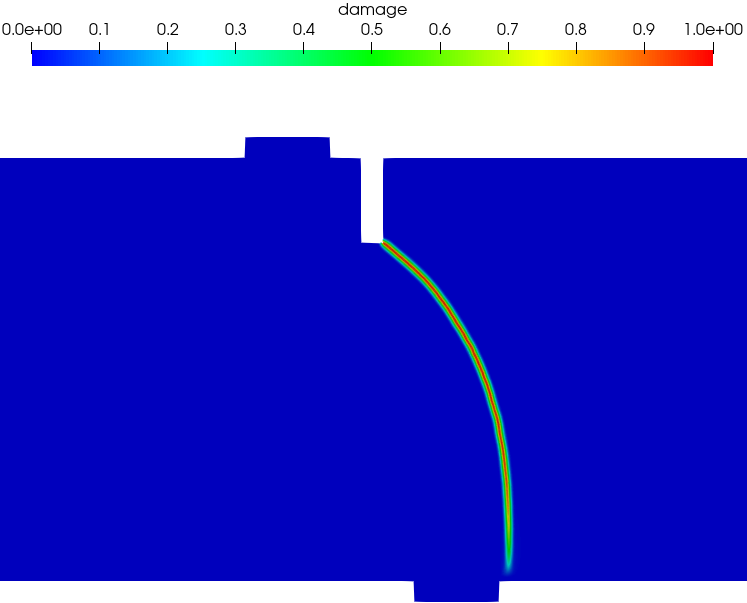}
  \label{fig:senb-crack-path-pfczm}}
	\caption{Single edge-notched beam: Numerical crack patterns predicted by the $\mu$\texttt{PF-CZM} and \texttt{PF$^{2}$-CZM}.}
	\label{fig:SENB-crack}
\end{figure}

\Cref{fig:SENB-crack} presents the predicted crack paths at the CMSD = 0.1 mm. As can be seen, the experimentally observed curved crack paths, propagating from the notch tip downward to the right hand side of the middle steel cap, were well captured by both the \texttt{$\mu$PF-CZM} and \texttt{PF$^{2}$-CZM}. Again, regarding the \texttt{$\mu$PF-CZM} the traction order parameter $p \ge 1$ does not affect the crack path, though the crack tip is a bit more sharp for a larger value $p \ge 1$.

\begin{figure}[htbp] \centering
	\includegraphics[width=0.55\textwidth]{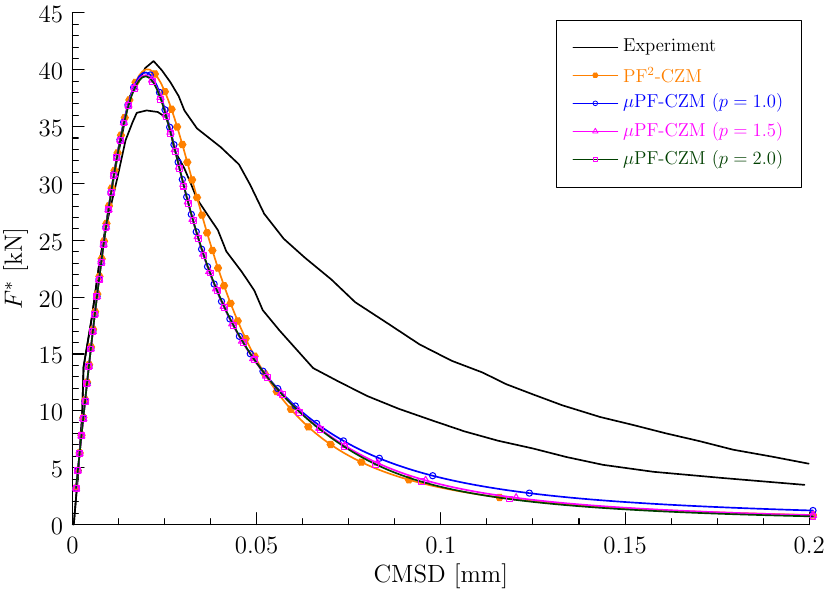}
	\caption{Single edge-notched beam test: Applied load--CMSD curves predicted by the $\mu$\texttt{PF-CZM} and \texttt{PF$^{2}$-CZM}.}
	\label{fig:senb-load-cmsd}
\end{figure}

The numerical force \textit{versus} CMSD curves are compared in \cref{fig:senb-load-cmsd} against the test data. As can be seen, the discrepancy between the numerical results given by the non-associated $\mu$\texttt{PF-CZM} and the associated \texttt{PF$^{2}$-CZM} is minor. The predicted peak load agrees well with the test results and the overall agreement of the post-peak regime is also satisfactory. Moreover, the value of $p \ge 1$ has negligible effects on the global responses, verifying that the non-associated \texttt{$\mu$PF-CZM} is indeed insensitive to the traction order parameter.


\subsection{Double edge-notched beam under proportional loading}

\begin{figure}[b!] \centering
	\includegraphics[width=0.70\textwidth]{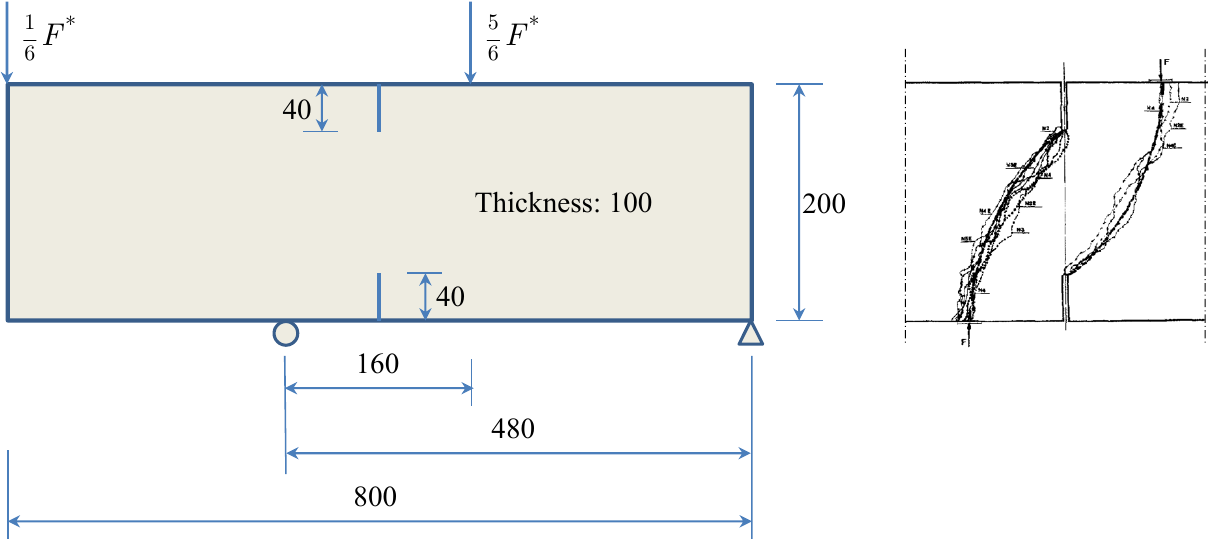}
	\caption{Double edge-notched beam \citep{BCV1990}. Left: Geometry (Unit of length: mm), loading and boundary conditions; Right: Experimentally observed crack paths.}
	\label{fig:denb-bocca-geometry}	
\end{figure}

The next example is the four-point shear test of double edge-notched beams (DENB) reported in \cite{BCV1990}. As shown in \cref{fig:denb-bocca-geometry}, the specimen was of dimensions 800 mm $\times$ 200 mm $\times$ 100 mm, with two notches of identical sizes 5 mm $\times$ 40 mm $\times$ 100 mm. Two eccentrically proportional downward forces $5F^{\ast} / 6$ and $F^{\ast} / 6$ were applied at the top of the beam. The CMSD of the notch was monitored to control the loading procedure. 

%
%
In the numerical simulation, the following mechanical parameters for concrete were adopted: Young's modulus $E_{0} = 2.7 \times 10^{4}$ MPa, Poisson's ratio $\nu_{0} = 0.18$, the failure strength $f_{\text{t}} = 2.0$ MPa and the fracture energy $G_{\text{f}} = 0.1$ N/mm. Again, the modified von Mises failure criterion \eqref{eq:equivalent-effective-stress}$_{2}$ with the strength ratio $\rho_{s} = 10.0$ was employed. The phase-field length scale $b = 2.0$ mm and the mesh size $h_{e} = 0.4$ mm around the FPZ were adopted. The CMSD-based indirect displacement control with an incremental size 0.001 mm was employed to track the equilibrium path. 

\begin{figure}[h!] \centering
  \subfigure[$\mu$PF-CZM ($p = 1.0$)]{
  \includegraphics[width=0.465\textwidth]{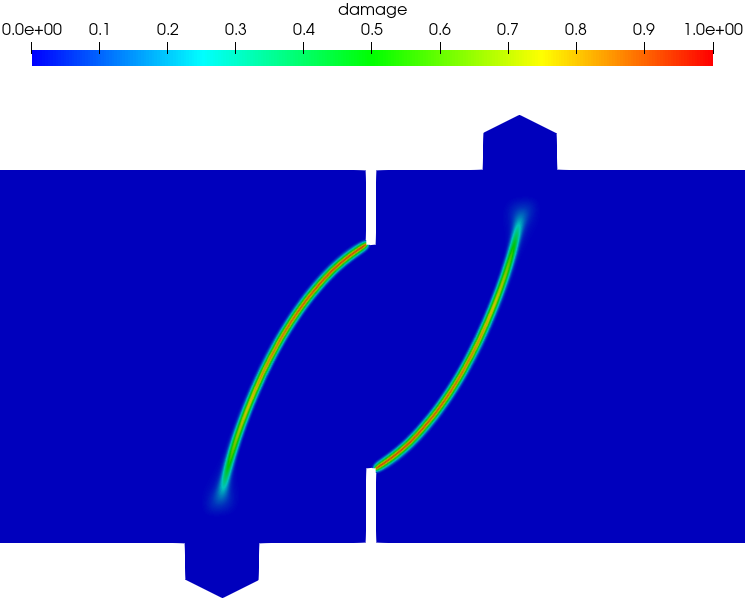}
  \label{fig:denb-crack-path-p10}} \hspace{3mm}
  \subfigure[$\mu$PF-CZM ($p = 1.5$)]{
  \includegraphics[width=0.465\textwidth]{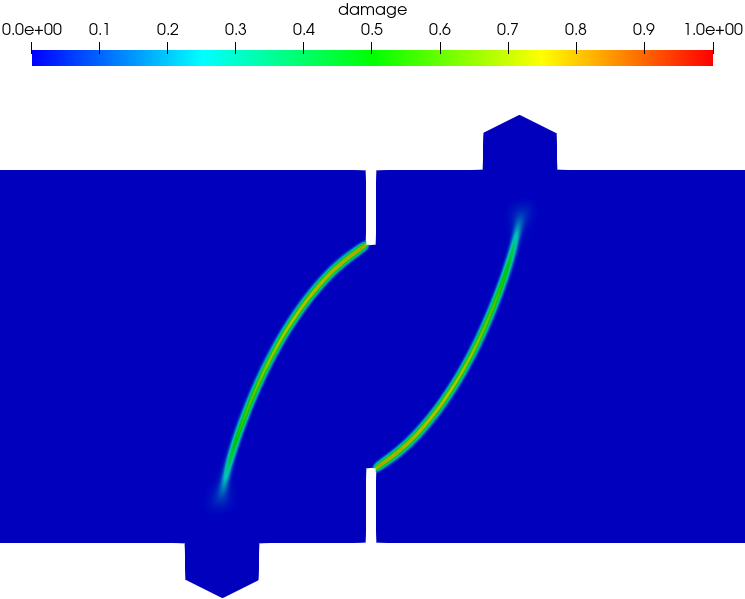}
  \label{fig:denb-crack-path-p15}} \\
  \subfigure[$\mu$PF-CZM ($p = 2.0$)]{
  \includegraphics[width=0.465\textwidth]{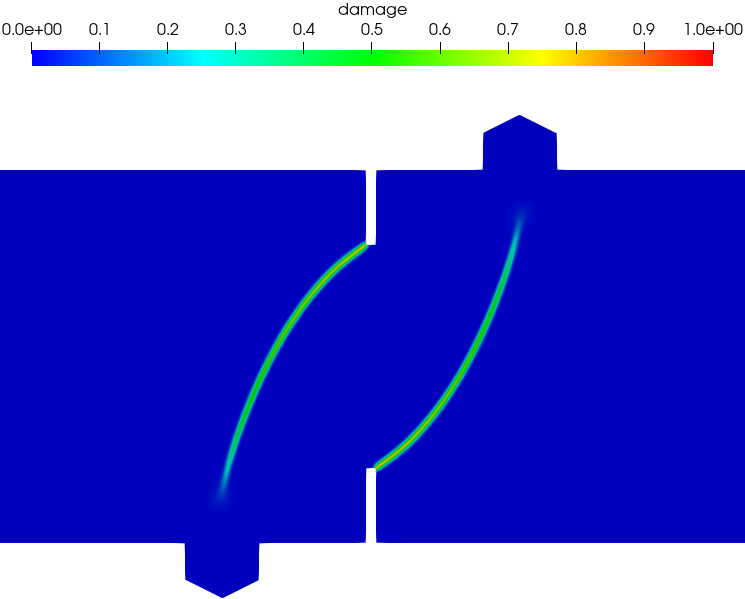}
  \label{fig:denb-crack-path-p20}} \hspace{3mm}
  \subfigure[\texttt{PF$^{2}$-CZM}]{
  \includegraphics[width=0.465\textwidth]{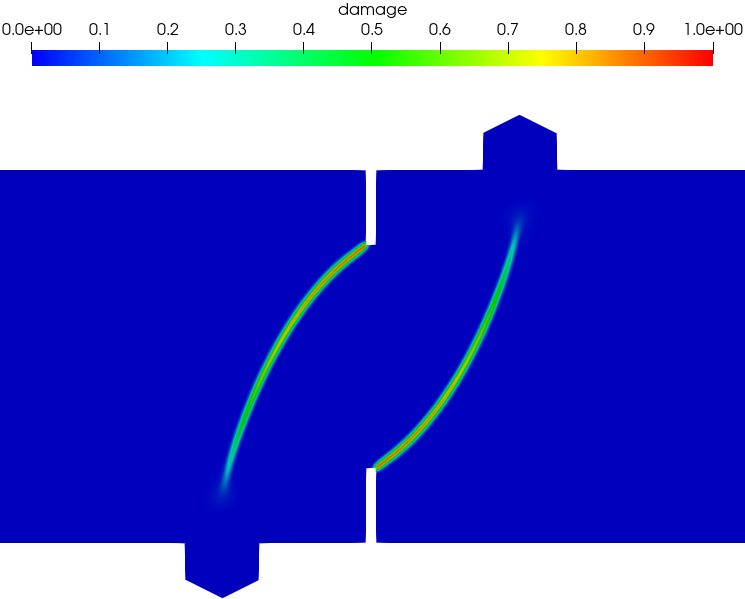}
  \label{fig:denb-crack-path-pfczm}}
	\caption{Double edge-notched beam: Numerical crack patterns predicted by the $\mu$\texttt{PF-CZM} and \texttt{PF$^{2}$-CZM}.}
	\label{fig:denb-cracks}
\end{figure}

\cref{fig:denb-cracks} presents the numerically predicted crack paths at the CMSD = 0.1 mm. As can be seen, two anti-symmetric cracks initiate at the notch tips and propagate along curved paths to the two interior supports. Both the associated \texttt{PF$^{2}$-CZM} and the non-associated $\mu$\texttt{PF-CZM} are able to capture the mixed-mode failure with non-negligible shear stresses. Moreover, for the non-associated $\mu$\texttt{PF-CZM} though the crack tip is a bit more sharp for a larger value of the traction order parameter $p \ge 1$, the predicted crack pattern is unaffected.

\begin{figure}[htbp] \centering
	\centering
	\includegraphics[width=0.55\textwidth]{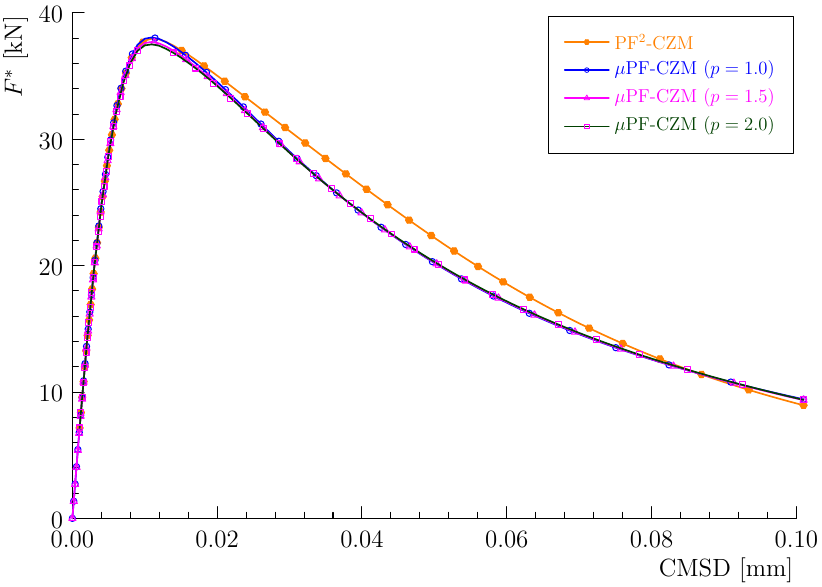}
	\caption{Double edge-notched beam: Applied load--CMSD curves predicted by the $\mu$\texttt{PF-CZM} and \texttt{PF$^{2}$-CZM}.}
	\label{fig:denb-bocca-load-cmsd}	
\end{figure}

The numerical load \textit{versus} CMSD curves are depicted in \cref{fig:denb-bocca-load-cmsd}. As can be seen, the associated \texttt{PF$^{2}$-CZM} and the non-associated $\mu$\texttt{PF-CZM} give rather close predictions, though the post-peak softening behavior exhibits minor discrepancies due to the different approximations of the \cite{CHR1986} softening curve. Again, the numerical results predicted by the non-associated $\mu$\texttt{PF-CZM} are also insensitive to the traction order parameter $p \ge 1$ for this example involving mixed-mode failure. 

\subsection{Double edge-notched specimens under non-proportional loading}

\begin{figure}[h!]\centering
	\includegraphics[width=0.95\textwidth]{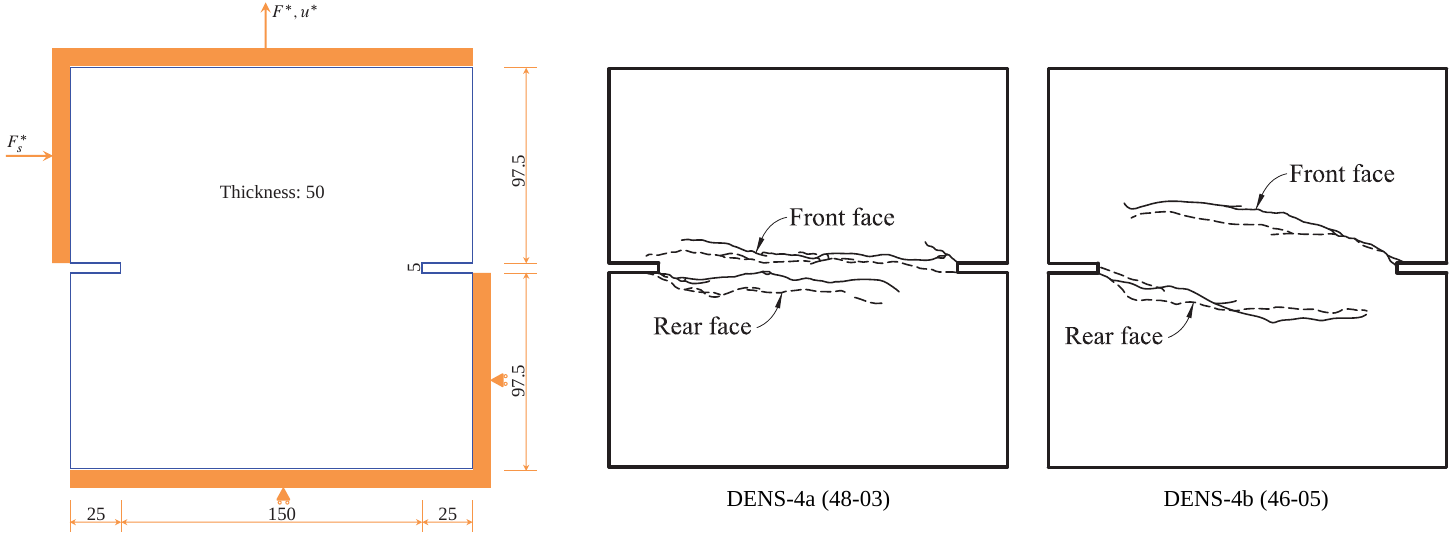}
	\caption{Double edge notched specimens \citep{Nooru1992}. Left: Geometry (Unit of length: mm), loading and boundary conditions; Right: Experimentally observed crack paths.}
	\label{fig:dens-model}
\end{figure}

The test on the double edge-notched specimens (DENS) under non-proportional loading \citep{Nooru1992} is another benchmark example of mixed fracture in concrete. As shown in \cref{fig:dens-model}, the specimen was of dimensions 200 mm $\times$ 200 mm $\times$ 50 mm, with two identical notches of sizes 25 mm $\times$ 5 mm $\times$ 50 mm at the center of left and right edges. The specimen was glued to a rigid steel frame supported at the bottom and the right edge below the notch. In the test, a horizontal ``shear'' force $F_{s}^{\ast}$ was first imposed along the left edge above the notch, and then a vertical normal force $F^{\ast}$ was applied at the top edge with a monotonically increasing displacement $u^{\ast}$ while the shear force $F_{s}^{\ast}$ was maintained fixed. The specimens DENS-4a (48-03) and DENS-4b (46-05) were considered here, with the applied shear force $F_{s}^{\ast}$ being 5 kN and 10 kN, respectively. Experimental observation shows that two nearly antisymmetric cracks propagate from the notches to the opposite sides, with the curvature dependent on the level of the applied shear force $F_{s}^{\ast}$. 

The following mechanical parameters for concrete were adopted in the numerical simulation: Young's modulus $E_{0} = 3.0 \times 10^{4}$ MPa, Poisson's ratio $\nu_{0} = 0.2$, the failure strength $f_{\text{t}} = 3.0$ MPa and the fracture energy $G_{\text{f}} = 0.11$ N/mm. As in the previous examples, the modified von Mises failure criterion \eqref{eq:equivalent-effective-stress}$_{2}$ with the strength ratio $\rho_{s} = 10.0$ was used to model mixed-mode failure. The phase-field length scale $b = 1.5$ mm and the mesh size $h = 0.3$ mm around the FPZ were considered. Ths simulation was performed using the force control (20 and 40 equal increments for the DENS-4a and DENS-4b, respectively) for the shear stage and the displacement control with an incremental size 0.001 mm for the tensile stage. 

\begin{figure}[h!] \centering
  \subfigure[$\mu$PF-CZM ($p = 1.0$)]{
  \includegraphics[width=0.225\textwidth]{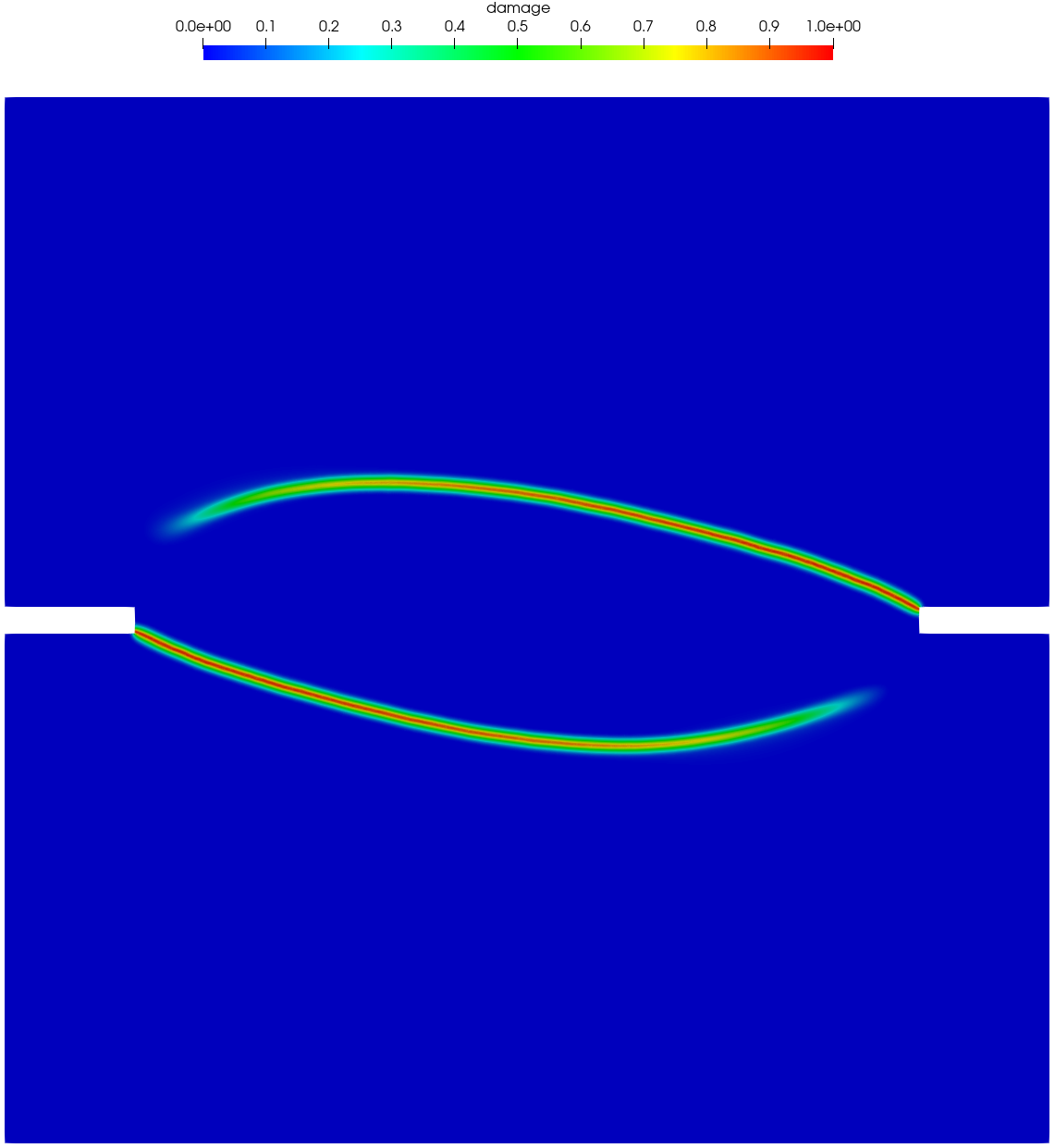}
  \label{fig:dens-4a-damage-p10}} \hfill
  \subfigure[$\mu$PF-CZM ($p = 1.5$)]{
  \includegraphics[width=0.225\textwidth]{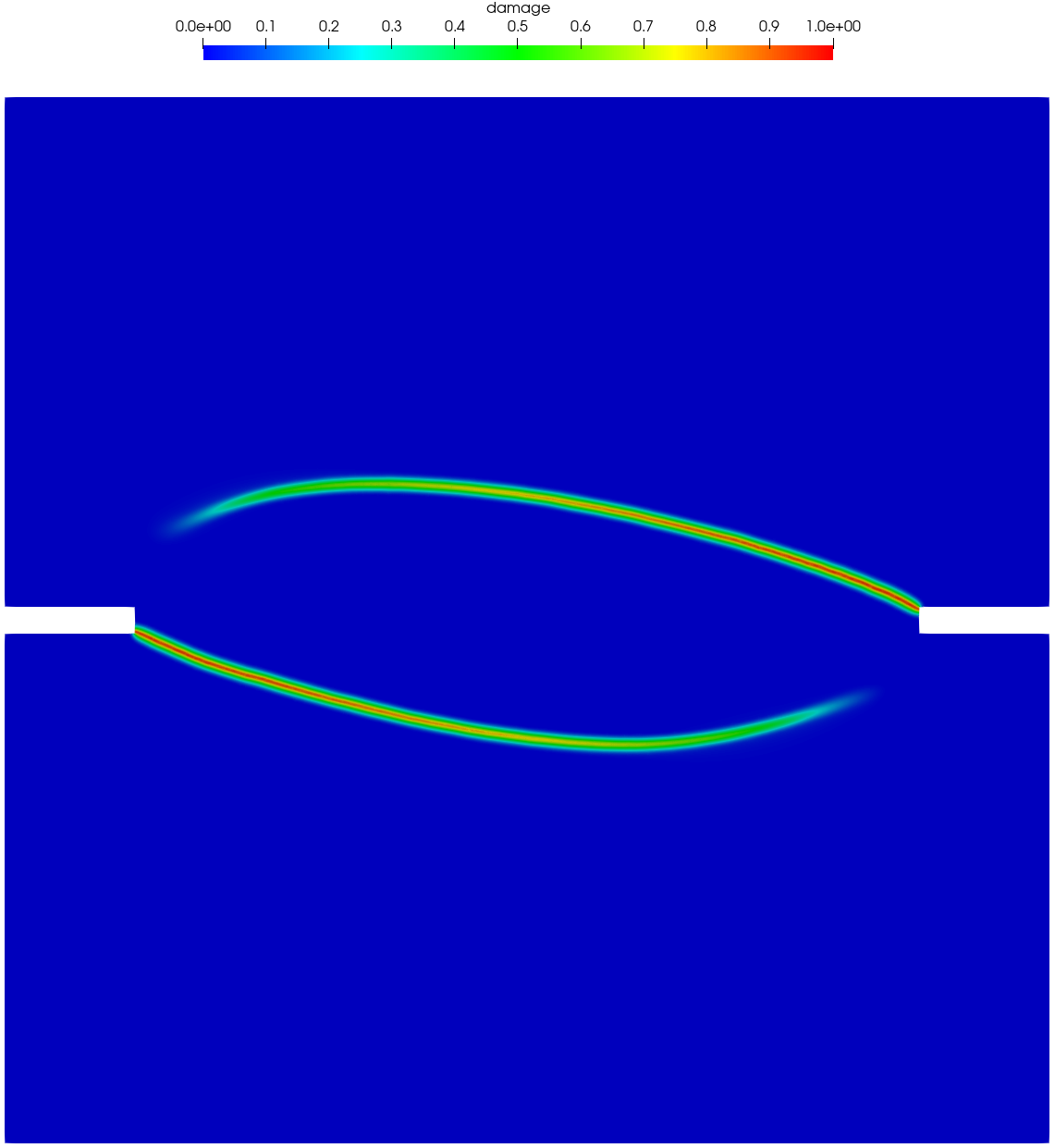}
  \label{fig:dens-4a-damage-p15}} \hfill
  \subfigure[$\mu$PF-CZM ($p = 2.0$)]{
  \includegraphics[width=0.225\textwidth]{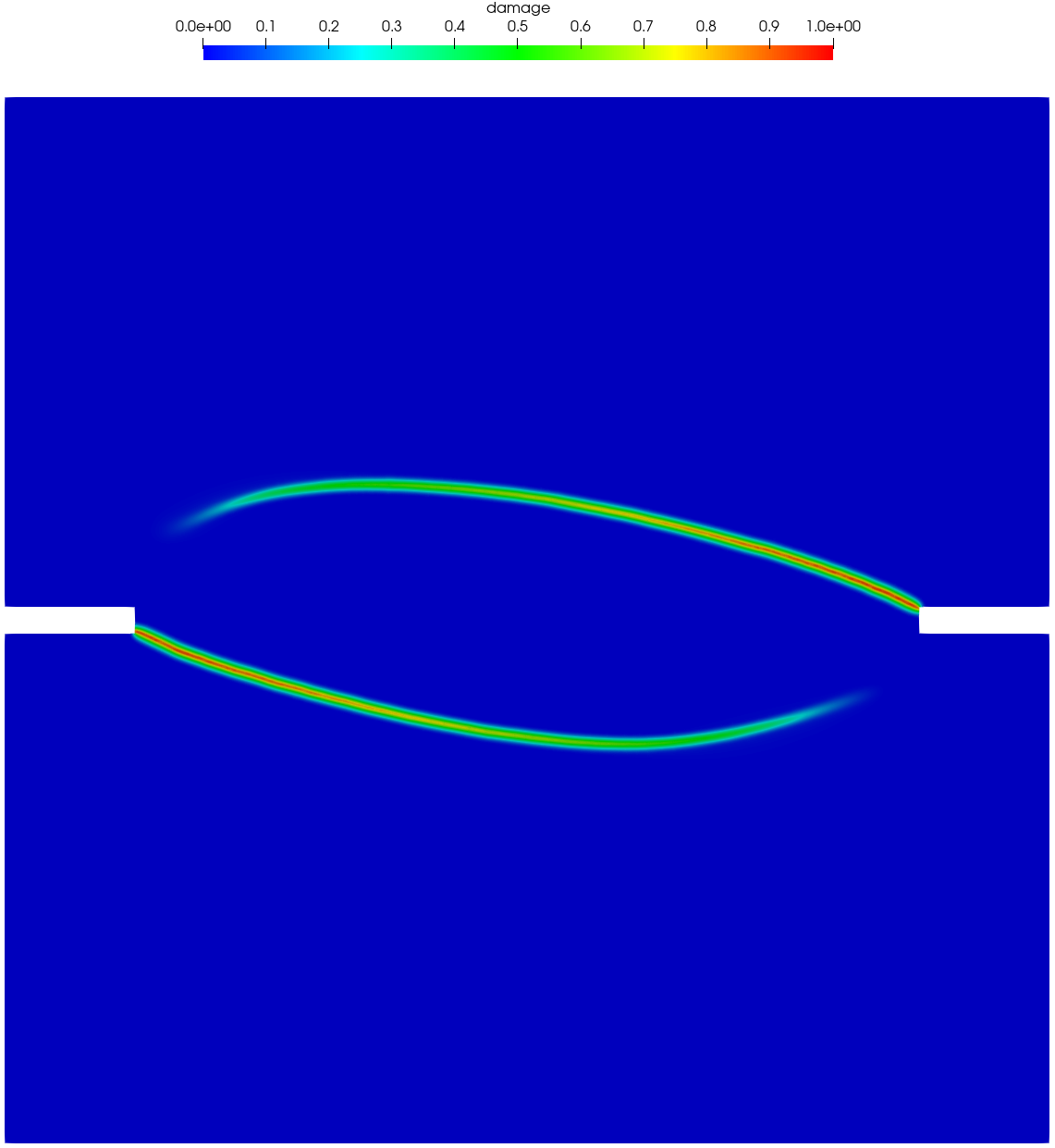}
  \label{fig:dens-4a-damage-p20}} \hfill
  \subfigure[PF$^{2}$-CZM]{
  \includegraphics[width=0.225\textwidth]{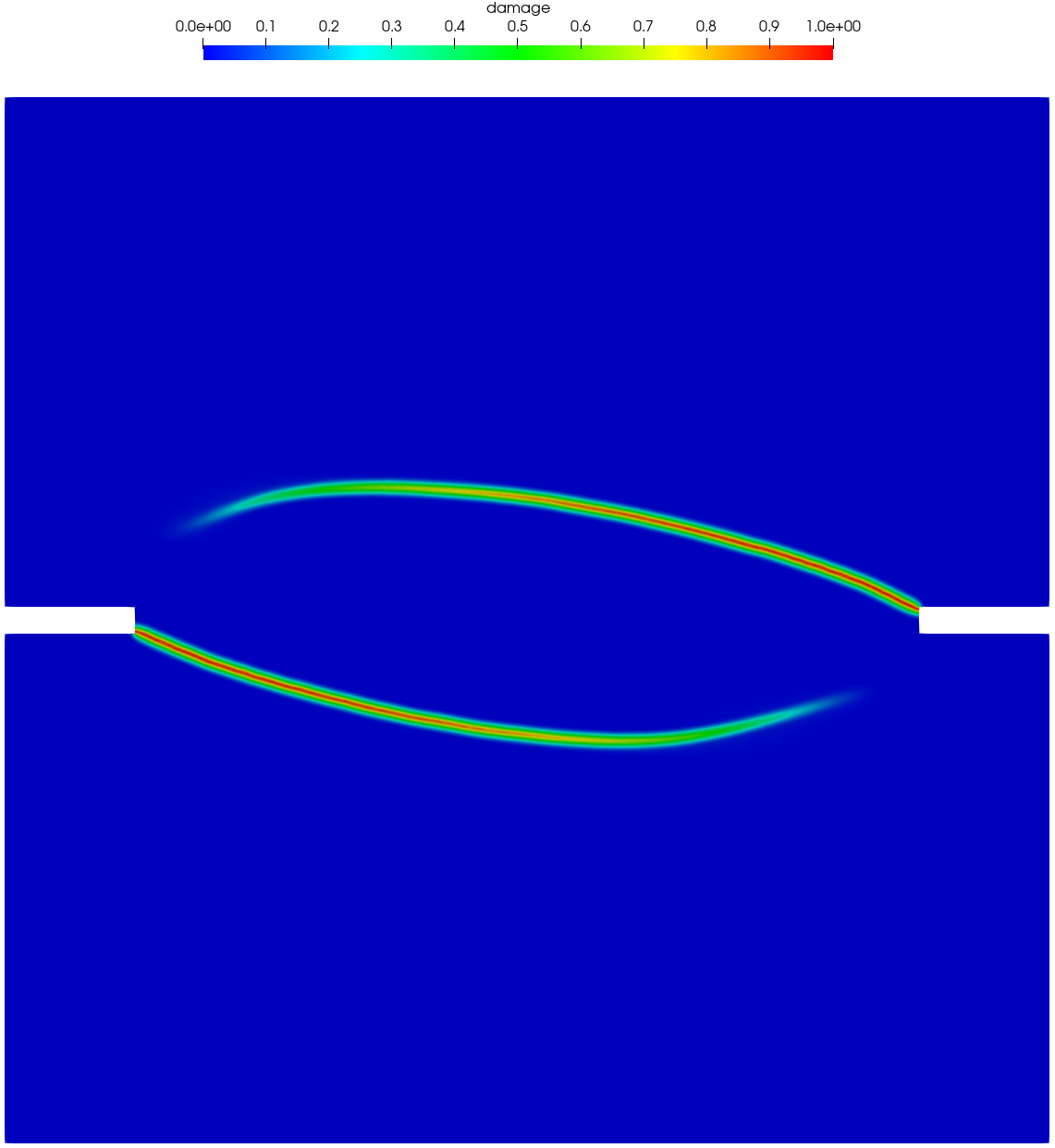}
  \label{fig:dens-4a-damage-pfczm}}
  \caption{Double edge notched specimen (DENS-4a): Numerically predicted damage profiles at $u^{\ast} = 0.2$ mm.}
	\label{fig:dens-4a-damage}
\end{figure}

\begin{figure}[h!] \centering
  \subfigure[$\mu$PF-CZM ($p = 1.0$)]{
  \includegraphics[width=0.225\textwidth]{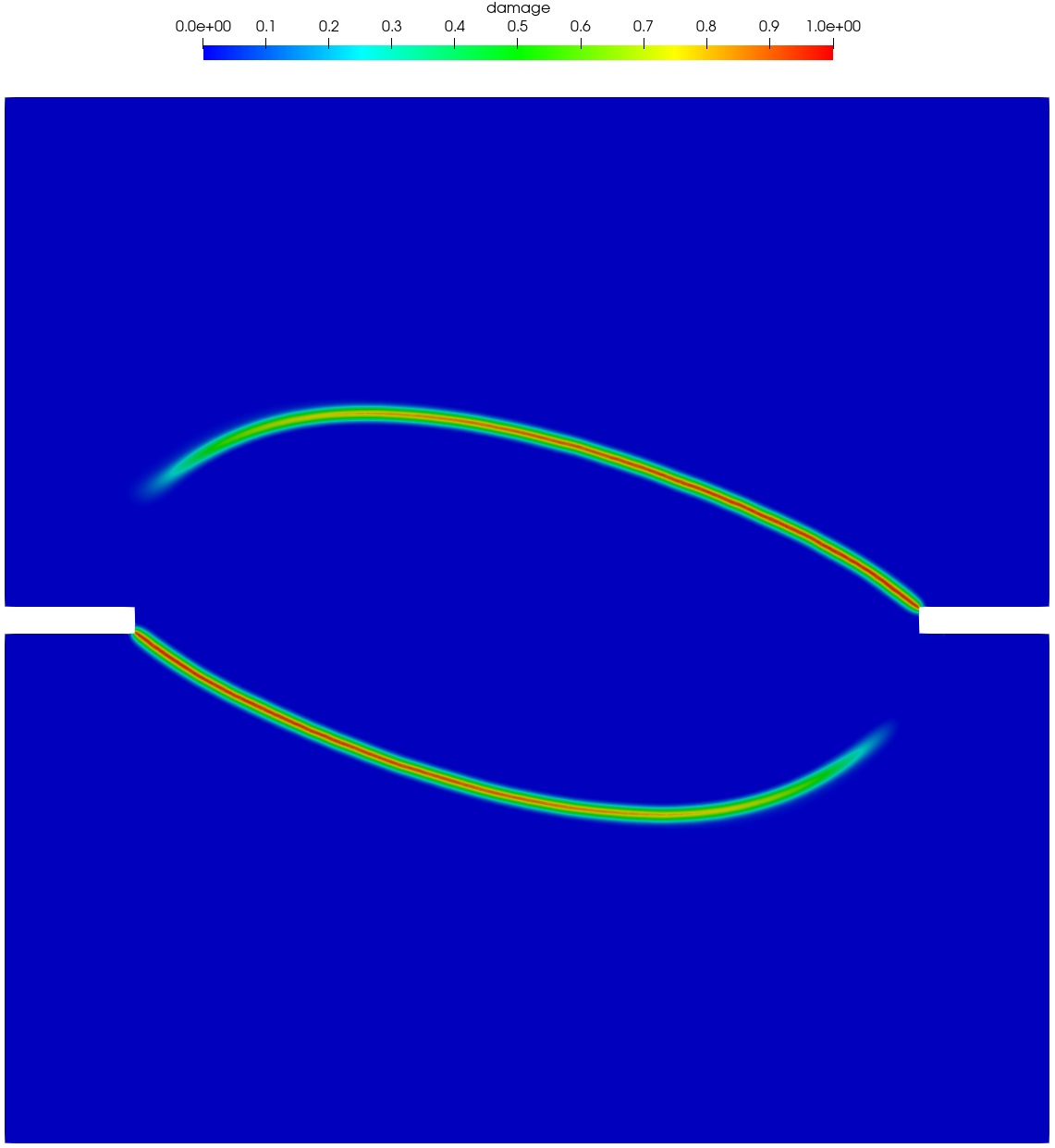}
  \label{fig:dens-4b-damage-p10}} \hfill
  \subfigure[$\mu$PF-CZM ($p = 1.5$)]{
  \includegraphics[width=0.225\textwidth]{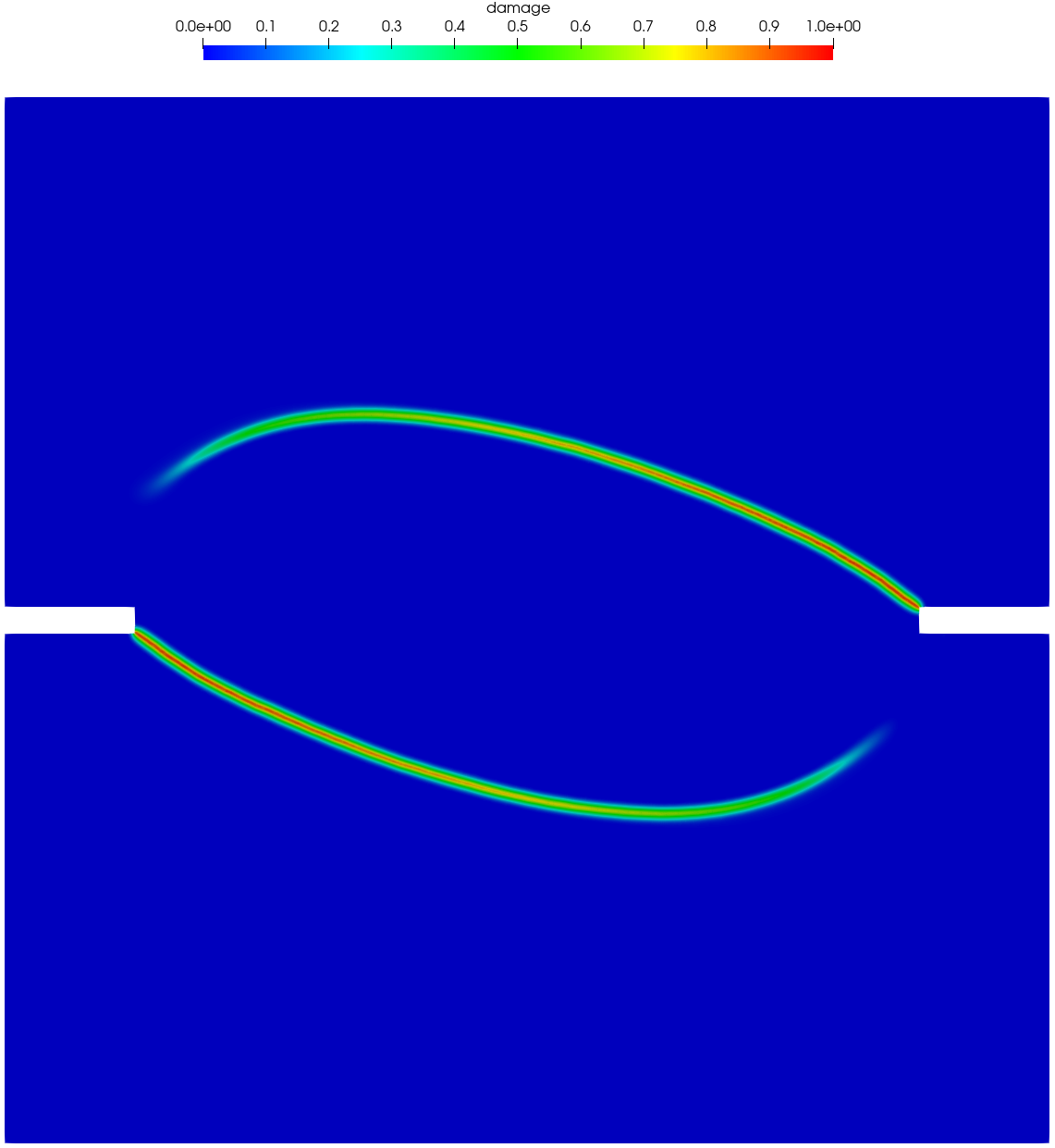}
  \label{fig:dens-4b-damage-p15}} \hfill
  \subfigure[$\mu$PF-CZM ($p = 2.0$)]{
  \includegraphics[width=0.225\textwidth]{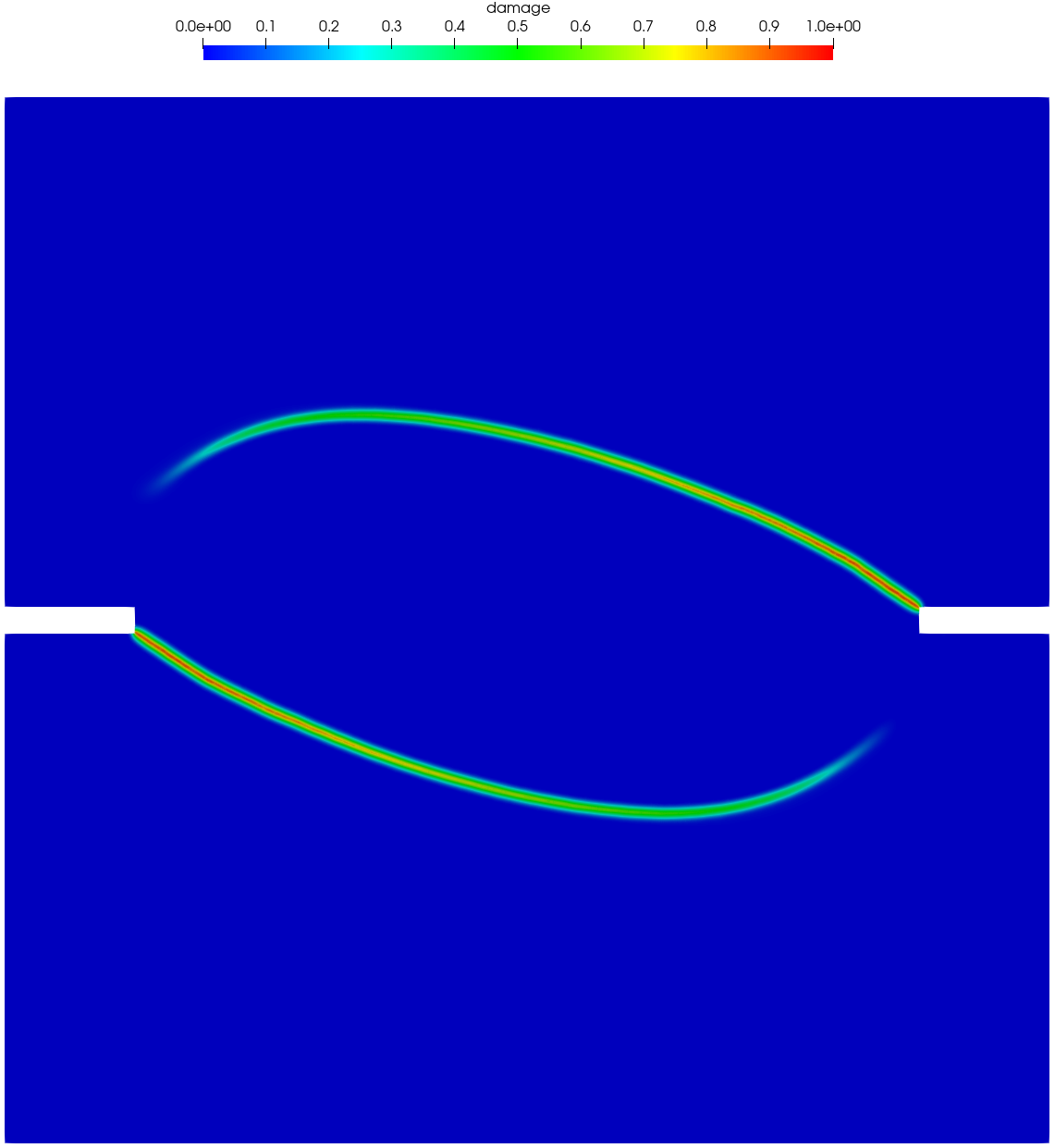}
  \label{fig:dens-4b-damage-p20}} \hfill
  \subfigure[PF$^{2}$-CZM]{
  \includegraphics[width=0.225\textwidth]{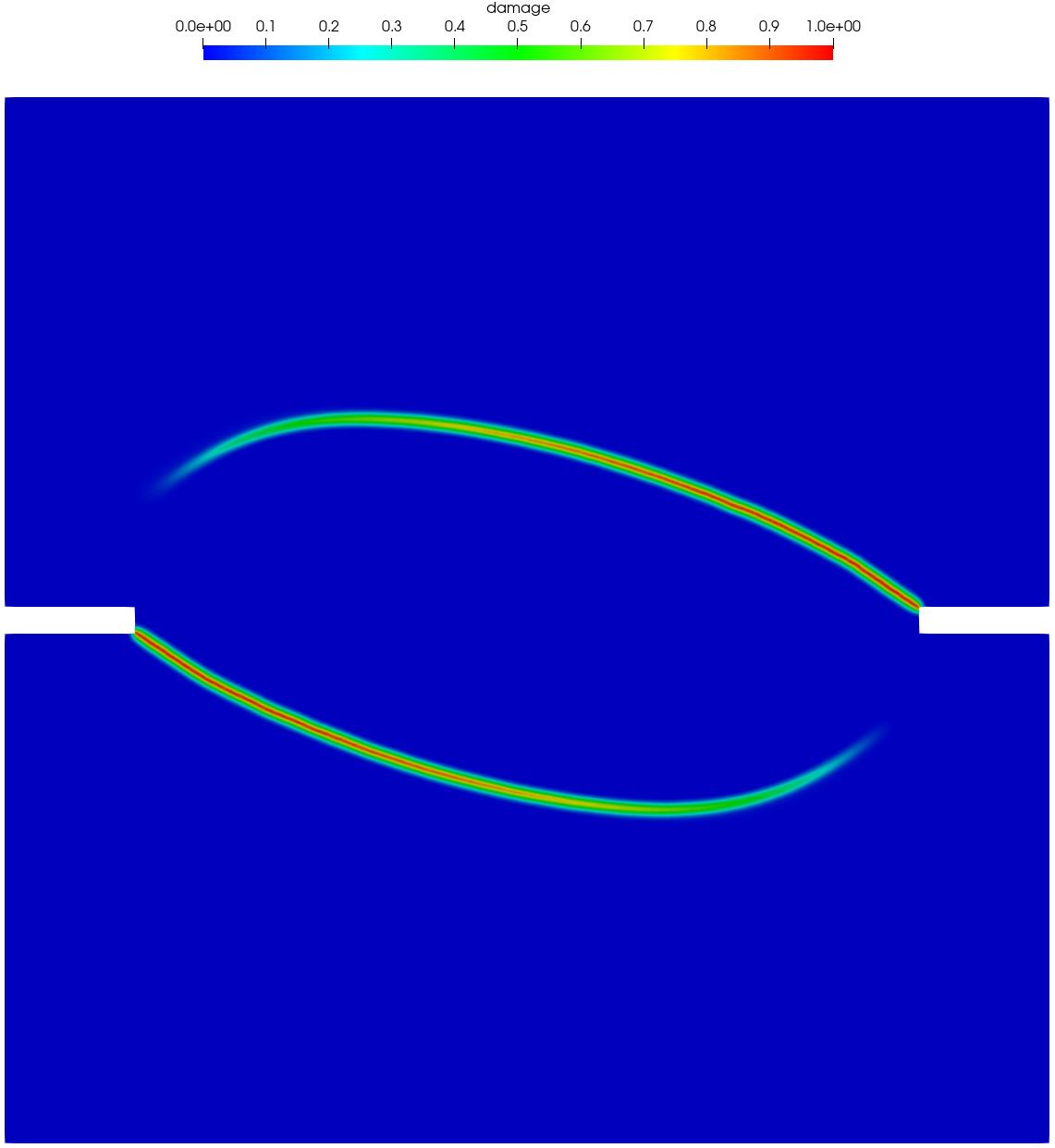}
  \label{fig:dens-4b-damage-pfczm}}
  \caption{Double edge notched specimen (DENS-4b): Numerically predicted damage profiles at $u^{\ast} = 0.2$ mm.}
	\label{fig:dens-4b-damage}
\end{figure}

\begin{figure}[h!] \centering
  \subfigure[DENS-4a]{
  \includegraphics[width=0.48\textwidth]{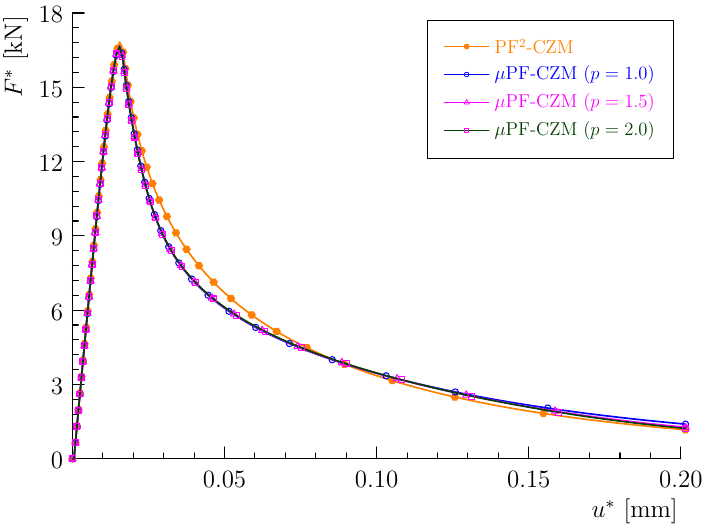}
  \label{fig:dens-4a-load-disp}} \hfill
  \subfigure[DENS-4b]{
  \includegraphics[width=0.48\textwidth]{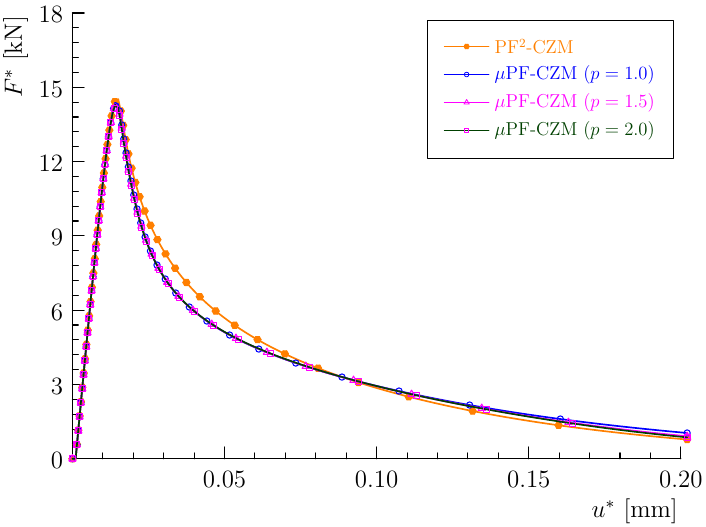}
  \label{fig:dens-4b-load-disp}}
  \caption{Double edge notched specimens: Applied load -- displacement curves predicted by the $\mu$\texttt{PF-CZM} and \texttt{PF$^{2}$-CZM}}
	\label{fig:dens-load-disp}
\end{figure}

\Cref{fig:dens-4a-damage} and \cref{fig:dens-4b-damage} present the numerically predicted crack patterns at the vertical displacement $u^{\ast} = 0.2$ mm. As can be seen, all the numerical results agree qualitatively well with the experimentally observed crack paths --- as the load level of the applied shear force $F_{s}^{\ast}$ increases, the curvature of the two anti-symmetric cracks becomes larger. Moreover, for the non-associated $\mu$\texttt{PF-CZM} the traction order parameter $p \ge 1$ has negligible effects on the predicted crack pattern, though the crack tip is more sharp for a larger value $p$.

The numerically predicted load \textit{versus} displacement curves are depicted in \cref{fig:dens-load-disp}. As expected, the associated \texttt{PF$^{2}$-CZM} and the non-associated $\mu$\texttt{PF-CZM} give rather close numerical results. In particular, the vertical load capacity decreases as the load level of the applied shear force increases. Once the failure mechanism is fully developed, the vertical force tends to vanish completely and no stress locking is exhibited. Again, the traction order parameter $p \ge 1$ does not affect the global responses predicted by the non-associated $\mu$\texttt{PF-CZM}. 

\subsection{Double cantilever beam test under mode-I failure}

\begin{figure}[h!] \centering
  \includegraphics[width=0.9\textwidth]{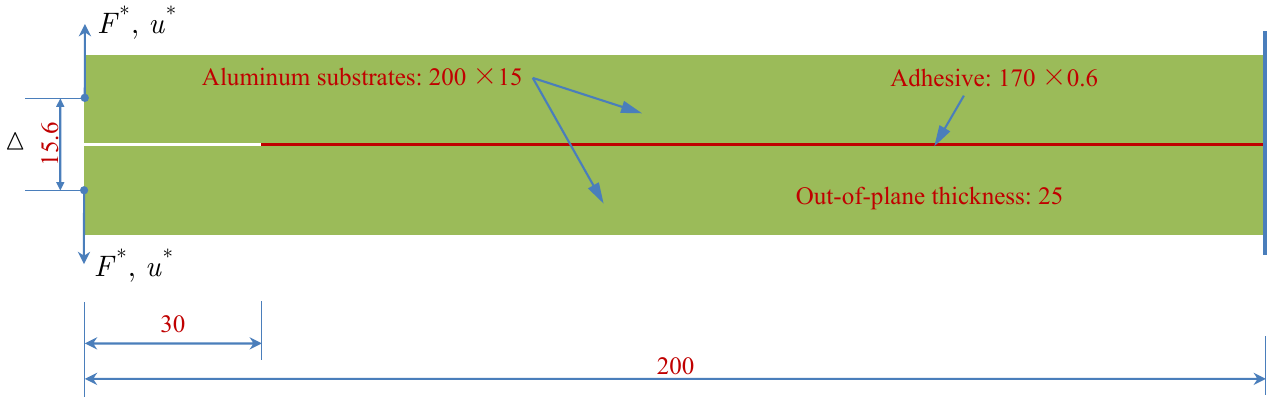}
  \caption{Double cantilever beam (DCB) test: Geometry (unit of length: mm), loading and boundary conditions.}
	\label{fig:dcb-geometry}
\end{figure}

This final example involves the double cantilever beam (DCB) test reported in \cite{AFLMP2009,AFPP2010}. The geometry, boundary and loading condition of the specimen is shown in \cref{fig:dcb-geometry}. Note that in  \cite{Wu2024}, a similar but different specimen was considered and it was shown that the associated \texttt{PF-CZM} cannot be used for cohesive fracture with concave softening curves.

The specimen consists of two aluminum alloy (AA6060-TA16) substrates partially bonded with a toughened epoxy adhesive interface. The dimensions of each aluminum substrate are 200 mm $\times$ 15 mm $\times$ 25 mm, and those of the adhesive interface are 170 mm $\times$ 0.6 mm $\times$ 25 mm, respectively. The initial notch is 30 mm long measured from the left edge of the beams. Plane strain was assumed.

The aluminum substrates were considered using a linear elastic material with Young's modulus 65.7 GPa and Poisson's ratio 0.33. The adhesive layer was modeled by the non-associated \texttt{$\mu$PF-CZM} with the following material parameters: Young's modulus $E_{0} = 1700$ MPa, Poisson's ratio $\nu_{0} = 0.35$, the failure strength $f_{\text{t}} = 10$ MPa and the fracture energy $G_{\text{f}} = 4.0$ N/mm. The Rankine failure criterion \eqref{eq:equivalent-effective-stress}$_{1}$ and the \cite{PPR2009} softening curve \eqref{eq:PPR-softening} with the exponent $m = 1.15$ were adopted for the adhesive layer. The phase-field length scale parameter $b = 0.1$ mm and the mesh size $h = 0.02$ mm were considered in the numerical simulation. 

\begin{figure}[h!] \centering
  \subfigure[$p = 1.0$]{
  \includegraphics[width=0.48\textwidth]{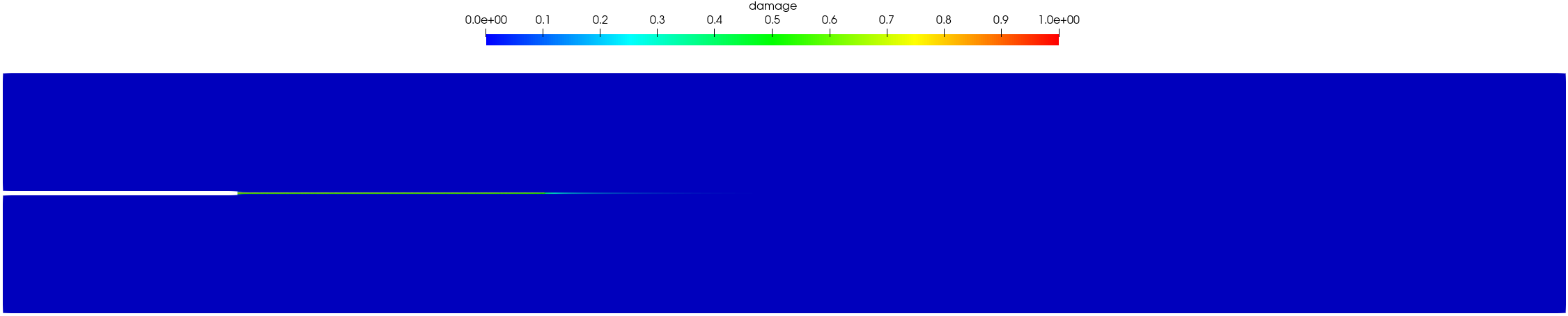}
  \label{fig:dcb-crack-profile-p10}} \hfill
  \subfigure[$p = 1.5$]{
  \includegraphics[width=0.48\textwidth]{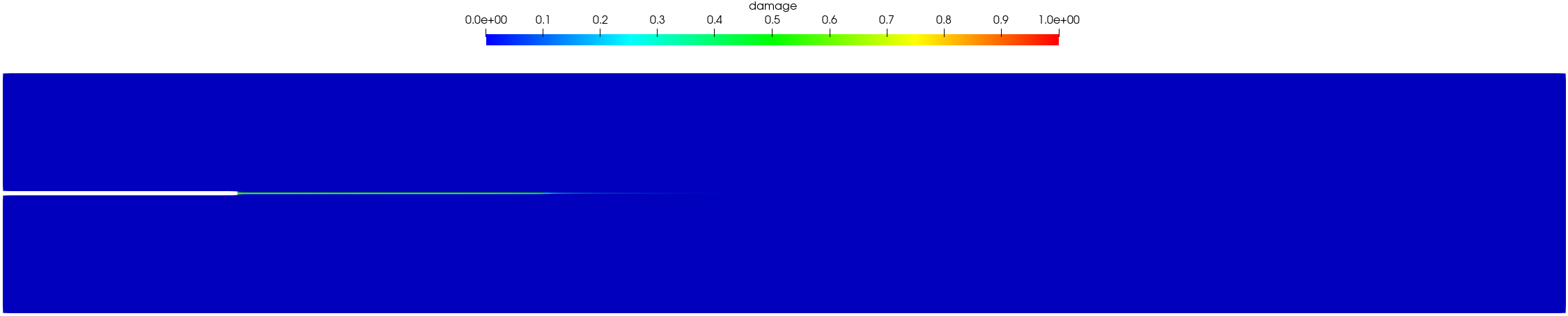}
  \label{fig:dcb-crack-profile-p15}}
  \subfigure[$p = 2.0$]{
  \includegraphics[width=0.48\textwidth]{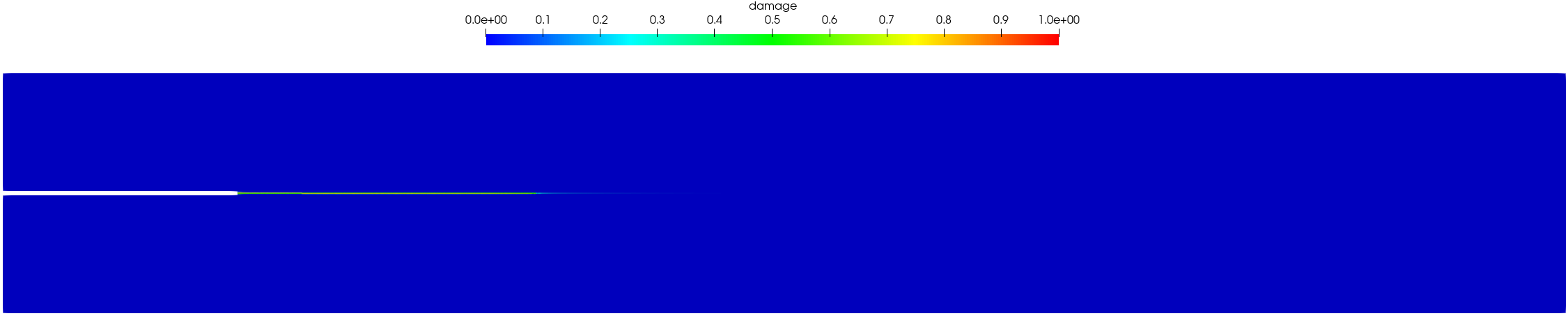}
  \label{fig:dcb-crack-profile-p20}} \hfill
  \subfigure[Zoomed crack profile around the notch]{
  \includegraphics[width=0.48\textwidth]{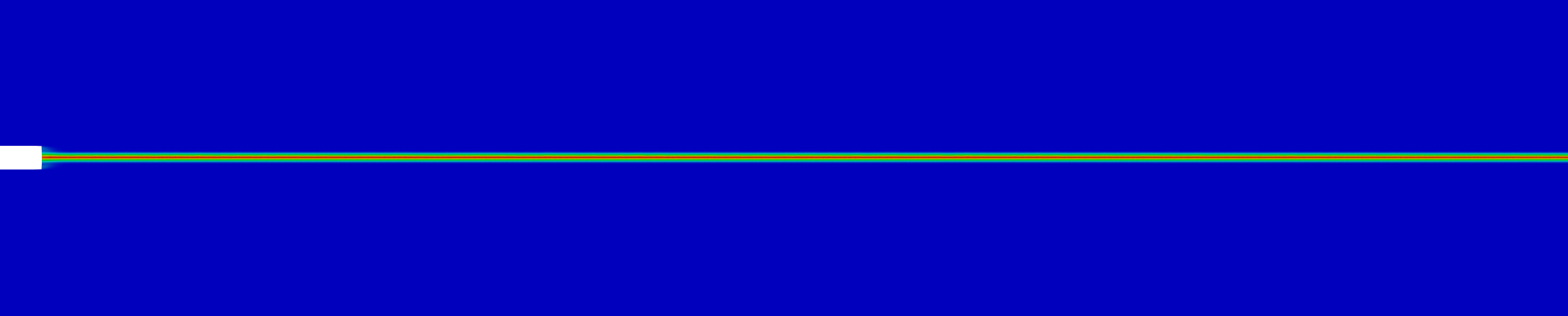}
  \label{fig:dcb-crack-profile-zoomed}}
  \caption{Double cantilever beam (DCB) test: Predicted crack profiles at CMOD = 2.5 mm.}
  \label{fig:dcb-crack-profiles}  
\end{figure}

\begin{figure}[h!] \centering
  \includegraphics[width=0.55\textwidth]{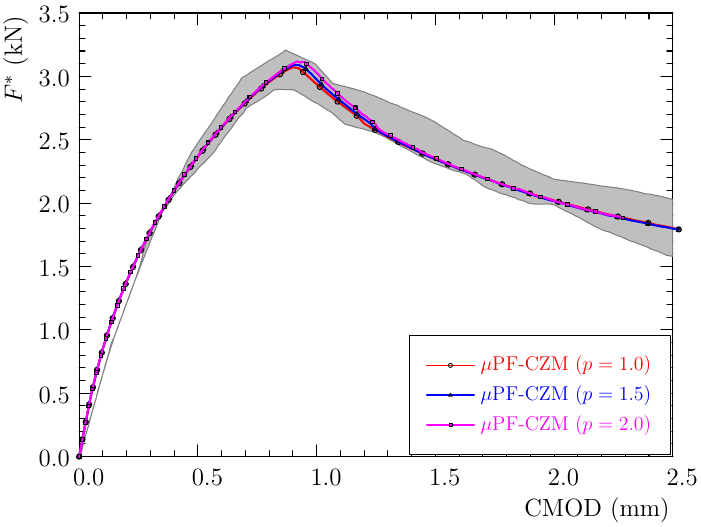}
  \caption{Double cantilever beam (DCB) test: Force--CMOD curves predicted by the non-associated $\mu$\texttt{PF-CZM}. The \cite{PPR2009} softening with $m = 1.15$ was used in the numerical simulations.}
	\label{fig:dcb-load-cmod}
\end{figure}

As shown in \cref{fig:dcb-crack-profiles}, the crack propagates horizontally within the adhesive layer between two substrates. The traction order parameter $p \ge 1$ has negligible effects on the crack profile. 

The numerical result of applied force \textit{versus} CMOD curve  is compared in \cref{fig:dcb-load-cmod} against the experimental test data. As can be seen, the numerical results from various traction order parameters almost coincide and all agree well with the test data. The insensitivity to the traction order parameter is also validated for concave softening behavior.

\section{Conclusions}
\label{sec:conclusions}

In this work unified analysis of the existing phase-field models for cohesive fracture is addressed within the extended framework of the unified phase-field theory for fracture \citep{Wu2017}. Specifically, the \cite{CFI2016,CFI2024} model and the improved versions \citep{FI2017,LCM2023}, the associated phase-field cohesive zone model (\texttt{PF-CZM}) \citep{Wu2018,WN2018} and its variants \citep{Lorentz2017,Wang2000,FFL2021}, and the non-associated generalized phase-field cohesive zone model ($\mu$\texttt{PF-CZM}) \citep{Wu2024} are systematically discussed. From the discussion the following conclusions can be drawn:
\begin{itemize}
\item All these models belong to the phase-field regularized counterpart of the \cite{Barenblatt1959} cohesive zone model (CZM) by means of only the displacement field and the crack phase-field. They are distinguished by the characteristic functions involved in the formulation, i.e., the geometric function for the crack profile, the degradation function for the constitutive relation and the dissipation function for the crack driving force. The latter two characteristic functions coincide in the associated formulation, while in the non-associated one they are distinct. This distinction decouples the traction--separation softening law from the crack bandwidth, greatly facilitating the determination of the degradation function.

\item In order for a phase-field model to be applicable to cohesive fracture, the length scale parameter has to be properly incorporated into the dissipation and/or degradation functions. Upon this incorporation, the resulting phase-field model converges to the \cite{Barenblatt1959} CZM for a vanishing length scale parameter $b \to 0$, with both the failure strength and the traction--separation softening curve being well-defined. Moreover, the resulting crack bandwidth needs to be non-decreasing during failure such that the target traction--separation softening law is recovered upon imposition of the crack irreversibility condition.


\item The \cite{CFI2016,CFI2024} model, to the best knowledge of the author, might be the first phase-field model for cohesive fracture with the mathematical proof of $\varGamma$-convergence available. With the degradation function being proportional to the length scale parameter, it converges in 1D case to the \cite{Barenblatt1959} CZM with a particular nonlinear softening curve. However, due to the truncation in the degradation function, the resulting failure strength of finite value can only be achieved for a vanishing length scale $b \to 0$. This deficiency not only limits its dealing with crack nucleation but also demands extremely fine mesh in the numerical simulation.

\item The aforesaid issue of the \cite{CFI2016,CFI2024} model is removed by a particular version of the associated \texttt{PF-CZM}. Specifically, a continuous degradation/dissipation function of rational fraction, with its limit upon a vanishing length scale coincident with the one adopted in the \cite{CFI2016,CFI2024} model, is introduced. In this way, the truncation is no longer needed and the $\varGamma$-convergence is preserved. Accordingly, the failure strength is independent of the incorporated length scale parameter and the traction--separation softening curve coincides with the limiting one of the \cite{CFI2016,CFI2024} model. The in-between correspondence justifies this particular version of the \texttt{PF-CZM}, though only a special softening curve can be considered.

\item More general versions of the \texttt{PF-CZM} are discussed in a unified framework. With respect to the manner how the degradation function is determined, two groups of \texttt{PF-CZM}s are identified. 

In the first group, only the associated formulation with identical degradation and dissipation functions is discussed. The dissipation/degradation function is assumed \textit{a priori} to be of rational fraction characterized by a parameterized polynomial. The involved parameters are determined from the failure strength, the initial slope and the ultimate crack opening in closed-form. Moreover, it is found that the geometric function $\alpha (d) = 2d - d^{2}$ is optimal in guaranteeing a non-shrinking crack band for the linear and general convex softening curves. 

In the second group, the degradation function is solved analytically upon a simple relationship between the dissipation and geometric functions. Both the associated and non-associated formulations are considered. For the associated formulation, all the involved characteristic functions are determined in closed-form. Nevertheless, the so-determined geometric function cannot always guarantee the condition for a non-shrinking crack band, leading to the incapability of reproducing the anticipated traction--separation softening curve upon imposition of the crack irreversibility condition. Comparatively, for the non-associated formulation only the degradation function is solved analytically from the given softening curve independent of the geometric function. In this case the geometric function $\alpha (d) = 2d - d^{2}$ is also optimal for any arbitrary softening curve and any traction order parameter $p \ge 1$.

\item Representative numerical examples indicate that the associated \texttt{PF-CZM} and the non-associated $\mu$\texttt{PF-CZM}, both with the optimal geometric function, give rather close numerical predictions for cohesive fracture with linear and convex softening behavior. Not only the crack pattern but also the global response can be well captured by both models. However, the non-associated $\mu$\texttt{PF-CZM} is advantageous since it applies to almost any arbitrary softening behavior including those concave ones. Moreover, it is insensitive not only to the phase-field length scale parameter, but also to the traction order exponent.

\end{itemize}

Currently the non-associated $\mu$\texttt{PF-CZM} has been applied only to quasi-static problems. Its extension to fatigue scenario and to fracture in inelastic solids, among many others, can be considered in the forthcoming studies.

\section*{Acknowledgments}
	
This work is supported by the National Natural Science Foundation of China (52125801) and Guangdong Provincial Key Laboratory of Modern Civil Engineering Technology (2021B1212040003) to the author (J.Y. Wu).


\bibliographystyle{elsarticle-harv}
\bibliography{references}

\clearpage
\begin{appendix}

\section{Softening curves}
\label{sec:softening-curves}

In this work, the following softening curves are considered.
\begin{itemize}
\item Linear softening
\begin{align}\label{eq:linear-softening-cohesive}
	\mathcal{G} (\itw)
		= \begin{cases}
				G_{\text{f}} \dfrac{\itw}{\itw_{\text{cL}}} \Big( 2 
					- \dfrac{\itw}{\itw_{\text{cL}}} \Big) & \qquad \text{if} \;
					\itw \le \itw_{\text{cL}} \\
				G_{\text{f}} & \qquad \text{if} \; \itw \ge \itw_{\text{cL}}
			\end{cases} \qquad
  \sigma (\itw)
		= f_{\text{t}} \max \Big( 1 - \dfrac{f_{\text{t}}}{2 G_{\text{f}}} 
			\itw, 0 \Big)
\end{align}
with the following initial slope $k_{0}$ and ultimate crack opening $\itw_{\text{c}}$
\begin{align}
	k_{0}
	  = k_{\text{0L}} 
	  =-\dfrac{f_{\text{t}}^{2}}{2 G_{\text{f}}}, \qquad
	\itw_{\text{c}}
	  = \itw_{\text{cL}}
		= \dfrac{2 G_{\text{f}}}{f_{\text{t}}}	  	
\end{align}

\item Exponential softening
\begin{align}\label{eq:softening-curve-exponential}
	\mathcal{G} (\itw)
		= G_{\text{f}} \bigg[ 1 - \exp \Big(-\dfrac{f_{\text{t}}}{G_{\text{f}}} 
			\itw \Big) \bigg], \qquad
	\sigma (\itw)
		= f_{\text{t}} \exp \Big(-\dfrac{f_{\text{t}}}{G_{\text{f}}} 
			\itw \Big)
\end{align}
with the following initial slope and ultimate crack opening
\begin{align}
	k_{0}
	  =-\dfrac{f_{\text{t}}^{2}}{G_{\text{f}}}, \qquad
	\itw_{\text{c}}
		=+\infty
\end{align}


\item \cite{CHR1986} softening
\begin{align}\label{eq:softening-curve-Cornelissen}
  \sigma (\itw)
    = f_{\text{t}} \Big[ \big( 1.0 + \eta_{1}^{3} r^{3} \big) 
	    \exp \big(-\eta_{2} r \big) 
    - r \big(1.0 + \eta_{1}^{3} \big) 
	    \exp \big(-\eta_{2} \big) \Big] \qquad \text{with} \qquad
	r :
		= \itw / \itw_{\text{c}}
\end{align}
for the initial slope $k_{0}$ and ultimate crack opening $\itw_{\text{c}}$
\begin{align}
  k_{0}
    =-1.3546 \dfrac{f_{\text{t}}^{2}}{G_{\text{f}}}, \qquad
  \itw_{\text{c}}
    = 5.1361 \dfrac{G_{\text{f}}}{f_{\text{t}}}
\end{align}
where the typical values $\eta_{1} = 3.0$ and $\eta_{2} = 6.93$ have been considered for normal concrete. 

For the 6th-order polynomial fitting of this softening curve, the coefficients $\bar{c}_{n}$ in \cref{eq:cracking-function-characteristics} are given by \citep{Wu2024}
\begin{subequations}
\begin{align}
	\bar{c}_{1}
	 &= \; 101.6763, \quad
	\bar{c}_{2}
		=-40.4105, \quad
	\bar{c}_{3}
	  =-129.1615 \\
	\bar{c}_{4}
	 &=-60.6300, \quad
	\bar{c}_{5}
		= \phantom{-}30.0532, \quad
	c_{6}
		= \; -0.2668
\end{align}
\end{subequations}

\item \cite{PPR2009} softening
\begin{align}\label{eq:PPR-softening}
	\mathcal{G} (\itw)
		= G_{\text{f}} \Bigg[ 1 - \bigg(1 - \dfrac{f_{\text{t}}}{
			m G_{\text{f}}} \itw \bigg)^{m} \Bigg], \qquad
	\sigma (\itw)
		= f_{\text{t}} \Big(1 - \dfrac{f_{\text{t}}}{m G_{\text{f}}} 
			\itw \Big)^{m - 1}
\end{align}
with the initial slope $k_{0}$ and ultimate crack opening $\itw_{\text{c}}$
\begin{align}
	k_{0}
		=-\dfrac{m-1}{m} \dfrac{f_{\text{t}}^{2}}{G_{\text{f}}}, \qquad
	\itw_{\text{c}}
		= m \dfrac{G_{\text{f}}}{f_{\text{t}}}
\end{align}
where the exponent $m > 1$ controls the shape of the softening curve, i.e., convex for $m > 2$, concave for $1 < m < 2$ and linear for $m = 2$.

In order to fit this softening curve, 6th-order polynomials can be considered with the coefficients $\bar{c}_{n}$ in \cref{eq:cracking-function-characteristics} given by \citep{Wu2024}
\begin{align}
\begin{cases}
	m = 1.15: & \bar{c}_{1} =-0.851, \; \bar{c}_{2}	= 0.069, \; \bar{c}_{3} = 3.322, \; \bar{c}_{4} = 1.804, \; \bar{c}_{5} =-1.896, \; \bar{c}_{6} 	= 2.812 \\
	m = 1.25: & \bar{c}_{1} = 1.563, \; \bar{c}_{2} = 0, \; 
	\bar{c}_{3} =-0.9375, \; \bar{c}_{4} = 0.9375, \; \bar{c}_{5} = 0, \; \bar{c}_{6} = 0 
		\\
	m = 1.50: & \bar{c}_{1} = 0.75, \; \bar{c}_{2}	= 0.75, \; \bar{c}_{3} = 0, \; \bar{c}_{4} = 0, \; \bar{c}_{5}	= 0, \; \bar{c}_{6}	= 0 \\
	m = 1.75: & \bar{c}_{1} =-8.060, \; \bar{c}_{2}	= 1.774, \; \bar{c}_{3} = 14.785, \; \bar{c}_{4} =6.108, \; \bar{c}_{5} =-5.850, \; \bar{c}_{6} 	= 1.346 \\
\end{cases}
\end{align}

\end{itemize}

\section{A generic stress-based failure criterion}
\label{sec:stress-based-failure-criterion}

In \cite{WC2015} the following generic stress-based failure criterion was proposed
\begin{align}\label{eq:general-stress-failure-criterion}
	\bar{J}_{2} - \frac{1}{6} A_{0} \bar{I}_{1}^{2} 
		+ \frac{1}{3} B_{0} \bar{I}_{1} \bar{\sigma}_{\text{eq}} 
		- C_{0} \bar{\sigma}_{\text{eq}}^{2} \le 0
\end{align}
where the non-negative parameters $B_{0}$ and $C_{0}$ are given by
\begin{align}\label{eq:parameters-failure-function-stress-general}
	B_{0}
		= \frac{2 - A_{0}}{2} \big(\rho_{s} - 1 \big) \ge 0, \qquad
	C_{0}
		= \frac{2 - A_{0}}{6} \rho_{s} \ge 0
\end{align}
in terms of the parameter $A_{0} \le B_{0}^{2} / (6C_{0}) < 2$. 

On the meridian $\bar{I}_{1} - \sqrt{\bar{J}_{2}}$ plane, the failure surface \eqref{eq:general-stress-failure-criterion} defines either an ellipse for $A_{0} < 0$, a parabola for $A_{0} = 0$, a hyperbola for $0 < A_{0} < B_{0}^{2} / (6C_{0})$ and two straight lines for $A_{0} = B_{0}^{2} / (6C_{0})$, respectively. The classical von Mises criterion for $A_{0} = 0$ and $\rho_{s} = 1$, the modified von Mises one for $A_{0} = 0$ and the Drucker-Prager one for $A_{0} = 2 (\rho_{s} - 1)^{2} / (\rho_{s} + 1)^{2}$ are recovered as particular examples.

In the plane stress condition ($\sigma_{3} = 0$), only the parameters satisfying $A_{0} \le 1/2 + B_{0}^{2} / (6C_{0})$ or $A_{0} \le 2 (\rho_{s}^{2} - \rho_{s} + 1) / (\rho_{s} + 1)^{2}$ are meaningful. The resulting failure criteria can be classified into four cases
\begin{align}\label{eq:consistent-parameter-three-cases}
\begin{cases}
	A_{0} < \frac{1}{2} & \qquad \text{Elliptical function} \\
	A_{0} = \frac{1}{2} & \qquad \text{Parabolic function} \\
	\frac{1}{2} < A_{0} < 2 (\rho_{s}^{2} - \rho_{s} + 1) / ( \rho_{s} + 1)^{2} 
		& \qquad \text{Hyperbolic function} \\
	A_{0} = 2 (\rho_{s}^{2} - \rho_{s} + 1) / ( \rho_{s} + 1)^{2}
		& \qquad \text{Two straight lines}
\end{cases}
\end{align}
Note that the last case corresponds to the classical Mohr-Coulomb criterion. 


\begin{figure}[h!] \centering 		
	\includegraphics[width=0.55\textwidth]{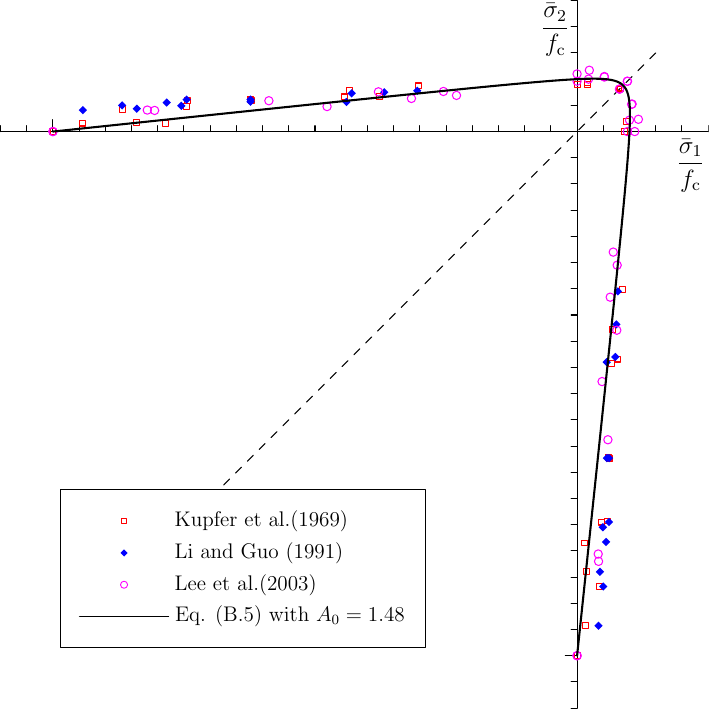}
  \caption{Biaxial strength envelope of normal concrete ($\rho_{s} = 10.0, A_{0} = 1.48$). The test data refer to \cite{KHR1969,LG1991,LSH2003}.}
	\label{fig:biaxial-strength-envolope}
\end{figure}

The equivalent uniaxial failure strength $\bar{\sigma}_{\text{eq}}$ is solved from \cref{eq:general-stress-failure-criterion} as 
\begin{align}
	\bar{\sigma}_{\text{eq}}
	  = \dfrac{\rho_{s} - 1}{2 \rho_{s}} \bar{I}_{1}
	 	+ \dfrac{1}{2 \rho_{s}} \dfrac{1}{\sqrt{2 - A_{0}}} \sqrt{ 
	 		\Big[ \big( 2 - A_{0} \big) \rho_{s}^{2}
	 	+ 2 \big( 3 A_{0} - 2 \big) \rho_{s} + 2 - A_{0} \Big] 
	 		\bar{I}_{1}^{2} + 24 \rho_{s} \bar{J}_{2} }
\end{align}
The modified von Mises criterion \eqref{eq:equivalent-effective-stress}$_{2}$ is recovered for the parameter $A_{0} = 0$. As an illustrative example, let us consider the typical strength ratio $\rho_{s} := f_{\text{c}} / f_{\text{t}} = 10$ for concrete. The biaxial strength envelope resulted from the following parameters
\begin{align}
	A_{0} 
		= 1.48 \qquad \Longrightarrow \qquad
	B_{0}
		= 2.34, \qquad
	C_{0}
		= 0.93
\end{align}
fits the test data of normal concrete under tension-tension and tension-compression rather well; see \cref{fig:biaxial-strength-envolope}.

%

\end{appendix}

\end{document}